\documentclass[11pt,epsfig]{article} 
\usepackage[utf8]{inputenc}
\usepackage[top=1in, left=0.95in, bottom=1.1in, right=0.95in]{geometry}
\usepackage[colorlinks=true,
linkcolor=blue,
urlcolor=red,
citecolor=red]{hyperref}

\usepackage[T1]{fontenc} 
\usepackage{times} 
\usepackage{mathptmx} 

\usepackage[vcentermath,enableskew]{youngtab}
\usepackage{ytableau}
\usepackage{tikz}
\usepackage{lscape}
\usepackage{tabularx}
\usepackage[toc]{appendix}
\usepackage{longtable}
\usepackage{enumerate}
\usepackage{float}
\usepackage{subfig}
\usepackage{color}
\usepackage{xcolor}
\usepackage{multirow}
\usepackage{cite}

\let\counterwithin\relax
\usepackage{chngcntr}
\usepackage{amssymb, amsmath,mathrsfs}
\usepackage{multicol,multirow}
\usepackage{ulem}
\usepackage{graphics}
\usepackage{graphicx}
\usepackage{epsf}
\usepackage{epsfig}
\usepackage{float}
\usepackage{makecell}
\usepackage{multirow}
\usepackage{color}
\usepackage{xcolor}
\usepackage{simplewick}
\usepackage{amsmath}
\usepackage{amsfonts}
\usepackage{makeidx}

\newcommand{\op}{{\cal O}}

\newcommand{\Op}{{\cal O}}		
\newcommand{\todo}[1]{{\color{red} \ifmmode\else[todo]\fi #1}}
\newcommand{\lrpartial}{\negthickspace\stackrel{\leftrightarrow}{\partial}\negthickspace{}}

\newcommand{\lrD}{\negthickspace\stackrel{\leftrightarrow}{D}\negthickspace{}}

\def\baselinestretch{1.12}

\begin{document}
\begin{center}


{\Large \textbf  {Systematic Spurion Matching between Low Energy EFT and Chiral Lagrangian}
}\\[10mm]


Chuan-Qiang Song $^{a,b,c}$\footnote{songchuanqiang21@mails.ucas.ac.cn}, Hao Sun $^{ b,c}$\footnote{sunhao@itp.ac.cn}, 
 Jiang-Hao Yu $^{a, b, c}$\footnote{jhyu@itp.ac.cn}\\[10mm]

\noindent 
$^a${ \small School of Fundamental Physics and Mathematical Sciences, Hangzhou Institute for Advanced Study, UCAS, Hangzhou 310024, China}  \\
$^b${ \small School of Physical Sciences, University of Chinese Academy of Sciences,   Beijing 100049, P.R. China}   \\
$^c${ \small Institute of Theoretical Physics, Chinese Academy of Sciences,   Beijing 100190, P. R. China} \\
[10mm]

\date{\today}   

\end{center}
\begin{abstract}

The hadronic chiral Lagrangian can be matched from the low energy effective field theory (LEFT) operators at the quark level. Traditionally, as the mass dimension of the LEFT operators increases, more and more external sources are necessarily introduced in the chiral perturbation theory (ChPT). In this work, we present a systematic matching procedure with the single spurion field $\mathbf{T}$ of the flavor $SU(3)_V$ octet, without the need of external sources. We present the complete sets of the LEFT operators, which have been reformulated using the flavor $SU(3)_V$ symmetry up to dimension 9. At the same time, the ChPT Lagrangian is also reformulated using the single spurion and lepton fields, instead of external sources. The spurion matching can be performed between LEFT and ChPT operators with the same flavor $SU(3)_V$ and $CP$ structures, and the same leptonic currents, using the naive dimensional analysis at the quark and hadronic levels.


\end{abstract}
\newpage
\setcounter{tocdepth}{3}
\setcounter{secnumdepth}{3}

\tableofcontents

\setcounter{footnote}{0}

\def\baselinestretch{1.5}
\counterwithin{equation}{section}

\newpage

\section{Introduction}

There are theoretical and experimental motivations for probing new physics (NP) beyond the standard model (SM). Usually, new physics particles are directly searched in the high-energy experiments, such as Large Hadron Collider (LHC). At the same time, given the null results of new physics signals at the LHC, many low-energy processes at the precision frontier can also be used to indirectly search for the new physics, which provides a complementary search on new physics. For example, the electron and neutrino scattering processes, the lepton number violating processes, the baryon number violating processes, etc could probe new physics up to several or even tens of TeV scale. Since these processes involve several different scales, it is necessary to utilize the tower of effective field theories (EFTs), including the standard model effective field theory (SMEFT), the low energy effective field theory (LEFT), the chiral perturbation theory (ChPT) from high to low scales, to avoid large logs between the high energy NP scale and the physical scale at low energy. In the EFT framework, the matching and running among different EFTs are needed, and among different scales, the matching between the LEFT at the quark level and the ChPT at the hadronic level is quite challenging because the degrees of freedom are very different.

For the low energy processes around GeV scales, only light quarks are involved, and all the heavy $W$ and $Z$ degrees of freedom should be integrated out. Thus the LEFT is introduced to describe non-renormalizable interactions among 5 light quarks, the leptons, the gluons, and the photon. It parametrizes various electroweak and new physics effects around/above the electroweak energy scale $\Lambda_{\text{EW}}\sim 100\text{ GeV}$ in a systematic way. The effective operators of the LEFT up to dimension-9 have been constructed~\cite{Jenkins:2017jig,Jenkins:2017dyc,Liao:2020zyx,Li:2020tsi,Murphy:2020cly}, among which there are various effective operators involved quarks including the strong interactions, the dipole momentum interactions, the leptonic weak interactions, and the non-leptonic weak interactions. In particular, the parity (P) and the charge conjugation (C) are not assumed to be conserved for the quark interactions in the weak processes. 

%

In fact, these low energy processes involve the leptonic and hadronic degrees of freedom, including meson and baryons, due to the quark condensation below the chiral symmetry breaking scale $\Lambda_\chi\sim 4 \pi f$. At such a low-energy scale, the LEFT is invalid and the appropriate theory characterizing the interaction among the quarks and the other fields such as the leptons and the photon is the ChPT~\cite{Weinberg:1968de,Weinberg:1978kz,Gasser:1983yg,Gasser:1984gg,Gasser:1987rb}. The ChPT describes the effective interactions of the hadrons composed by the 3 lightest quarks $u,d,s$ and other fields in terms of the chiral symmetry $SU(3)_\mathbf{L} \times SU(3)_\mathbf{R}$~\cite{Callan:1969sn,Coleman:1969sm}. Usually, the ChPT is utilized to describe the strong interaction, however, it can also describe the electroweak interaction via the external sources. Nowadays, the effective operators of pure meson sector have been constructed up to $\mathcal{O}(p^8)$~\cite{Gasser:1983yg,Gasser:1984gg,Wess:1971yu,Witten:1983tw,Fearing:1994ga,Bijnens:1999sh,Ebertshauser:2001nj,Bijnens:2001bb,Cata:2007ns,Bijnens:2018lez,Bijnens:2023hyv,Li:2024ghg}, and the operators of the meson-baryon sector have been constructed up to $\mathcal{O}(p^5)$~\cite{Krause:1990xc,Ecker:1995rk,Fettes:1998ud,Fettes:2000gb,Oller:2006yh,Frink:2006hx,Jiang:2016vax,Song:2024fae}, in both of which the external sources are considered.

To implement the new high-energy physics in the low-energy processes, the matching of the operators between the LEFT and the ChPT is necessary. Such a matching has been performed in various weak processes. For example, the Kaon decay is an important non-leptonic procedure violating CP, and the matching of the LEFT operators to the ChPT operators has been discussed in Ref.~\cite{Bernard:1985wf,Grinstein:1985ut,Cheng:1987dk,Kambor:1989tz,Pich:1990mw,Kambor:1992he,Cirigliano:2003gt,Akdag:2022sbn,Pich:2021yll,Cornella:2023kjq}. The neutrinoless double beta decay ($0\nu\beta\beta$) not only violates CP but also violates the lepton number. The matching between the dimension-9 LEFT operators and the ChPT operators corresponding the $0\nu\beta\beta$ has been discussed in Ref.~\cite{Savage:1998yh,Prezeau:2003xn,Graesser:2016bpz,Cirigliano:2017ymo,Cirigliano:2017djv,Cirigliano:2017tvr,Pastore:2017ofx,Cirigliano:2018yza,Cirigliano:2019vdj}. Besides, the matching of other lepton number-violating procedures has also been discussed~\cite{Weinberg:1979sa,Cvetic:2010rw,Chun:2019nwi,Liao:2019gex,He:2021mrt,Li:2021fvw,Liao:2021qfj}. In addition, the matching has also been applied to other situations such as the dark matter direct interaction~\cite{Fitzpatrick:2012ix,Cirigliano:2012pq,Bishara:2016hek,Bishara:2017pfq,Bishara:2017nnn,Brod:2017bsw,Korber:2017ery,deVries:2023sux}, the $\mu$-$e$ conversion processes~\cite{Crivellin:2017rmk,Cirigliano:2017azj,Bartolotta:2017mff,Dekens:2018pbu,Rule:2021oxe,Cirigliano:2022ekw,Haxton:2022piv,Haxton:2024lyc}, the non-standard neutrino interactions~\cite{Lindner:2016wff,AristizabalSierra:2018eqm,Farzan:2018gtr,Altmannshofer:2018xyo,Bischer:2019ttk,Hoferichter:2020osn,Du:2020dwr,Li:2024iij}, and the nuclear electric dipole moments (EDM)~\cite{deVries:2011an,Engel:2013lsa,Seng:2014pba,Pitschmann:2014jxa,Bsaisou:2014zwa,Bsaisou:2014oka,deVries:2015gea,Seng:2016pfd,deVries:2020iea,Froese:2021civ}.

The common-used technique for the matching is the external source method, in which the lepton bilinear or the photon field is identified as a scalar, pseudo scalar, vector, axial vector, or tensor source in the LEFT Lagrangian
\begin{equation}
\begin{split}
\label{external}
{\cal L}={\cal L}_{\rm QCD}^{0}+\bar q  \gamma^\mu \big[v_\mu+\gamma_5 a_\mu\big] q-\bar q \big[ s-i \gamma_5 p\big] q+\bar q\sigma_{\mu\nu}\Bar{t}^{\mu\nu}q \,.
\end{split}
\end{equation}
Up to dimension 6, these external sources are related to the charged leptons or the photon as~\cite{Bishara:2016hek}
\begin{align}
\label{eq:external_sources_1}
    v^\mu \rightarrow& A^\mu + C_1\bar e\gamma^\mu e+ C_2\bar e\gamma^5\gamma^\mu e\,,\quad a^\mu \rightarrow C_3\bar e\gamma^\mu e+C_4\bar e\gamma^5\gamma^\mu e\,,\quad\notag\\ 
    s \rightarrow& C_5\bar ee+C_6\bar e\gamma^5 e\,,\quad p \rightarrow C_7\bar e e+C_8\bar e\gamma^5 e \,,\quad \bar t^{\mu\nu}\rightarrow C_9\bar e\sigma^{\mu\nu}e+C_{10}\bar e\gamma^5\sigma^{\mu\nu}e\,,
\end{align}
where $C_i$ are the Wilson coefficients.
For example, both the dimension-6 operators $(\bar q\gamma^\mu q)(\bar e\gamma_\mu e)$ and $(\bar q\gamma^\mu q)(\bar e\gamma^5\gamma_\mu e)$ can contribute to the vector external source and we use the different Wilson coefficients $C_1$ and $C_2$ to distinguish them. When considering the higher dimensional operators of the LEFT, the external source method encounters problems. There will be more external sources for the higher-dimension LEFT operators, and the existence of the derivatives makes the matching more complicated. For example, the dimension-7 and dimension-8 LEFT operators 
\begin{equation}
    (\Bar{q}\lrpartial^\mu q)(\Bar{e}\gamma_\mu e)\,,\quad(\Bar{q}\lrpartial^\mu\gamma^\nu q)(\Bar{e}\lrpartial_\mu\gamma_\nu e)\,,
\end{equation}
as well as the dimension-9 operators with more than 1 quark bilinear such as $(\bar q q)(\bar qq)(\bar e e^c)$ can not be recognized in the Lagrangian in Eq.~\eqref{external}, which means we need to define more and more external sources when considering higher-dimension operators. The five types of the external sources in Eq.~\eqref{eq:external_sources_1} are matched to the ones in the ChPT
\begin{align}
u_\mu^\prime &\sim i\{u^\dagger[\partial_\mu -i(v_\mu+a_\mu)]u-u[\partial_\mu -i(v_\mu-a_\mu)]u^\dagger\}  \,,\label{eq:ext_1_int} \\
    \chi^+ &\sim u^\dagger (s+ip)u^\dagger + u (s+ip)^\dagger u \,,\quad
    \chi^- \sim u^\dagger (s+ip)u^\dagger - u (s+ip)^\dagger u \,,\\
    t_{\mu\nu}^+&\sim u^\dagger \Bar{t}_{\mu\nu}u^\dagger + u {\Bar{t}_{\mu\nu}}^\dagger u\,,\quad t_{\mu\nu}^-\sim u^\dagger \Bar{t}_{\mu\nu}u^\dagger - u {\Bar{t}_{\mu\nu}}^\dagger u\label{eq:ext_2_int} 
\end{align}
where $\chi^\pm\,,u_\mu^\prime\,,t_{\mu\nu}^+$ are the ChPT building blocks. The operators are the invariants composed of these building blocks, and some relations are used to reduce the redundant ones. Such a matching is dependent on the redundant relations of the LEFT operators such as the equation of motion (EOM) and the integrating-by-part (IBP) and makes the power counting scheme unclear. As for the matching of the ChPT operators beyond the leading order with more derivatives and hadrons, there will be considerable relations such as the Cayley-Hamilton relations, which makes it difficult to obtain the independent hadron operators.

In this paper, we would propose a new spurion method to do the matching from the LEFT operators to the ChPT operators generally without consulting the external sources, by which the matching of higher-dimension operators is systematic. The main idea is that we consider the leptons and the photon as degrees of freedom of both the LEFT and the ChPT instead of embedding them in the external sources, at the same time, we parameterize the LEFT operators by adding only one spurion $\mathbf{T}$ and make the LEFT Lagrangian formally invariant under the symmetry $SU(3)_V$. The Lagrangian of the two methods are equivalent, and the external sources of dimension 6 in Eq.~\eqref{eq:external_sources_1} and the spurions are related by
\begin{align}
v^\mu &= \mathbf{T}(A^\mu+C_1\overline{e}\gamma^\mu e+C_2\overline{e}\gamma^5\gamma^\mu e)   \,,\\
     a^\mu &= \mathbf{T}(C_3\overline{e}\gamma^\mu e+C_4\overline{e}\gamma^5\gamma^\mu e)   \,,\\
    s &= \mathbf{T}(C_5\overline{e}e+C_6\overline{e}\gamma^5e)  \,,\\
    p &= \mathbf{T}(C_7\overline{e}e+C_8\overline{e}\gamma^5e)  \,,\\
    \overline{t}^{\mu\nu} &= \mathbf{T} (C_9\overline{e}\sigma^{\mu\nu} e+C_{10}\overline{e}\gamma^5\sigma^{\mu\nu} e)  \,,
\end{align}
The leptons and the photon have been extracted from the external sources and have been regarded as independent degrees of freedom as argued before. 
On the other hand, the effective operators of the ChPT can also be constructed by the $SU(3)_V$ symmetry~\cite{Gasser:1983yg,Gasser:1984gg,Wess:1971yu,Witten:1983tw,Fearing:1994ga,Bijnens:1999sh,Bijnens:2001bb,Ebertshauser:2001nj,Bijnens:2018lez,Bijnens:2023hyv,Li:2024ghg,Krause:1990xc,Ecker:1995rk,Fettes:1998ud,Fettes:2000gb,Song:2024fae,Oller:2006yh,Frink:2006hx,Jiang:2016vax} and introducing the spurion with different dressing forms, 
\begin{align}
    \Sigma_{\pm} &= u^\dagger \textbf{T} u^\dagger \pm u \textbf{T}^\dagger u \,,\notag\\
    Q_{\pm}&=u^\dagger\textbf{T}u\pm u\textbf{T}^\dagger u^\dagger\,,
\end{align}
which are related to the external sources from Eq.~\eqref{eq:ext_1_int} to Eq.~\eqref{eq:ext_2_int} by
\begin{align}
u_\mu^\prime &\sim i(u^\dagger\partial_\mu -u-u\partial_\mu u^\dagger )+Q_+(C_3\overline{e}\gamma_\mu e+ C_4\overline{e}\gamma^5\gamma_\mu e)+Q_-(A_\mu+C_1\overline{e}\gamma_\mu e+ C_2\overline{e}\gamma^5\gamma_\mu e) \,,\label{eq:spurion_redef_1_int}\\
    \chi^+ &\sim \Sigma_+(C_5\overline{e}e+C_6\overline{e}\gamma^5e) + i \Sigma_- (C_7\overline{e}e+C_8\overline{e}\gamma^5e)\,,\\
    \chi^- &\sim \Sigma_-(C_5\overline{e}e+C_6\overline{e}\gamma^5e) + i \Sigma_+ (C_7\overline{e}e+C_8\overline{e}\gamma^5e)\,,\\
    t^{\mu\nu}_+ &\sim \Sigma_+(C_9\overline{e}\sigma^{\mu\nu} e+C_{10}\overline{e}\gamma^5\sigma^{\mu\nu} e)\,,\\
    t^{\mu\nu}_- &\sim \Sigma_-(C_9\overline{e}\sigma^{\mu\nu} e+C_{10}\overline{e}\gamma^5\sigma^{\mu\nu} e)\label{eq:spurion_redef_2_int}\,.
\end{align}
Therefore, in this proposed spurion method, the LEFT Lagrangian respects the $SU(3)_V$ symmetry and is organized in terms of a single spurion $\mathbf{T}$ of the $SU(3)_V$ adjoint representation.
The Lagrangian is then matched to the ChPT operators by mapping all the quarks to the ChPT building blocks simultaneously.

The spurion method is general and has been used in other theories such as the Higgs effective field theory~\cite{Sun:2022ssa,Sun:2022snw}. Even in the matching of the LEFT and the ChPT, similar methods have been discussed in Ref.~\cite{Savage:1998yh,Prezeau:2003xn,Graesser:2016bpz,Cirigliano:2017ymo,Liao:2019gex,He:2021mrt,Akdag:2022sbn}, in which the leptons or the photon are extracted from the external sources and the remaining Wilson coefficients are prompted to be spurions. However, there are three important differences from the spurion method we propose here.

Firstly, in the old-fashioned spurion method, the adopted symmetry is $SU(3)_\mathbf{L}\times SU(3)_\mathbf{R}$, and the spurions introduced there are of different irreducible representations of the group $SU(3)_\mathbf{L}\times SU(3)_\mathbf{R}$, thus for the high-dimension operators with more quarks, more spurions would arise due to the decompositions of the representation tensor products. 
Take the quark bilinear as example, the quark bilinear has the following group decompositions:
\begin{eqnarray}
(\mathbf{3}_L\,,\mathbf{1}_R) \otimes (\mathbf{3}_L\,,\mathbf{1}_R) = 
(\mathbf{1}_L\,,\mathbf{1}_R) \oplus (\mathbf{8}_L\,,\mathbf{1}_R), \quad 
(\mathbf{3}_L\,,\mathbf{1}_R) \otimes (\mathbf{1}_L\,,\mathbf{3}_R)
(\mathbf{3}_L\,,\mathbf{3}_R),
\end{eqnarray}
which indicates that there would exist at least three spurions: $(\mathbf{3}_L\,,\overline{\mathbf{3}}_R)$, $(\mathbf{8}_L\,,\mathbf{1}_R)$, and $(\mathbf{1}_L\,,\mathbf{8}_R)$.
For the operators with 4 or more quarks, there would be more spurions of $(\mathbf{8}_L\,,\mathbf{8}_R)$, $(\mathbf{27}_L\,,\mathbf{1}_R)$, for example, the two quark bilinear has 
\begin{eqnarray}
(\mathbf{8}_L\,,\mathbf{1}_R) \otimes (\mathbf{8}_L\,,\mathbf{1}_R) = (\mathbf{27},\mathbf{1})\oplus(\mathbf{10},\mathbf{1})\oplus(\overline{\mathbf{10}},\mathbf{1})\oplus4\times(\mathbf{8},\mathbf{1})\oplus2\times(\mathbf{1},\mathbf{1}),
\end{eqnarray}
and so on, which causes complications similar to the external source method. 

On the other hand, in our new spurion technique, these spurions introduced above can be reduced to a single spurion of the $SU(3)_V$ adjoint representation $\mathbf{T}$, which can be realized by the product of singlet and octet representations of the $SU(3)_V\subset SU(3)_\mathbf{L}\times SU(3)_\mathbf{R}$, for example,
\begin{align}
    \text{2 quarks:} \quad& (\mathbf{3}_L\,,\overline{\mathbf{3}}_R) \rightarrow \mathbf{1} \oplus \mathbf{8}\,, \notag \\
    \text{4 quarks:} \quad& (\mathbf{3}_L\,,\overline{\mathbf{3}}_R)\otimes (\mathbf{3}_L\,,\overline{\mathbf{3}}_R) \rightarrow \mathbf{1} \oplus \mathbf{8} \oplus \mathbf{8} \oplus (\mathbf{8}\otimes \mathbf{8}) \,,\label{eq:spurion_lr_method}
\end{align}
Here if we do not expand the $(\mathbf{8}\otimes \mathbf{8})$ further, all the quark bilinear can be written by two building blocks: the trivial $1$ and the spurion $\mathbf{T}$.
Similar decomposition exist for the operators with any number of bilinear quarks: we can always decompose product of quark bilinear into direct sum of the product of the trivial $1$ and the spurion $\mathbf{T}$. The operators written in this way (organized by the $SU(3)_V$ symmetry) is equivalent to the operators written by $SU(3)_\mathbf{L}\times SU(3)_\mathbf{R}$ symmetry in literature~\footnote{This can be seen once the spurions take their vacuum expectation values, and would be discussed in App.~\ref{app:comparison}.}. 



Secondly, in the old-fashioned spurion method, the matching of quarks to hadrons is one by one as 
\begin{align}
    &\bar q_{\mathbf{L}}\rightarrow u^\dagger\,, q_{\mathbf{L}}\rightarrow u\,,\bar q_{\mathbf{R}}\rightarrow u\,,q_{\mathbf{R}}\rightarrow u^\dagger\,,\notag\\
    &\bar q_{\mathbf{L}}\rightarrow D^\mu u^\dagger\,, q_{\mathbf{L}}\rightarrow D^\mu u\,,\bar q_{\mathbf{R}}\rightarrow D^\mu u\,,q_{\mathbf{R}}\rightarrow D^\mu u^\dagger\,,\label{eq:correspondence_1}\\
    & \dots\,, \notag 
\end{align}
where $u=\text{exp}\left (\frac{i\Pi}{\sqrt{2}f}\right)$, $f$ is the decay constants, and $\Pi$ is hadron degrees of freedom. Because $u$ is not the fundamental building block of the ChPT, the constructions of the invariant operators are complicated, and in many cases, this correspondence could give rise to wrong results.  
In our proposed spurion method, the matching is done in the operator level, instead of the field level in eq.~\eqref{eq:correspondence_1}. We have reformulated the LEFT and the ChPT, by the $SU(3)_V $ symmetry, and in the spurion $\mathbf
T$, then we can establish the matching rules that keep the spurions, the non-quark fields, and the CP property at both the quark and the hadronic level, while replace the quarks with the building blocks of the ChPT, then constructing the independent operators from them by more efficient methods such as the Young tensor method~\cite{Li:2020gnx,Li:2022tec,Li:2020xlh}.

Finally as mentioned above, different from old-fashioned spurion method, the non-quark fields are extracted out from the non-quark current at the quark level, or external sources at the hadronic level. These non-quark fields are treated as the building blocks of both the LEFT and the ChPT. Therefore the non-quark matching are thus one-to-one. 

\begin{figure}[htbp]
    \begin{centering}
    \includegraphics[scale=0.5]{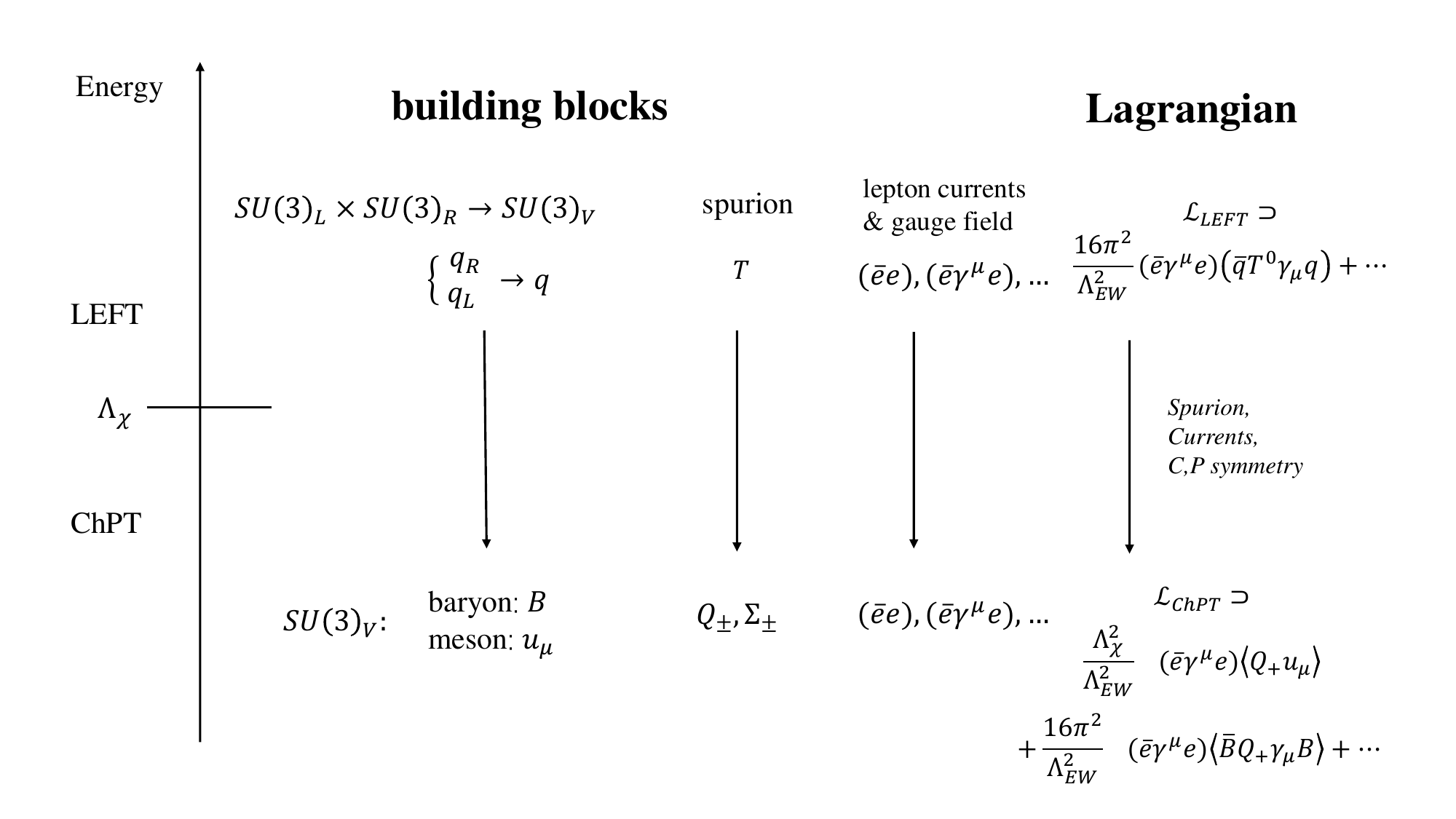}
    \caption{A diagrammatic illustration of the spurion method in this paper. }
    \label{fig:illustration}
    \end{centering}  
\end{figure}

Because of the above three differences, the matching between LEFT and ChPT is more systematic by power counting rules. 
We use the naive dimension analysis (NDA)~\cite{Manohar:1983md,Gavela:2016bzc,Jenkins:2013sda,Panico:2015jxa,Buchalla:2013eza} to assess the importance of the ChPT operators from the LEFT operators via the matching, which is consistent with both the chiral power-counting scheme of the ChPT and the canonical power-counting scheme of the LEFT, and makes the matching of high-dimension operators systematic. The matching NDA formulae is 
\begin{align}
     \frac{\Lambda_{\text{EW}}^4}{16\pi^2}\left[\frac{\partial}{\Lambda_{\text{EW}}}\right]^{N_p} \left[\frac{4\pi F}{\Lambda_{\text{EW}}^2}\right]^{N_F} \left[\frac{4\pi \psi}{\Lambda_{\text{EW}}^{3/2}}\right]^{N_\psi} \notag \sim \left[\frac{\Lambda_\chi}{\Lambda_{\text{EW}}}\right]^{\mathcal{D}} \left(f^2\Lambda_\chi^2 \left[\frac{\partial}{\Lambda_\chi}\right]^{N_p} \left[\frac{\psi}{f\sqrt{\Lambda_\chi}}\right]^{N_\psi} \left[\frac{F}{\Lambda_\chi f}\right]^{N_A} \right)\,, \label{eq:NDA_2}
\end{align}
where $\Lambda_{\rm EW}$ and $\lambda_\chi$ are the electroweak symmetry and QCD chiral symmetry breaking scales.
By these matching rules, every LEFT operator has counterparts in the ChPT. In particular, the forms of the high-dimension operators of the LEFT are not unique because of the redundancies such as the EOM and the IBP. These redundancy relations should not affect the matching results of the ChPT operators, which means that the equivalent LEFT operators related by these redundancies should matched to the same ChPT operators. 
Overall, this matching procedure has been illustrated in Fig.~\ref{fig:illustration}. 

Let us finally summarize the comparison. Compared to the external source method and the old-fashioned spurion method, we outline the advantages of the spurion method here
\begin{itemize}
    \item Only one spurion of the $SU(3)_V$ adjoint representation is needed in this spurion method, while in other two methods, the external sources or the spurions are of different forms for different operators. As the higher dimensional LEFT operators considered, there would be infinite kinds of external sources or spurions.
    \item Utilizing the spurion, the matching from the high-dimension LEFT operators to the high-dimension ChPT operators is systematic. At the same time, the NDA can be used to organize different order of contributions. Thus the equivalences such as the IBP and EOM that relate different LEFT operators to each other can be illustrated transparently after considering NDA.
    \item Given all the spurion and non-quark fields, the complete set of the LEFT and ChPT operators can be obtained systematically via the Young tensor method~\cite{Li:2020gnx,Li:2022tec,Li:2020xlh}, rather than reducing a redundant set by complicated relations such as the Cayley-Hamilton relations.
\end{itemize}
Therefore, this systematic spurion method can be used in the matching of the LEFT and the ChPT operators up to arbitrary high dimensions.

This paper is organized as follows. In Sec.~\ref{sec:LEFT} we reformulate the LEFT operators by the spurion and set our notations. In particular, the equivalence of the $SU(3)_V$ symmetry adopted here and the conventional $SU(3)_{\mathbf{L}}\times SU(3)_{\mathbf{R}}$ symmetry is highlighted, of which the detailed discussion is presented in App.~\ref{app:comparison}. In Sec.~\ref{sec:chpt}. we review the ChPT and present the effective operator basis by spurions instead of the external sources, the equivalence between which is highlighted. Then we lay out the matching rules in Sec.~\ref{sec:matching} . We present some examples and list the matching results from the LEFT operators up to dimension-9 in Sec.~\ref{sec:matching}. Finally, we conclude the paper in Sec.~\ref{sec:con}.

\section{The Low-Energy EFT with $SU(3)_V$ Spurion}
\label{sec:LEFT}

The Low-Energy EFT (LEFT) can describe the theory that below the electroweak scale $\Lambda_\text{EW}$ with an expansion in powers of the inverse $1/\Lambda_{\text{EW}}$. The LEFT operators are obtained by integrating out the heavy degrees of freedom, thus there are only 5 types of the quark fields and 6 types of the lepton fields in the LEFT operators. Then the LEFT operators can be organized as
\begin{equation}
\mathcal{L}_{LEFT}=\sum_{d,a}\Tilde{\mathcal{C}}_a^{(d)}\mathcal{O}_a^{(d)}\,,\quad\Tilde{\mathcal{C}}_a^{(d)}=\frac{\mathcal{C}_a^{(d)}}{{\Lambda_{\text{EW}}}^{d-4}}\,,
\end{equation}
where $d$ is the canonical dimension of the effective operators and ${\mathcal{C}}_a^{(d)}$ is the dimensionless Wilson coefficients. 

The effective Lagrangian of the LEFT has been obtained up to dimension-9 ($1/\Lambda_{\text{EW}}^5$)~\cite{Jenkins:2017jig,Jenkins:2017dyc,Liao:2020zyx,Li:2020tsi}, among which only the ones with the 3 lightest quarks $u,d,s$ are of our interest. These effective operators are referred to as relevant operators in this paper sometimes.
With these relevant operators, we can introduce a spurion to re-organize them in terms of the $SU(3)_V$ rather than the $SU(3)_\mathbf{L}\times SU(3)_\mathbf{R}$. Such a reorganization is helpful for the matching to the ChPT operators, while the two organizations of different symmetries are equivalent, which is illustrated by some examples in App.~\ref{app:comparison}. 

\subsection{The $SU(3)_V$ Spurion Field}

When matching to the ChPT, only the three lightest $u$, $d$, $s$ quarks are needed since $\Lambda_{\chi}\gg m_u\,,m_d\,,m_s$ and the lepton part and the gauge boson part are the same as the LEFT. The relevant leading-order (LO) Lagrangian is
\begin{align}
    \mathcal{L}_{LEFT}^{(4)} \supset \sum_{q=u,d,s} \left(i\overline{q}_\mathbf{L}\gamma^\mu D_\mu q_\mathbf{L} + i\overline{q}_\mathbf{R}\gamma^\mu D_\mu q_\mathbf{R} + \overline{q}_\mathbf{L} m_q q_\mathbf{R} + \overline{q}_\mathbf{R} m_q^\dagger q_\mathbf{L} \right) \,. \label{eq:left_lagrangian_1}
\end{align}
To make the chiral symmetry $SU(3)_\mathbf{L}\times SU(3)_\mathbf{R}$ transparent we define the quark triplet as 
\begin{equation}
\label{eq:lrquarks}
    q_\mathbf{L}=\left(\begin{array}{c}
         u_\mathbf{L}  \\
         d_\mathbf{L} \\
         s_\mathbf{L}
    \end{array}\right)\,,\quad q_\mathbf{R}=\left(\begin{array}{c}
         u_\mathbf{R}  \\
         d_\mathbf{R} \\
         s_\mathbf{R}
    \end{array}\right)\,,\quad \bar q_\mathbf{L}=\left(\bar u_\mathbf{L},\,\bar d_\mathbf{L},\,\bar s_\mathbf{L}\right)\,,\quad\bar q_\mathbf{R}=\left(\bar u_\mathbf{R},\,\bar d_\mathbf{R},\,\bar s_\mathbf{R}\right)\,,
\end{equation}
and let the left- and right-handed quark triplets transform as
\begin{equation}
    q_\mathbf{L}\rightarrow \mathbf{L}q_\mathbf{L}\,,\quad q_\mathbf{R}\rightarrow \mathbf{R}q_\mathbf{R}\,\quad \text{where }\,\mathbf{L}\in SU(3)_\mathbf{L}\,\text{ and }\,\mathbf{R}\in SU(3)_\mathbf{R}\,,
\end{equation}
under the chiral symmetry $SU(3)_\mathbf{L}\times SU(3)_\mathbf{R}$, then the LO Lagrangian becomes
\begin{align}
    \mathcal{L}_{LEFT}^{(4)} \supset \left(i\overline{q}_\mathbf{L}\gamma^\mu D_\mu q_\mathbf{L} + i\overline{q}_\mathbf{R}\gamma^\mu D_\mu q_\mathbf{R} + \overline{q}_\mathbf{L} M_q q_\mathbf{R} + \overline{q}_\mathbf{R} M_q^\dagger q_\mathbf{L} \right) \notag \,,
\end{align}
where $M_q = \text{diag}(m_u,m_d,m_s)$. This Lagrangian is invariant under the $SU(3)_\mathbf{L}\times SU(3)_\mathbf{R}$ if $M_q$ transforms as $M_q \rightarrow \mathbf{L} M_q \mathbf{R}^\dagger$, which means $M_q$ becomes a  spurion field.

Furthermore, we want to parameterize the Lagrangian via $SU(3)_V$ instead of $SU(3)_\mathbf{L}\times SU(3)_\mathbf{R}$, for which parity transformation is needed. The parity of the LEFT operator can link the left- and right-handed quarks, then we can define that
\begin{equation}
    u=\left(\begin{array}{c}
         u_\mathbf{L}  \\
         u_\mathbf{R} 
    \end{array}\right)\,,\quad d=\left(\begin{array}{c}
         d_\mathbf{L}  \\
         d_\mathbf{R} 
    \end{array}\right)\,,\quad s=\left(\begin{array}{c}
         s_\mathbf{L}  \\
         s_\mathbf{R} 
    \end{array}\right)\,,
\end{equation}
and the three quarks would also be combined as
\begin{equation}
\label{eq:trans_r}
    q=u\oplus d\oplus s=(u_\mathbf{L},u_\mathbf{R},d_\mathbf{L},d_\mathbf{R},s_\mathbf{L},s_\mathbf{R})^{\rm T}\simeq q_\mathbf{L}\oplus q_\mathbf{R}\,,
\end{equation}
where the last equivalence is realized by a similar transformation
\begin{equation}
\label{eq:R_tras}
    \mathcal{R}\cdot q = q_\mathbf{L}\oplus q_\mathbf{R}\,,
\end{equation}
with the transformation matrix
\begin{equation}
\label{eq:R}
    \mathcal{R} = \left(
\begin{array}{cccccc}
 1 & 0 & 0 & 0 & 0 & 0 \\
 0 & 0 & 0 & 1 & 0 & 0 \\
 0 & 1 & 0 & 0 & 0 & 0 \\
 0 & 0 & 0 & 0 & 1 & 0 \\
 0 & 0 & 1 & 0 & 0 & 0 \\
 0 & 0 & 0 & 0 & 0 & 1 \\
\end{array}
\right)\,.
\end{equation}
In particular, the inverse transformation is
\begin{equation}
    q_L = \frac{1-\gamma^5}{2}\mathcal{R}\cdot q\,,\quad q_R = \frac{1+\gamma^5}{2}\mathcal{R}\cdot q\,.
\end{equation}
With parity considered, the $SU(3)_\mathbf{L}\times SU(3)_\mathbf{R}$ symmetry of the Lagrangian breaks down to the $SU(3)_V$ symmetry,
\begin{equation}
    q \rightarrow V q\,,\quad V\in SU(3)_V \subset SU(3)_\mathbf{L}\times SU(3)_\mathbf{R}\,,
\end{equation}
thus we can reorganize the LEFT operators by the $SU(3)_V$ symmetry directly. In terms of $q$, the LO Lagrangian becomes
\begin{align}
    \mathcal{L}_{LEFT}^{(4)} \supset \left(i\overline{q}\gamma^\mu D_\mu q + \overline{q} \mathcal{M}_q q\right) \,,
\end{align}
where $\mathcal{R} \mathcal{M}_q \mathcal{R}^{-1} = M^\dagger_q\oplus M_q$. This Lagrangian is invariant under the $SU(3)_V$ if the mass matrix $\mathcal{M}_q$ transforms as $\mathcal{M}_q\rightarrow V \mathcal{M}_q V^\dagger$, which means it can be decomposed into the trivial representation and the adjoint representation of the $SU(3)_V$
\begin{equation}
    \mathcal{M}_q \in \mathbf{1} \oplus \mathbf{8}\,,
\end{equation}
thus the mass matrix $\mathcal{M}_q$ can be divided into a trace term and a traceless term
\begin{equation}
    \mathcal{M}_q = \frac{1}{3}\text{Tr}(\mathcal{M}_q) I + \overline{\mathcal{M}}_q\,,
\end{equation}
where $\overline{\mathcal{M}}_q$ is the traceless part of the adjoint representation of the $SU(3)_V$ and is the spurion field to form formally invariant Lagrangian under the $SU(3)_V$. 
Thus we introduce the spurion that
\begin{equation}
\label{eq:spu_intro_new}
    \mathbf{T} \rightarrow V \mathbf{T} V^\dagger \,,\quad V \in SU(3)_V\,,
\end{equation}
and the LO Lagrangian becomes
\begin{equation}
\label{eq:left_lagrangian_2}
    \mathcal{L}_{LEFT}^{(4)} \supset \left(i\overline{q}\gamma^\mu D_\mu q + \frac{1}{3}\text{Tr}(\mathcal{M}_q)\overline{q}  q + m\overline{q}\mathbf{T} q\right)\,,
\end{equation}
where $m$ is needed to compensate the mass dimension so that the spurion $\mathbf{T}$ is dimensionless.

The vacuum expectation value (VEV) of the spurion $\mathbf{T}$ is just the Wilson coefficient and can be expanded by the Gell-mann matrices
\begin{equation}
\label{eq:mass_expansion}
    \mathbf{T} \xrightarrow{\text{VEV}} \sum_{a=1}^8 c_a \lambda^a\,,
\end{equation}
where the Gell-mann matrices are dimensionless and of the form 
\begin{align}\lambda_1=&\left(\begin{array}{ccc}
         0& 1&0 \\
         1& 0&0\\
         0& 0&0\\
    \end{array}\right)\,,\quad\lambda_2=\left(\begin{array}{ccc}
         0& -i&0 \\
         i& 0&0\\
         0& 0&0\\
    \end{array}\right)\,,\quad\lambda_3=\left(\begin{array}{ccc}
         1& 0&0 \\
         0& -1&0\\
         0& 0&0\\
    \end{array}\right)\,,\quad\lambda_4=\left(\begin{array}{ccc}
         0& 0&1 \\
         0& 0&0\\
         1& 0&0\\
    \end{array}\right)\,,\notag\\
    \lambda_5=&\left(\begin{array}{ccc}
         0& 0&-i \\
         0& 0&0\\
         i& 0&0\\
    \end{array}\right)\,,\quad\lambda_6=\left(\begin{array}{ccc}
         0& 0&0 \\
         0& 0&1\\
         0& 1&0\\
    \end{array}\right)\,,\quad\lambda_7=\left(\begin{array}{ccc}
         0& 0&0 \\
         0& 0&-i\\
         0& i&0\\
    \end{array}\right)\,,\quad\lambda_8=\frac{1}{\sqrt{3}}\left(\begin{array}{ccc}
         1& 0&0 \\
         0& 1&0\\
         0& 0&-2\\
    \end{array}\right)\,.
    \label{spurion2}
\end{align}
For the matching, we adopt a more convenient basis of the Lie algebra of the $SU(3)_V$, which is obtained from the Gell-mann matrice via a linear transformation that
\begin{align}\textbf{t}^1=&\frac{1}{2}(\lambda_1+i\lambda_2)=\left(\begin{array}{ccc}
         0& 1&0 \\
         0& 0&0\\
         0& 0&0\\
    \end{array}\right)\,,&\textbf{t}^2=\frac{1}{2}(\lambda_1-i\lambda_2)=\left(\begin{array}{ccc}
         0& 0&0 \\
         1& 0&0\\
         0& 0&0\\
    \end{array}\right)\,,\notag\\
    \textbf{t}^3=&\frac{1}{2}(\lambda_4+i\lambda_5)=\left(\begin{array}{ccc}
         0& 0&1 \\
         0& 0&0\\
         0& 0&0\\
    \end{array}\right)\,,&\textbf{t}^4=\frac{1}{2}(\lambda_4-i\lambda_5)=\left(\begin{array}{ccc}
         0& 0&0 \\
         0& 0&0\\
         1& 0&0\\
    \end{array}\right)\,,\notag\\
    \textbf{t}^5=&\frac{1}{2}(\lambda_6+i\lambda_7)=\left(\begin{array}{ccc}
         0& 0&0 \\
         0& 0&1\\
         0& 0&0\\
    \end{array}\right)\,,&\textbf{t}^6=\frac{1}{2}(\lambda_6-i\lambda_7)=\left(\begin{array}{ccc}
         0& 0&0 \\
         0& 0&0\\
         0& 1&0\\
    \end{array}\right)\,,\notag\\
    \textbf{t}^7=&\lambda_3=\left(\begin{array}{ccc}
         1& 0&0 \\
         0& -1&0\\
         0& 0&0\\
    \end{array}\right)\,,&\textbf{t}^8=\lambda_8=\frac{1}{\sqrt{3}}\left(\begin{array}{ccc}
         1& 0&0 \\
         0& 1&0\\
         0& 0&-2\\
    \end{array}\right)\,,
    \label{spurion1}
\end{align}
or $\mathbf{t}^a = X^a_b \lambda^b$ with the transformation matrix
\begin{equation}
    X = \left(\begin{array}{cccccccc}
\frac{1}{2} & \frac{i}{2} & 0 & 0 & 0 & 0 & 0 & 0 \\
\frac{1}{2} & -\frac{i}{2} & 0 & 0 & 0 & 0 & 0 & 0 \\
0 & 0 & 0 & \frac{1}{2} & \frac{i}{2} & 0 & 0 & 0 \\
0 & 0 & 0 & \frac{1}{2} & -\frac{i}{2} & 0 & 0 & 0 \\
0 & 0 & 0 & 0 & 0 & \frac{1}{2} & \frac{i}{2} & 0 \\
0 & 0 & 0 & 0 & 0 & \frac{1}{2} & -\frac{i}{2} & 0 \\
0 & 0 & 1 & 0 & 0 & 0 & 0 & 0 \\
0 & 0 & 0 & 0 & 0 & 0 & 0 & 1 
    \end{array}\right)\,.
\end{equation}

Although only one spurion $\mathbf{T}$ is sufficient, it can be divided into 3 different kinds according to their electric charges, $\mathbf{T}^0$ and $\mathbf{T}^\pm$, which can be distinguished by their VEVs,
\begin{equation}
\label{eq:3_spurion}
\begin{aligned}
    \mathbf{T}^0 &\rightarrow  span\{\mathbf{t}^5,\mathbf{t}^6,\mathbf{t}^7,\mathbf{t}^8\} \,, \\
    \mathbf{T}^+ &\rightarrow span\{\mathbf{t}^2,\mathbf{t}^4\}\,,\\
    \mathbf{T}^- &\rightarrow span\{\mathbf{t}^1,\mathbf{t}^3\}\,,
\end{aligned}
\end{equation}
where $\mathbf{T}^\pm$ is of $\pm 1$ electric charge and $\mathbf{T}^0$ is free from electric charge. In this paper, we will write a general spurion as $\mathbf{T}$ consistently and indicate their electric charges by the superscript when necessary.



\subsection{The Effective Operators Up to Dimension 9}

Including the spurion $\mathbf{T}$, the independent LEFT operators can be obtained through the Young tensor technique~\cite{Li:2020gnx,Li:2022tec,Li:2020xlh}. Next, we review this method. 

For the Lorentz structure of the operators, the complexification of the Lie algebra of the Lorentz group $SO(3,1)$ is the direct sum Lie algebra $sl(2,\mathbb{C})_l\oplus sl(2,\mathbb{C})_r$, thus all the irreducible representations of the Lorentz group is labeled by two half integers $(j_l,j_r)$, where $j_l$ and $j_r$ label the irreducible representations of $sl(2,\mathbb{C})$.

If we define $\lambda_\alpha\in (\frac{1}{2},0)$ as a left-handed spinor and $\Tilde{\lambda}_{\dot{\alpha}}\in (0,\frac{1}{2})$ as a right-handed spinor, the other fields of different irreducible representations of the Lorentz group can be expressed as their products,
\begin{align}
\label{define}
    \text{Scalar field: }&\phi\in (0,0) \sim 1\notag \,,\\
    \text{Left-handed spinor field: } & \psi \in (\frac{1}{2},0) \sim \lambda_\alpha\notag \,,\\
    \text{Right-handed spinor field: } & \psi^\dagger \in (0,\frac{1}{2}) \sim \Tilde{\lambda}_{\dot{\alpha}}\notag \,,\\
    \text{Left-handed field strength tensor: } & F_L=\frac{F-i\tilde{F}}{2} \in (1,0) \sim \lambda_\alpha\lambda_\beta \notag \,,\\
    \text{Right-handed field strength tensor: } & F_R=\frac{F+i\tilde{F}}{2} \in (0,1) \sim \Tilde{\lambda}_{\dot{\alpha}} \Tilde{\lambda}_{\dot{\beta}}\notag \,,\\
    \text{Derivative: } & D \in (1,1) \sim \lambda_\alpha\Tilde{\lambda}_{\dot{\alpha}} \,.
\end{align}
Thus an operator can be expressed as a product of spinors with all the left- and right-handed indices contracted by the asymmetric tensors $\epsilon^{\alpha\beta}\,,\epsilon^{\dot{\alpha}\dot{\beta}}$. Defining 
\begin{equation}
    \epsilon^{\alpha\beta} \lambda^i_\beta\lambda^j_\alpha = \langle ij \rangle\,,\quad \tilde{\lambda}^i_{\dot{\alpha}} \tilde{\lambda}^j_{\dot{\beta}}\epsilon^{\dot{\beta}\dot{\alpha}} = [ij]\,,
\end{equation}
a general operator takes the form that
\begin{equation}
    \mathcal{O} = \prod^n \langle ij\rangle \prod^{\tilde{n}} [kl]\,,
\end{equation}
where $n$ and $\tilde{n}$ are the half numbers of left- and right-handed spinor indices carried by the operator respectively. 

In addition, the global $SU(3)_V$ should also be considered. According to the group theory, every irreducible representation of the $SU(N)$ group corresponds to a standard Young diagram. The first several irreducible representations of $SU(3)$ are presented in Tab.~\ref{tab: young diagrams}. The effective operators are $SU(3)_V$ invariants, which means they are of the trivial representation $\mathbf{1}$, and corresponds to the Young diagram of the form
\begin{equation}
    \mathcal{O}\sim \yng(2,2,2)\dots\yng(1,1,1)\,.
\end{equation}
Thus the $SU(3)_V$ invariant operators correspond to the Young diagrams of the shape above during the outer product of the Young diagrams corresponding to the fields, where the outer product respects the Littlewood-Richardson rule.
\begin{table}
\ytableausetup
{boxsize=1em}
\ytableausetup
{aligntableaux=center}
    \centering
    \begin{tabular}{|c|c|c|c|c|}
    \hline
 \multirow{2}{*}{$SU(3)$} & $\mathbf{1}$ & $\mathbf{3}$ & $\overline{\mathbf{3}}$ & $\mathbf{8}$ \\
 \cline{2-5}
                & \ydiagram{1,1,1} & \ydiagram{1} & \ydiagram{1,1} & \ydiagram{2,1} \\
    \hline
    \end{tabular}
    \caption{The corresponding Young diagrams of the first several irreducible representations of the $SU(3)$ group.}
    \label{tab: young diagrams}
\end{table}



To apply the Young tensor method reviewed above to the LEFT operators, we must identify its building blocks first.
Including the spurion introduced in Eq.~\eqref{eq:spu_intro_new}, the building blocks of the LEFT become
\begin{equation}
 (q_L\,,\quad q_R\,,\quad e_L\,,\quad e_R\,,\quad\nu_L\,,\quad F_L^{\mu\nu}\,,\quad F_R^{\mu\nu}\,,\quad \mathbf{T})^T \,,  
\end{equation}
which transform under the $SU(3)_V$ symmetry as
\begin{equation}
    \left(\begin{array}{c}
          q_L  \\
          q_R\\
          e_L\\
          e_R\\
          \nu_L\\
          F_L^{\mu\nu}\\
          F_R^{\mu\nu}\\
          \mathbf{T}\\
    \end{array}\right)\rightarrow\left(\begin{array}{c}
         Vq_L \\
          Vq_R\\
          e_L\\
          e_R\\
          \nu_L\\
           F_L^{\mu\nu}\\
          F_R^{\mu\nu}\\
           V\mathbf{T}V^\dagger\\
    \end{array}\right)\,,\quad V\in SU(3)_V\,,
\end{equation}
where the spurion could be $\mathbf{T}^0$ and $\mathbf{T}^\pm$ according to their electric charges. It should be noted that the $L$ and $R$ correspond to the Lorentz group which is different from the $\mathbf{L}$ and $\mathbf{R}$ in $SU(3)_\mathbf{L}\times SU(3)_\mathbf{R}$. The representations of the Lorentz group and the spinor forms of these building blocks can be found in Eq.~\eqref{define}. 
 The building blocks of the LEFT in our notation and their group representations under the $SU(3)_V$ symmetry as well as the gauge symmetry have been listed in Tab.~\ref{lorentz_internal}. 
 Because we will classify the LEFT operators in terms of the $CP$ properties of the quarks, the $CP$ properties for the quark bilinears can also be obtained in Tab.~\ref{tab:quarkbilinear}.

 In the LO Lagrangian, there is at most one spurion, but up to dimension 9, there are cases where the operators have two spurions. For example, when the non-quark fields are free of electric charges there are only 3 combinations of the two spurions $\mathbf{T}_1$ and $\mathbf{T}_2$,
\begin{equation}
\label{eq:2_spurions}
    \mathbf{T}_1\times \mathbf{T}_2 = \left\{\begin{array}{l}
\mathbf{T}^0\times \mathbf{T}^0  \\
\mathbf{T}^+\times \mathbf{T}^-\\
\mathbf{T}^-\times \mathbf{T}^+ \\
    \end{array}\right..
\end{equation}
On the other hand, when the non-quark fields are not free of charges, for example, a lepton bilinear $(\overline{e
}\Gamma e^c)$ appearing in the dimension 9 operators, which is of electric charge +2, there is only one combination
\begin{equation}
\label{eq:t_1t_3}
    \mathbf{T}'_1\times \mathbf{T}'_2 = \mathbf{T}^-\times \mathbf{T}^-\,.
\end{equation}
In this paper, it is not necessary to always distinguish these combinations, so we write all the spurions as $\mathbf{T}$ generally even for the operators with more than one spurion. The management of the spurions in the Young tensor method can also be found in Ref.~\cite{Sun:2022snw,Sun:2022ssa}.


In addition, the operators generated by the Young tensor technique~\cite{Li:2020gnx,Li:2022tec,Li:2020xlh} need to be transformed to the $CP$ eigenstates. For example, the four-quark dimension-6 operators obtained by the Young tensor method take the form
\begin{align}
    \mathcal{B}_1=&( q_R^\dagger\Bar{\sigma}^\mu \mathbf{T} q_R)( q_R^\dagger\Bar{\sigma}_\mu \mathbf{T}  q_R)\,,\\
    \mathcal{B}_2=&( q_R^\dagger\Bar{\sigma}_\mu \mathbf{T} q_R)(q_L^\dagger\sigma^\mu \mathbf{T}  q_L)\,,\\
    \mathcal{B}_3=&(q_L^\dagger\sigma^\mu \mathbf{T}  q_L)( q_R^\dagger\Bar{\sigma}_\mu \mathbf{T}  q_R)\,,\\
    \mathcal{B}_4=&(q_L^\dagger\sigma^\mu \mathbf{T}  q_L)(q_L^\dagger\sigma_\mu \mathbf{T}  q_L)\,,\\
    \mathcal{B}_5=&(q_L^\dagger \mathbf{T} q_R)(q_L^\dagger \mathbf{T} q_R)\,,\\
    \mathcal{B}_6=&(q_L^\dagger \mathbf{T} q_R)(q_R^\dagger \mathbf{T} q_L)\,,\\
    \mathcal{B}_7=&(q_R^\dagger \mathbf{T} q_L)(q_L^\dagger \mathbf{T} q_R)\,,\\
    \mathcal{B}_8=&(q_R^\dagger \mathbf{T} q_L)(q_R^\dagger \mathbf{T} q_L)\,.
\end{align}
Recalling the combinations of Weyl spinors in Eq.~\eqref{define} to form the Dirac spinors and the Dirac matrices
\begin{equation}
    q=\left(\begin{array}{c}
         q_L  \\
         q_R 
    \end{array}\right)\,,\quad \bar q=\left(\bar q_L,\bar q_R\right)\,,\quad\gamma^\mu=\left(\begin{array}{cc}
         0&\bar \sigma^\mu  \\
        \sigma^\mu &0 
    \end{array}\right)\,,
\end{equation}
we obtain the corresponding operators with specific $CP$ eigenvalues,
\begin{align}
    \mathcal{B}_1+\mathcal{B}_2+\mathcal{B}_3+\mathcal{B}_4&=(\bar q\gamma^\mu\mathbf{T}q)(\bar q\gamma^\mu\mathbf{T}q)\,,\\
    \mathcal{B}_1+\mathcal{B}_2-\mathcal{B}_3-\mathcal{B}_4&=(\bar q\gamma^5\gamma^\mu\mathbf{T}q)(\bar q\gamma^\mu\mathbf{T}q)\,,\\
    \mathcal{B}_1-\mathcal{B}_2+\mathcal{B}_3-\mathcal{B}_4&=(\bar q\gamma^\mu\mathbf{T}q)(\bar q\gamma^5\gamma^\mu\mathbf{T}q)\,,\\
    \mathcal{B}_1-\mathcal{B}_2-\mathcal{B}_3+\mathcal{B}_4&=(\bar q\gamma^5\gamma^\mu\mathbf{T}q)(\bar q\gamma^5\gamma^\mu\mathbf{T}q)\,,\\
    \mathcal{B}_5+\mathcal{B}_6+\mathcal{B}_7+\mathcal{B}_8&=(\bar q\mathbf{T}q)(\bar q\mathbf{T}q)\,,\\
    \mathcal{B}_5+\mathcal{B}_6-\mathcal{B}_7-\mathcal{B}_8&=(\bar q\gamma^5\mathbf{T}q)(\bar q\mathbf{T}q)\,,\\
    \mathcal{B}_5-\mathcal{B}_6+\mathcal{B}_7-\mathcal{B}_8&=(\bar q\mathbf{T}q)(\bar q\gamma^5\mathbf{T}q)\,,\\
    \mathcal{B}_5-\mathcal{B}_6-\mathcal{B}_7+\mathcal{B}_8&=(\bar q\gamma^5\mathbf{T}q)(\bar q\gamma^5\mathbf{T}q)\,,
\end{align}
which will be divided into $C$+$P$+, $C$+$P$- and $C$-$P$- types.


\begin{table}[]
    \centering
    \begin{tabular}{|c|c|c|c|c|}
\hline
\multirow{2}{*}{building blocks} & \multicolumn{2}{c|}{gauge symmetry} & \multicolumn{2}{c|}{global symmetry} \\
\cline{2-5}
& $SU(3)_c$ & $U(1)_e$ & $U(1)_B$ & $SU(3)_V$\\
\hline
$q$ & $\mathbf{3}$ & 0 & $\frac{1}{3}$ & $\mathbf{3}$ \\
\hline
$e_L$ & $\mathbf{1}$ & -1 & 0 & $\mathbf{1}$ \\
$e_R$ & $\mathbf{1}$ & -1 & 0 & $\mathbf{1}$ \\
$\nu_L$ & $\mathbf{1}$ & 0 & 0 & $\mathbf{1}$ \\
$F_{\mu\nu}$ & $\mathbf{1}$ & 0 & 0 & $\mathbf{1}$ \\
$\tilde{F}_{\mu\nu}$ & $\mathbf{1}$ & 0 & 0 & $\mathbf{1}$ \\
\hline
$\mathbf{T}^0$ & $\mathbf{1}$ & 0 & 0 & $\mathbf{8}$ \\
$\mathbf{T}^+$ & $\mathbf{1}$ & +1 & 0 & $\mathbf{8}$ \\
$\mathbf{T}^-$ & $\mathbf{1}$ & -1 & 0 & $\mathbf{8}$ \\
\hline
    \end{tabular}
    \caption{The building blocks of the LEFT and their group representations under the $SU(3)_V$ symmetry and gauge symmetries. We distinguish the 3 kinds of spurion via their electric charges explicitly. We introduce an additional abelian symmetry $U(1)_B$ to ensure electric charge conservation of the quark parts.}
    \label{lorentz_internal}
\end{table}

\begin{table}
    \centering
    \begin{tabular}{|l|ccccccccc|}
    \hline
           & $\Bar{q}q$ & $\Bar{q}\gamma^5q$ & $\Bar{q}\gamma^\mu q$ & $\Bar{q}\gamma^5\gamma^\mu q$ & $\Bar{q}\sigma^{\mu\nu}q$ & $\Bar{q}\lrpartial^\mu q$&$\Bar{q}\gamma^5\lrpartial^\mu q$&$\Bar{q}\gamma^\mu\lrpartial^\nu q$&$\Bar{q}\gamma^5\gamma^\mu\lrpartial^\nu q$\\
          \hline
$C$ & $+$ & $+$ & $-$ & $+$ & $-$&$-$&$-$&$+$&$-$\\
$P$ & $+$ & $-$ & $+$ & $-$ & $+$&$+$&$-$&$+$&$-$\\
\hline
    \end{tabular}
    \caption{The $CP$ properties of all the quark bilinears up to dimension-9 in the LEFT.}
    \label{tab:quarkbilinear}
\end{table}



As discussed before, if we constrain ourselves on the dimension-6 operators of the LEFT with only the charged leptons and the photon, the equivalence of the external sources in Eq.~\eqref{eq:external_sources_1} to the spurion is clear, 
\begin{align}
    v^\mu &\sim \mathbf{T}(A^\mu+C_1\overline{e}\gamma^\mu e+C_2\overline{e}\gamma^5\gamma^\mu e) \,,\\
     a^\mu &\sim \mathbf{T}(C_3\overline{e}\gamma^\mu e+C_4\overline{e}\gamma^5\gamma^\mu e) \,,\\
    s &\sim \mathbf{T}(C_5\overline{e}e+C_6\overline{e}\gamma^5e)  \,,\\
    p &\sim \mathbf{T}(C_7\overline{e}e+C_8\overline{e}\gamma^5e)  \,,\\
    \overline{t}^{\mu\nu} &\sim \mathbf{T} (C_9\overline{e}\sigma^{\mu\nu} e+C_{10}\overline{e}\gamma^5\sigma^{\mu\nu} e)  \,,
\end{align}
where the $\mathbf{T}$ on the right-hand is the spurion. The spurion is related to the Wilson coefficients $T$ in Eq.~\eqref{eq:external_sources_1}.
Although the equivalence above is straightforward, once we consider the LEFT operators beyond dimension 6, the external sources in Eq.~\eqref{eq:external_sources_1} are inadequate. On the contrary, the spurion method can take all the lepton bilinears and the photon into account naturally with only one spurion $\mathbf{T}$.

\subsubsection*{Effective Operator Basis Up to Dimension 9}

In this section, we present the higher-dimension LEFT operators in terms of the building blocks in Tab.~\ref{lorentz_internal}. 
When considering the four quark operators, the $SU(3)_c$ color symmetry should be considered and $T^A$ is the $SU(3)_c$ generator while the color symmetry will not affect the matching. In particular, we classify the effective operators by the $CP$ properties of their quark parts, which is also important for matching (see Sec.~\ref{sec:matching}). Besides, we specify the electric charges of the spurions by the superscripts for the operators with only one spurion.

\subsubsection*{Dimension-5}
The dimension-5 operators with one quark bilinear can be classified by CP properties of the quark bilinear as
\begin{align}
C-P+:\notag\\
\Op_1^{(5)}=&F_{\mu\nu} (\bar q\sigma^{\mu\nu}q)\,,
&
\Op_2^{(5)}=&F_{\mu\nu} (\bar q\textbf{T}^0\sigma^{\mu\nu}q)\,,\notag\\
\Op_3^{(5)}=&\Tilde{F}_{\mu\nu} (\bar q\sigma^{\mu\nu}q)\,,
&
\Op_4^{(5)}=&\Tilde{F}_{\mu\nu} (\bar q\textbf{T}^0\sigma^{\mu\nu}q)\,.
\end{align}

\subsubsection*{Dimension-6}

The dimension-6 operators with one quark bilinear are
\begin{align}
C+P+:\notag\\
\Op^{(6)}_1 &= (\bar e_Le_R) (\bar{q} \textbf{T}^0 q )+h.c.\,,
&
\Op^{(6)}_2 &= (\bar e_Le_R)  (\bar q  q )+h.c.\,,\notag
\\
\Op^{(6)}_3 &= (\nu_L^TC\nu_L) (\bar{q} \textbf{T}^0 q )+h.c.\,,
&
\Op^{(6)}_4 &= (\nu_L^TC\nu_L)  (\bar q  q )+h.c.\,,\notag
\\
\Op^{(6)}_5 &= (\bar e_R\nu_L)(\bar{q} \textbf{T}^- q )+h.c.\,,
&
\Op^{(6)}_6 &= ( \nu_L^TCe_L) (\bar{q} \textbf{T}^+ q )+h.c.\,,
\end{align}
\begin{align}
C+P-:\notag\\
\Op^{(6)}_7 &= (\bar e_Le_R) (\bar{q} \textbf{T}^0i\gamma^5 q )+h.c.\,,
&
\Op^{(6)}_8 &= (\bar e_Le_R)  (\bar q i\gamma^5 q )+h.c.\,,\notag
\\
\Op^{(6)}_9 &= (\nu_L^TC\nu_L) (\bar{q} \textbf{T}^0i\gamma^5 q )+h.c.\,,
&
\Op^{(6)}_{10} &= (\nu_L^TC\nu_L)  (\bar q i\gamma^5 q )+h.c.\,,\notag
\\
\Op^{(6)}_{11} &= (\bar e_R\nu_L) (\bar{q} \textbf{T}^-i\gamma^5 q )+h.c.\,,
&
\Op^{(6)}_{12} &= ( \nu_L^TCe_L) (\bar{q} \textbf{T}^+i\gamma^5 q\notag )+h.c.\,,
\\
\Op^{(6)}_{13} &= (\bar\nu_L\gamma^\mu\nu_L) (\bar{q} \textbf{T}^0\gamma^5\gamma_\mu q )\,,
&
\Op^{(6)}_{14} &= (\bar e_R\gamma^\mu e_R)  (\bar{q} \textbf{T}^0\gamma^5\gamma_\mu q )\,,\notag
\\
\Op^{(6)}_{15} &= (\bar e_L\gamma^\mu e_L)(\bar{q} \textbf{T}^0\gamma^5\gamma_\mu q )\,,
&
\Op^{(6)}_{16} &= (\bar\nu_L\gamma^\mu\nu_L)  (\bar q \gamma^5\gamma_\mu q )\,,\notag
\\
\Op^{(6)}_{17} &= (\bar e_R\gamma^\mu e_R)  (\bar q \gamma^5\gamma_\mu q )\,,
&
\Op^{(6)}_{18} &= (\bar e_L\gamma^\mu e_L)(\bar q \gamma^5\gamma_\mu q )\,,\notag
\\
\Op^{(6)}_{19} &= (\bar e_L\gamma^\mu \nu_L)(\bar{q} \textbf{T}^-\gamma^5\gamma_\mu q )+h.c.\,,
&
\Op^{(6)}_{20} &= (\nu_L^TC\gamma^\mu e_R)  (\bar{q} \textbf{T}^+\gamma^5\gamma_\mu q )+h.c.\,,
\end{align}
\begin{align}
C-P+:\notag\\
\Op^{(6)}_{21} &= (\bar\nu_L\gamma^\mu\nu_L)  (\bar{q} \textbf{T}^0\gamma_\mu q )\,,
&
\Op^{(6)}_{22} &= (\bar e_R\gamma^\mu e_R)  (\bar{q} \textbf{T}^0\gamma_\mu q )\,,
\notag\\
\Op^{(6)}_{23} &= (\bar e_L\gamma^\mu e_L)  (\bar{q} \textbf{T}^0\gamma_\mu q )\,,
&
\Op^{(6)}_{24} &= (\bar\nu_L\gamma^\mu\nu_L)  (\bar q \gamma_\mu q )\,,
\notag\\
\Op^{(6)}_{25} &= (\bar e_R\gamma^\mu e_R)  (\bar q \gamma_\mu q )\,,
&
\Op^{(6)}_{26} &= (\bar e_L\gamma^\mu e_L)  (\bar q \gamma_\mu q )\,,
\notag\\
\Op^{(6)}_{27} &= (\bar e_L\gamma^\mu \nu_L)  (\bar{q} \textbf{T}^-\gamma_\mu q )+h.c.\,,
&
\Op^{(6)}_{28} &= (\nu_L^TC\gamma^\mu e_R)  (\bar{q} \textbf{T}^+\gamma_\mu q )+h.c.\,,
\notag\\
\Op^{(6)}_{29} &= (\bar e_L\sigma^{\mu\nu}e_R)(\bar{q} \textbf{T}^0\sigma^{\mu\nu} q )+h.c.\,,
&
\Op^{(6)}_{30} &= (\nu_L^TC\sigma^{\mu\nu}\nu_L)  (\bar{q} \textbf{T}^0\sigma^{\mu\nu} q )+h.c.\,,
\notag\\
\Op^{(6)}_{31} &= (\bar e_L\sigma^{\mu\nu}e_R)(\bar q \sigma^{\mu\nu} q )+h.c.\,,
&
\Op^{(6)}_{32} &= (\nu_L^TC\sigma^{\mu\nu}\nu_L)  (\bar q \sigma^{\mu\nu} q )+h.c.\,,
\notag\\
\Op^{(6)}_{33} &= (\bar e_R\sigma^{\mu\nu}\nu_L) (\bar{q} \textbf{T}^-\sigma^{\mu\nu} q )+h.c.\,,
&
\Op^{(6)}_{34} &= (\nu_L^TC\sigma^{\mu\nu}e_R)  (\bar{q} \textbf{T}^+\sigma^{\mu\nu} q )+h.c.\,,
\end{align}
and the dimension-6 operators with two quark bilinears are
\begin{align}
C+P+:\notag\\
\Op^{(6)}_{35}&= (\bar q q )( \bar q  q )\,,
&
\Op^{(6)}_{36}&= (\bar q q )( \bar q \textbf{T}^0q )\,,
\notag\\
\Op^{(6)}_{37}&= (\bar q\textbf{T} q )( \bar q \textbf{T} q )\,,
&
\Op^{(6)}_{38}&= (\bar qi\gamma^5 q )( \bar q i\gamma^5 q )\,,
\notag\\
\Op^{(6)}_{39}&= (\bar qi\gamma^5 q )( \bar q \textbf{T}^0i\gamma^5 q )\,,
&
\Op^{(6)}_{40}&= (\bar q\textbf{T}i\gamma^5 q )( \bar q \textbf{T}i\gamma^5 q )\,,
\notag\\
\Op^{(6)}_{41}&= (\bar q\gamma^\mu q )( \bar q \gamma_\mu q )\,,
&
\Op^{(6)}_{42}&= (\bar q\gamma^\mu q )( \bar q \textbf{T}^0\gamma_\mu q )\,,
\notag\\
\Op^{(6)}_{43}&= (\bar q\textbf{T}\gamma^\mu q )( \bar q \textbf{T}\gamma_\mu q )\,,
&
\Op^{(6)}_{44}&= (\bar q\gamma^5\gamma^\mu q )( \bar q \gamma^5\gamma_\mu q )\,,
\notag\\
\Op^{(6)}_{45}&= (\bar q\gamma^5\gamma^\mu q )( \bar q \textbf{T}^0\gamma^5\gamma_\mu q )\,,
&
\Op^{(6)}_{46}&= (\bar q\textbf{T}\gamma^5\gamma^\mu q )( \bar q \textbf{T}\gamma^5\gamma_\mu q )\,,
\notag\\
\Op^{(6)}_{47}&= (\bar qT^A q )( \bar q T^A q )\,,
&
\Op^{(6)}_{48}&= (\bar q T^Aq )( \bar q T^A\textbf{T}^0q )\,,
\notag\\
\Op^{(6)}_{49}&= (\bar qT^A\textbf{T} q )( \bar q T^A\textbf{T} q )\,,
&
\Op^{(6)}_{50}&= (\bar qT^Ai\gamma^5 q )( \bar q T^Ai\gamma^5 q )\,,
\notag\\
\Op^{(6)}_{51}&= (\bar qT^Ai\gamma^5 q )( \bar q T^A\textbf{T}^0i\gamma^5 q )\,,
&
\Op^{(6)}_{52}&= (\bar qT^A\textbf{T}i\gamma^5 q )( \bar q T^A\textbf{T}i\gamma^5 q )\,,
\notag\\
\Op^{(6)}_{53}&= (\bar q T^A\gamma^\mu q )( \bar q T^A\gamma_\mu q )\,,
&
\Op^{(6)}_{54}&= (\bar q T^A\gamma^\mu q )( \bar q T^A\textbf{T}^0\gamma_\mu q )\,,
\notag\\
\Op^{(6)}_{55}&= (\bar q T^A\textbf{T}\gamma^\mu q )( \bar q T^A\textbf{T}\gamma_\mu q )\,,
&
\Op^{(6)}_{56}&= (\bar q T^A\gamma^5\gamma^\mu q )( \bar q T^A\gamma^5\gamma_\mu q )\,,
\notag\\
\Op^{(6)}_{57}&= (\bar q T^A\gamma^5\gamma^\mu q )( \bar q T^A\textbf{T}^0\gamma^5\gamma_\mu q )\,,
&
\Op^{(6)}_{58}&= (\bar q T^A\textbf{T}\gamma^5\gamma^\mu q )( \bar q T^A\textbf{T}\gamma^5\gamma_\mu q )\,,
\end{align}
\begin{align}
C+P-:\notag\\
\Op^{(6)}_{59}&= (\bar q q )( \bar qi\gamma^5 q )\,,
&
\Op^{(6)}_{60}&= (\bar q q )( \bar q \textbf{T}^0i\gamma^5 q )\,,
\notag\\
\Op^{(6)}_{61}&= (\bar q\textbf{T}^0 q )( \bar q i\gamma^5 q )\,,
&
\Op^{(6)}_{62}&= (\bar q {\textbf{T}} q )( \bar q {\textbf{T}}i\gamma^5 q )\,,
\notag\\
\Op^{(6)}_{63}&= (\bar q T^Aq )( \bar qT^Ai\gamma^5 q )\,,
&
\Op^{(6)}_{64}&= (\bar q T^Aq )( \bar q T^A\textbf{T}^0i\gamma^5 q )\,,
\notag\\
\Op^{(6)}_{65}&= (\bar qT^A\textbf{T}^0 q )( \bar q T^Ai\gamma^5 q )\,,
&
\Op^{(6)}_{66}&= (\bar qT^A\textbf{T} q )( \bar q T^A\textbf{T}i\gamma^5 q )\,,
\\
C-P-:\notag\\
\Op^{(6)}_{67}&= (\bar q\gamma^\mu q )( \bar q\gamma^5\gamma_\mu q )\,,
&
\Op^{(6)}_{78}&= (\bar q \gamma^\mu q )( \bar q \textbf{T}^0\gamma^5\gamma_\mu q )\,,
\notag\\
\Op^{(6)}_{69}&= (\bar q\textbf{T}^0\gamma^\mu q )( \bar q \gamma^5\gamma_\mu q )\,,
&
\Op^{(6)}_{70}&= (\bar q\gamma^\mu\textbf{T} q )( \bar q \textbf{T}\gamma^5\gamma_\mu q )\,,
\notag\\
\Op^{(6)}_{71}&= (\bar qT^A\gamma^\mu q )( \bar qT^A\gamma^5\gamma_\mu q )\,,
&
\Op^{(6)}_{72}&= (\bar q T^A\gamma^\mu q )( \bar q T^A\textbf{T}^0\gamma^5\gamma_\mu q )\,,
\notag\\
\Op^{(6)}_{73}&= (\bar qT^A\textbf{T}^0\gamma^\mu q )( \bar q T^A\gamma^5\gamma_\mu q )\,,
&
\Op^{(6)}_{74}&= (\bar qT^A\gamma^\mu\textbf{T} q )( \bar q T^A\textbf{T}\gamma^5\gamma_\mu q )\,,
\end{align}
where for the operators with two spurions, the 3 different spurions structures in Eq.~\eqref{eq:2_spurions} are left implicitly.

\subsubsection*{Dimension-7}

The dimension-7 operators with one quark bilinear can also be classified
\begin{align}
C+P+:\notag\\
\Op^{(7)}_1 &= F^{\mu\nu}F_{\mu\nu}(\bar q\textbf{T}^0q)\,,
&
\Op^{(7)}_2 &= F^{\mu\nu}F_{\mu\nu}(\bar q  q )\,, 
\notag\\
\Op^{(7)}_3 &= \Tilde{F}^{\mu\nu}F_{\mu\nu}(\bar q\textbf{T}^0q)\,,
&
\Op^{(7)}_4 &= \Tilde{F}^{\mu\nu}F_{\mu\nu}(\bar q  q )\,, 
\\
C+P-:\notag\\
\Op^{(7)}_5 &= F^{\mu\nu}F_{\mu\nu}(\bar q\textbf{T}^0i\gamma^5q)\,,
&
\Op^{(7)}_6 &= F^{\mu\nu}F_{\mu\nu}(\bar q i\gamma^5 q )\,, 
\notag\\
\Op^{(7)}_7 &= \Tilde{F}^{\mu\nu}F_{\mu\nu}(\bar q\textbf{T}^0i\gamma^5q)\,,
&
\Op^{(7)}_8 &= \Tilde{F}^{\mu\nu}F_{\mu\nu}(\bar q i\gamma^5 q )\,, 
\notag\\
\Op^{(7)}_9 &= (\bar e_Li\lrpartial^\mu e_R) (\bar{q} \textbf{T}^0\gamma^5\gamma_\mu q )+h.c.\,,
&
\Op^{(7)}_{10} &= (\nu_L^TCi\lrpartial^\mu \nu_L)  (\bar{q} \textbf{T}^0\gamma^5\gamma_\mu q )+h.c.\,,
\notag\\
\Op^{(7)}_{11} &= (\bar e_Li\lrpartial^\mu e_R)  (\bar q \gamma^5\gamma_\mu q )+h.c.\,,
&
\Op^{(7)}_{12} &= (\nu_L^TCi\lrpartial^\mu \nu_L)  (\bar q \gamma^5\gamma_\mu q )+h.c.\,,
\notag\\
\Op^{(7)}_{13} &= (\bar e_Ri\lrpartial^\mu\nu_L)(\bar{q} \textbf{T}^-\gamma^5\gamma_\mu q )+h.c.\,,
&
\Op^{(7)}_{14} &= (\nu_L^TCi\lrpartial^\mu e_L)  (\bar{q} \textbf{T}^+\gamma^5\gamma_\mu q )+h.c.\,,
\\
C-P+:\notag\\
\Op^{(7)}_{15} &= (\bar\nu_L\gamma^\mu\nu_L) (\bar{q} \textbf{T}^0i\lrpartial_\mu q )\,,
&
\Op^{(7)}_{16} &=(\bar e_L\gamma^\mu e_L)  (\bar{q} \textbf{T}^0i\lrpartial_\mu q )\,,
\notag\\
\Op^{(7)}_{17} &= (\bar e_R\gamma^\mu e_R)  (\bar{q} \textbf{T}^0i\lrpartial_\mu q )\,,
&
\Op^{(7)}_{18} &= (\bar\nu_L\gamma^\mu\nu_L)  (\bar q i\lrpartial_\mu q )\,,
\label{eq:ops1}
\notag\\
\Op^{(7)}_{19} &= (\bar e_L\gamma^\mu e_L) (\bar q i\lrpartial_\mu q )\,,
&
\Op^{(7)}_{20} &= (\bar e_R\gamma^\mu e_R)  (\bar q i\lrpartial_\mu q )\,,
\notag\\
\Op^{(7)}_{21} &= (\bar e_L\gamma^\mu \nu_L) (\bar{q} \textbf{T}^-i\lrpartial_\mu q )+h.c.\,,
&
\Op^{(7)}_{22} &= (\nu_L^TC\gamma^\mu e_R)  (\bar{q} \textbf{T}^-i\lrpartial_\mu q )+h.c.\,,
\notag\\
\Op^{(7)}_{23} &= (\bar e_Li\lrpartial^\mu e_R)(\bar{q} \textbf{T}^0\gamma_\mu q )+h.c.\,,
&
\Op^{(7)}_{24} &= (\nu_L^TCi\lrpartial^\mu \nu_L) (\bar{q} \textbf{T}^0\gamma_\mu q )+h.c.\,,
\notag\\
\Op^{(7)}_{25} &= (\bar e_Li\lrpartial^\mu e_R)  (\bar q \gamma_\mu q )+h.c.\,,
&
\Op^{(7)}_{26} &= (\nu_L^TCi\lrpartial^\mu \nu_L)  (\bar q \gamma_\mu q )+h.c.\,,
\notag\\
\Op^{(7)}_{27} &= (\bar e_Ri\lrpartial^\mu\nu_L)  (\bar{q} \textbf{T}^-\gamma_\mu q )+h.c.\,,
&
\Op^{(7)}_{28} &= (\nu_L^TCi\lrpartial^\mu e_L)  (\bar{q} \textbf{T}^+\gamma_\mu q )+h.c.\,,
\end{align}
\begin{align}
C-P-:\notag\\
\Op^{(7)}_{29} &= (\bar\nu_L\gamma^\mu\nu_L) (\bar{q} \textbf{T}^0i\gamma^5i\lrpartial_\mu q )\,,
&
\Op^{(7)}_{30} &= (\bar e_L\gamma^\mu e_L)  (\bar{q} \textbf{T}^0i\gamma^5i\lrpartial_\mu q )\,,
\notag\\
\Op^{(7)}_{31} &=  (\bar e_R\gamma^\mu e_R)  (\bar{q} \textbf{T}^0i\gamma^5i\lrpartial_\mu q )\,,
&
\Op^{(7)}_{32} &= (\bar\nu_L\gamma^\mu\nu_L)  (\bar q i\gamma^5i\lrpartial_\mu q )\,,
\label{eq:ops2}\notag\\
\Op^{(7)}_{33} &= (\bar e_L\gamma^\mu e_L)  (\bar q i\gamma^5i\lrpartial_\mu q )\,,
&
\Op^{(7)}_{34} &= (\bar e_R\gamma^\mu e_R) (\bar q i\gamma^5i\lrpartial_\mu q )\,,
\notag\\
\Op^{(7)}_{35} &= (\bar e_L\gamma^\mu \nu_L) (\bar{q} \textbf{T}^-i\gamma^5i\lrpartial_\mu q )+h.c.\,,
&
\Op^{(7)}_{36} &= (\nu_L^TC\gamma^\mu e_R)  (\bar{q} \textbf{T}^+i\gamma^5i\lrpartial_\mu q )+h.c.\,,
\end{align}
and the dimension-7 operators with two quark bilinear would be
\begin{align}
C+P+:\notag\\
\Op^{(7)}_{37}&= (\bar q\gamma^\mu q )( \bar qi\lrpartial_\mu q )\,,
&
\Op^{(7)}_{38}&= (\bar q \gamma^\mu q )( \bar q \textbf{T}^0i\lrpartial_\mu q )\,,
\notag\\
\Op^{(7)}_{39}&= (\bar q\gamma^\mu\textbf{T}^0 q )( \bar qi\lrpartial_\mu q )\,,
&
\Op^{(7)}_{40}&= (\bar q\gamma^\mu{\textbf{T}} q )( \bar q {\textbf{T}}i\lrpartial_\mu q )\,,
\notag\\
\Op^{(7)}_{41}&= (\bar qT^A\gamma^\mu{\textbf{T}} q )( \bar q T^A{\textbf{T}}i\lrpartial_\mu q )\,,
&
\Op^{(7)}_{42}&= (\bar qT^A\gamma^\mu q )( \bar qT^Ai\lrpartial_\mu q )\,,
\notag\\
\Op^{(7)}_{43}&= (\bar q T^A\gamma^\mu q )( \bar qT^A \textbf{T}^0i\lrpartial_\mu q )\,,
&
\Op^{(7)}_{44}&= (\bar qT^A\gamma^\mu\textbf{T}^0 q )( \bar qT^Ai\lrpartial_\mu q )\,,
\\
C+P-:\notag\\
\Op^{(7)}_{45}&= (\bar q\gamma^\mu q )( \bar qi\gamma^5i\lrpartial_\mu q )\,,
&
\Op^{(7)}_{46}&= (\bar q \gamma^\mu q )( \bar q \textbf{T}^0i\gamma^5i\lrpartial_\mu q )\,,
\notag\\
\Op^{(7)}_{47}&= (\bar q\gamma^\mu\textbf{T}^0 q )( \bar qi\gamma^5i\lrpartial_\mu q )\,,
&
\Op^{(7)}_{48}&= (\bar q\gamma^\mu{\textbf{T}} q )( \bar q {\textbf{T}}i\gamma^5i\lrpartial_\mu q )\,,
\notag\\
\Op^{(7)}_{49}&= (\bar qT^A\gamma^\mu{\textbf{T}} q )( \bar q T^A{\textbf{T}}i\gamma^5i\lrpartial_\mu q )\,,
&
\Op^{(7)}_{50}&= (\bar qT^A\gamma^\mu q )( \bar qT^Ai\gamma^5i\lrpartial_\mu q )\,,
\notag\\
\Op^{(7)}_{51}&= (\bar q T^A\gamma^\mu q )( \bar qT^A \textbf{T}^0i\gamma^5i\lrpartial_\mu q )\,,
&
\Op^{(7)}_{52}&= (\bar qT^A\gamma^\mu\textbf{T}^0 q )( \bar qT^Ai\gamma^5i\lrpartial_\mu q )\,,
\\
C-P+:\notag\\
\Op^{(7)}_{53}&= (\bar q\gamma^5\gamma^\mu q )( \bar qi\gamma^5i\lrpartial_\mu q )\,,
&
\Op^{(7)}_{54}&= (\bar q \gamma^5\gamma^\mu q )( \bar q \textbf{T}^0i\gamma^5i\lrpartial_\mu q )\,,
\notag\\
\Op^{(7)}_{55}&= (\bar q\gamma^5\gamma^\mu\textbf{T}^0 q )( \bar qi\gamma^5i\lrpartial_\mu q )\,,
&
\Op^{(7)}_{56}&= (\bar q\gamma^5\gamma^\mu{\textbf{T}} q )( \bar q {\textbf{T}}i\gamma^5i\lrpartial_\mu q )\,,
\notag\\
\Op^{(7)}_{57}&= (\bar qT^A\gamma^5\gamma^\mu{\textbf{T}} q )( \bar q T^A{\textbf{T}}i\gamma^5i\lrpartial_\mu q )\,,
&
\Op^{(7)}_{58}&= (\bar qT^A\gamma^5\gamma^\mu q )( \bar qT^Ai\gamma^5i\lrpartial_\mu q )\,,
\notag\\
\Op^{(7)}_{59}&= (\bar q T^A\gamma^5\gamma^\mu q )( \bar q T^A\textbf{T}^0i\gamma^5i\lrpartial_\mu q )\,,
&
\Op^{(7)}_{60}&= (\bar qT^A\gamma^5\gamma^\mu\textbf{T}^0 q )( \bar qT^Ai\gamma^5i\lrpartial_\mu q )\,,
\\
C-P-:\notag\\ 
\Op^{(7)}_{61}&= (\bar q\gamma^5\gamma^\mu q )( \bar qi\lrpartial_\mu q )\,,
&
\Op^{(7)}_{62}&= (\bar q \gamma^5\gamma^\mu q )( \bar q \textbf{T}^0i\lrpartial_\mu q )\,,
\notag\\
\Op^{(7)}_{63}&= (\bar q\gamma^5\gamma^\mu\textbf{T}^0 q )( \bar q i\lrpartial_\mu q )\,,
&
\Op^{(7)}_{64}&= (\bar q\gamma^5\gamma^\mu{\textbf{T}} q )( \bar q {\textbf{T}}i\lrpartial_\mu q )\,,
\notag\\
\Op^{(7)}_{65}&= (\bar qT^A\gamma^5\gamma^\mu{\textbf{T}} q )( \bar q T^A{\textbf{T}}i\lrpartial_\mu q )\,,
&
\Op^{(7)}_{66}&= (\bar qT^A\gamma^5\gamma^\mu q )( \bar qT^Ai\lrpartial_\mu q )\,,
\notag\\
\Op^{(7)}_{67}&= (\bar q T^A\gamma^5\gamma^\mu q )( \bar q T^A\textbf{T}^0i\lrpartial_\mu q )\,,
&
\Op^{(7)}_{68}&= (\bar qT^A\gamma^5\gamma^\mu\textbf{T}^0 q )( \bar q T^Ai\lrpartial_\mu q )\,.
\end{align}

\subsubsection*{Dimension-8}

The operators with one quark bilinear are 
\begin{align}
C+P+:\notag\\
\Op^{(8)}_1 &= F_{\mu\rho}F^\rho_\nu  (\bar{q} \textbf{T}^0\gamma^\mu i\lrpartial^\nu q )\,,
&
\Op^{(8)}_2 &= F_{\mu\rho}F^\rho_\nu  (\bar{q} \gamma^\mu \lrpartial^\nu q )\,,
\notag\\
\Op^{(8)}_3 &= F_{\mu\nu}(\bar e_L\sigma^{\mu\nu}e_R)(\bar{q} \textbf{T}^0 q )+h.c.\,,
&
\Op^{(8)}_4 &= F_{\mu\nu}(\nu_L^TC\sigma^{\mu\nu}\nu_L)(\bar{q} \textbf{T}^0 q )+h.c.\,,
\notag\\
\Op^{(8)}_5 &= F_{\mu\nu}(\bar e_L\sigma^{\mu\nu}e_R)(\bar{q}  q )+h.c.\,,
&
\Op^{(8)}_6 &= F_{\mu\nu}(\nu_L^TC\sigma^{\mu\nu}\nu_L)(\bar{q}  q )+h.c.\,,
\notag\\
\Op^{(8)}_7 &= F_{\mu\nu}(\bar e_R\sigma^{\mu\nu}\nu_L)(\bar{q} \textbf{T}^- q )+h.c.\,,
&
\Op^{(8)}_8 &= F_{\mu\nu}(\nu_L^TC\sigma^{\mu\nu}e_L)(\bar{q} \textbf{T}^+ q )+h.c.\,,
\notag\\
\Op^{(8)}_{9} &= \partial^2(\bar e_Le_R) (\bar{q} \textbf{T}^0 q )+h.c.\,,
&
\Op^{(8)}_{10} &= \partial^2(\nu_L^TC\nu_L) (\bar{q} \textbf{T}^0 q )+h.c.\,,
\notag\\
\Op^{(8)}_{11} &= \partial^2(\bar e_Le_R)  (\bar q  q )+h.c.\,,
&
\Op^{(8)}_{12} &= \partial^2(\nu_L^TC\nu_L)  (\bar q  q )+h.c.\,,
\notag\\
\Op^{(8)}_{13} &= \partial^2(\bar e_R\nu_L) (\bar{q} \textbf{T}^- q )+h.c.\,,
&
\Op^{(8)}_{14} &= \partial^2( \nu_L^TCe_L) (\bar{q} \textbf{T}^+ q )+h.c.\,,
\notag\\
\Op^{(8)}_{15} &=  (\bar e_L\gamma^\mu \lrpartial^\nu e_L) (\bar{q} \textbf{T}^0\gamma_\mu \lrpartial_\nu q )\,,
&
\Op^{(8)}_{16} &=  (\bar e_R\gamma^\mu i\lrpartial^\nu e_R) (\bar{q} \textbf{T}^0\gamma_\mu \lrpartial_\nu q )\,,
\notag\\
\Op^{(8)}_{17} &=  (\bar\nu_L\gamma^\mu\lrpartial^\nu\nu_L) (\bar{q} \textbf{T}^0\gamma_\mu\lrpartial_\nu q )\,,
&
\Op^{(8)}_{18} &=  (\bar e_L\gamma^\mu\lrpartial^\nu e_L)(\bar{q} \gamma_\mu\lrpartial_\nu q )\,,
\notag\\
\Op^{(8)}_{19} &=  (\bar e_R\gamma^\mu\lrpartial^\nu e_R) (\bar{q} \gamma_\mu\lrpartial_\nu q )\,,
&
\Op^{(8)}_{20} &=  (\bar\nu_L\gamma^\mu\lrpartial^\nu\nu_L) (\bar{q} \gamma_\mu\lrpartial_\nu q )\,,
\notag\\
\Op^{(8)}_{21} &=  (\bar e_L\gamma^\mu\lrpartial^\nu\nu_L)(\bar{q} \textbf{T}^-\gamma_\mu\lrpartial_\nu q )+h.c.\,,
&
\Op^{(8)}_{22} &=  (\nu_L^TC\gamma^\mu\lrpartial^\nu e_R) (\bar{q} \textbf{T}^+\gamma_\mu\lrpartial_\nu q )+h.c.\,,
\\
C+P-:\notag\\
\Op^{(8)}_{23} &= F_{\mu\nu}(\bar\nu_L\gamma^\mu\nu_L)(\bar{q} \textbf{T}^0\gamma^5\gamma^\nu q )\,,
&
\Op^{(8)}_{24} &= F_{\mu\nu}(\bar e_L\gamma^\mu e_L)(\bar{q} \textbf{T}^0\gamma^5\gamma^\nu q )\,,
\notag\\
\Op^{(8)}_{25} &= F_{\mu\nu}(\bar e_R\gamma^\mu e_R)(\bar{q} \textbf{T}^0\gamma^5\gamma^\nu q )\,,
&
\Op^{(8)}_{26} &= F_{\mu\nu}(\bar\nu_L\gamma^\mu\nu_L)(\bar{q} \gamma^5\gamma^\nu q )\,,
\notag\\
\Op^{(8)}_{27} &= F_{\mu\nu}(\bar e_L\gamma^\mu e_L)(\bar{q} \gamma^5\gamma^\nu q )\,,
&
\Op^{(8)}_{28} &= F_{\mu\nu}(\bar e_R\gamma^\mu e_R)(\bar{q} \gamma^5\gamma^\nu q )\,,
\notag\\
\Op^{(8)}_{29} &= F_{\mu\nu}(\bar e_L\gamma^\mu \nu_L)(\bar{q} \textbf{T}^-\gamma^5\gamma^\nu q )+h.c.\,,
&
\Op^{(8)}_{30} &= F_{\mu\nu}(\nu_L^TC\gamma^\mu e_R)(\bar{q} \textbf{T}^+\gamma^5\gamma^\nu q )+h.c.\,,
\notag\\
\Op^{(8)}_{31} &= \Tilde{F}_{\mu\nu}(\bar\nu_L\gamma^\mu\nu_L)(\bar{q} \textbf{T}^0\gamma^5\gamma^\nu q )\,,
&
\Op^{(8)}_{32} &= \Tilde{F}_{\mu\nu}(\bar e_L\gamma^\mu e_L)(\bar{q} \textbf{T}^0\gamma^5\gamma^\nu q )\,,
\notag\\
\Op^{(8)}_{33} &= \Tilde{F}_{\mu\nu}(\bar e_R\gamma^\mu e_R)(\bar{q} \textbf{T}^0\gamma^5\gamma^\nu q )\,,
&
\Op^{(8)}_{34} &= \Tilde{F}_{\mu\nu}(\bar\nu_L\gamma^\mu\nu_L)(\bar{q} \gamma^5\gamma^\nu q )\,,
\notag\\
\Op^{(8)}_{35} &= \Tilde{F}_{\mu\nu}(\bar e_L\gamma^\mu e_L)(\bar{q} \gamma^5\gamma^\nu q )\,,
&
\Op^{(8)}_{36} &= \Tilde{F}_{\mu\nu}(\bar e_R\gamma^\mu e_R)(\bar{q} \gamma^5\gamma^\nu q )\,,
\notag\\
\Op^{(8)}_{37} &= \Tilde{F}_{\mu\nu}(\bar e_L\gamma^\mu \nu_L)(\bar{q} \textbf{T}^-\gamma^5\gamma^\nu q )+h.c.\,,
&
\Op^{(8)}_{38} &= \Tilde{F}_{\mu\nu}(\nu_L^TC\gamma^\mu e_R)(\bar{q} \textbf{T}^+\gamma^5\gamma^\nu q )+h.c.\,,
\notag\\
\Op^{(8)}_{39} &= F_{\mu\nu}(\bar e_L\sigma^{\mu\nu}e_R)(\bar{q} \textbf{T}^0i\gamma^5 q )+h.c.\,,
&
\Op^{(8)}_{40} &= F_{\mu\nu}(\nu_L^TC\sigma^{\mu\nu}\nu_L)(\bar{q} \textbf{T}^0i\gamma^5 q )+h.c.\,,
\notag\\
\Op^{(8)}_{41} &= F_{\mu\nu}(\bar e_L\sigma^{\mu\nu}e_R)(\bar{q} i\gamma^5 q )+h.c.\,,
&
\Op^{(8)}_{42} &= F_{\mu\nu}(\nu_L^TC\sigma^{\mu\nu}\nu_L)(\bar{q}i\gamma^5  q )+h.c.\,,
\notag\\
\Op^{(8)}_{43} &= F_{\mu\nu}(\bar e_R\sigma^{\mu\nu}\nu_L)(\bar{q} \textbf{T}^-i\gamma^5 q )+h.c.\,,
&
\Op^{(8)}_{44} &= F_{\mu\nu}(\nu_L^TC\sigma^{\mu\nu}e_L)(\bar{q} \textbf{T}^+i\gamma^5 q )+h.c.\,,
\notag\\
\Op^{(8)}_{45} &= \partial^2(\bar e_Le_R) (\bar{q} \textbf{T}^0i\gamma^5 q )+h.c.\,,
&
\Op^{(8)}_{46} &= \partial^2(\nu_L^TC\nu_L) (\bar{q} \textbf{T}^0i\gamma^5 q )+h.c.\,,
\notag\\
\Op^{(8)}_{47} &= \partial^2(\bar e_Le_R)  (\bar q i\gamma^5 q )+h.c.\,,
&
\Op^{(8)}_{48} &= \partial^2(\nu_L^TC\nu_L)  (\bar q i\gamma^5 q )+h.c.\,,
\notag\\
\Op^{(8)}_{49} &= \partial^2(\bar e_R\nu_L) (\bar{q} \textbf{T}^-i\gamma^5 q )+h.c.\,,
&
\Op^{(8)}_{50} &= \partial^2( \nu_L^TCe_L) (\bar{q} \textbf{T}^-i\gamma^5 q )+h.c.\,,
\notag\\
\Op^{(8)}_{51} &= \partial^2(\bar\nu_L\gamma^\mu\nu_L) (\bar{q} \textbf{T}^0\gamma^5\gamma_\mu q )\,,
&
\Op^{(8)}_{52} &= \partial^2(\bar e_L\gamma^\mu e_L)  (\bar{q} \textbf{T}^0\gamma^5\gamma_\mu q )\,,
\notag\\
\Op^{(8)}_{53} &= \partial^2(\bar e_R\gamma^\mu e_R)  (\bar{q} \textbf{T}^0\gamma^5\gamma_\mu q )\,,
&
\Op^{(8)}_{54} &= \partial^2(\bar\nu_L\gamma^\mu\nu_L) (\bar q \gamma^5\gamma_\mu q )\,,
\notag\\
\Op^{(8)}_{55} &= \partial^2(\bar e_L\gamma^\mu e_L) (\bar q \gamma^5\gamma_\mu q )\,,
&
\Op^{(8)}_{56} &= \partial^2(\bar e_R\gamma^\mu e_R)  (\bar q \gamma^5\gamma_\mu q )\,,
\notag\\
\Op^{(8)}_{57} &= \partial^2(\bar e_L\gamma^\mu \nu_L) (\bar{q} \textbf{T}^-\gamma^5\gamma_\mu q )+h.c.\,,
&
\Op^{(8)}_{58} &= \partial^2(\nu_L^TC\gamma^\mu e_R)  (\bar{q} \textbf{T}^+\gamma^5\gamma_\mu q )+h.c.\,,
\end{align}

\begin{align}
C-P+:\notag\\
\Op^{(8)}_{59} &= F_{\mu\nu}(\bar\nu_L\gamma^\mu\nu_L)(\bar{q} \textbf{T}^0\gamma^\nu q )\,,
&
\Op^{(8)}_{60} &= F_{\mu\nu}(\bar e_L\gamma^\mu e_L)(\bar{q} \textbf{T}^0\gamma^\nu q )\,,
\notag\\
\Op^{(8)}_{61} &= F_{\mu\nu}(\bar e_R\gamma^\mu e_R)(\bar{q} \textbf{T}^0\gamma^\nu q )\,,
&
\Op^{(8)}_{62} &= F_{\mu\nu}(\bar\nu_L\gamma^\mu\nu_L)(\bar{q} \gamma^\nu q )\,,
\notag\\
\Op^{(8)}_{63} &= F_{\mu\nu}(\bar e_L\gamma^\mu e_L)(\bar{q} \gamma^\nu q )\,,
&
\Op^{(8)}_{64} &= F_{\mu\nu}(\bar e_R\gamma^\mu e_R)(\bar{q} \gamma^\nu q )\,,
\notag\\
\Op^{(8)}_{65} &= F_{\mu\nu}(\bar e_L\gamma^\mu \nu_L)(\bar{q} \textbf{T}^-\gamma^\nu q )+h.c.\,,
&
\Op^{(8)}_{66} &= F_{\mu\nu}(\nu_L^TC\gamma^\mu e_R)(\bar{q} \textbf{T}^+\gamma^\nu q )+h.c.\,,
\notag\\
\Op^{(8)}_{67} &= \Tilde{F}_{\mu\nu}(\bar\nu_L\gamma^\mu\nu_L)(\bar{q} \textbf{T}^0\gamma^\nu q )\,,
&
\Op^{(8)}_{68} &= \Tilde{F}_{\mu\nu}(\bar e_L\gamma^\mu e_L)(\bar{q} \textbf{T}^0\gamma^\nu q )\,,
\notag\\
\Op^{(8)}_{69} &= \Tilde{F}_{\mu\nu}(\bar e_R\gamma^\mu e_R)(\bar{q} \textbf{T}^0\gamma^\nu q )\,,
&
\Op^{(8)}_{70} &= \Tilde{F}_{\mu\nu}(\bar\nu_L\gamma^\mu\nu_L)(\bar{q} \gamma^\nu q )\,,
\notag\\
\Op^{(8)}_{71} &= \Tilde{F}_{\mu\nu}(\bar e_L\gamma^\mu e_L)(\bar{q} \gamma^\nu q )\,,
&
\Op^{(8)}_{72} &= \Tilde{F}_{\mu\nu}(\bar e_R\gamma^\mu e_R)(\bar{q} \gamma^\nu q )\,,
\notag\\
\Op^{(8)}_{73} &= \Tilde{F}_{\mu\nu}(\bar e_L\gamma^\mu \nu_L)(\bar{q} \textbf{T}^-\gamma^\nu q )+h.c.\,,
&
\Op^{(8)}_{74} &= \Tilde{F}_{\mu\nu}(\nu_L^TC\gamma^\mu e_R)(\bar{q} \textbf{T}^+\gamma^\nu q )+h.c.\,,
\notag\\
\Op^{(8)}_{75} &= F_{\mu\nu}(\bar e_Le_R)(\bar{q} \textbf{T}^0\sigma^{\mu\nu} q )+h.c.\,,
&
\Op^{(8)}_{76} &= F_{\mu\nu}(\nu_L^TC\nu_L)(\bar{q} \textbf{T}^0\sigma^{\mu\nu} q )+h.c.\,,
\notag\\
\Op^{(8)}_{77} &= F_{\mu\nu}(\bar e_Le_R)(\bar{q} \sigma^{\mu\nu} q )+h.c.\,,
&
\Op^{(8)}_{78} &= F_{\mu\nu}(\nu_L^TC\nu_L)(\bar{q} \sigma^{\mu\nu} q )+h.c.\,,
\notag\\
\Op^{(8)}_{79} &= F_{\mu\nu}(\bar e_R\nu_L)(\bar{q} \textbf{T}^-\sigma^{\mu\nu} q )+h.c.\,,
&
\Op^{(8)}_{80} &= F_{\mu\nu}( \nu_L^TCe_L)(\bar{q} \textbf{T}^+\sigma^{\mu\nu} q )+h.c.\,,
\notag\\
\Op^{(8)}_{81} &= \Tilde{F}_{\mu\nu}(\bar e_Le_R)(\bar{q} \textbf{T}^0\sigma^{\mu\nu} q )+h.c.\,,
&
\Op^{(8)}_{82} &= \Tilde{F}_{\mu\nu}(\nu_L^TC\nu_L)(\bar{q} \textbf{T}^0\sigma^{\mu\nu} q )+h.c.\,,
\notag\\
\Op^{(8)}_{83} &= \Tilde{F}_{\mu\nu}(\bar e_Le_R)(\bar{q} \sigma^{\mu\nu} q )+h.c.\,,
&
\Op^{(8)}_{84} &= \Tilde{F}_{\mu\nu}(\nu_L^TC\nu_L)(\bar{q} \sigma^{\mu\nu} q )+h.c.\,,
\notag\\
\Op^{(8)}_{85} &= \Tilde{F}_{\mu\nu}(\bar e_R\nu_L)(\bar{q} \textbf{T}^-\sigma^{\mu\nu} q )+h.c.\,,
&
\Op^{(8)}_{86} &= \Tilde{F}_{\mu\nu}( \nu_L^TCe_L)(\bar{q} \textbf{T}^+\sigma^{\mu\nu} q )+h.c.\,,
\notag\\
\Op^{(8)}_{87} &= F_{\nu}^\rho (\bar e_L\sigma^{\mu\nu}e_R)(\bar{q} \textbf{T}^0\sigma_{\mu\rho} q )+h.c.\,,
&
\Op^{(8)}_{88} &= F_{\nu}^\rho (\nu_L^TC\sigma^{\mu\nu}\nu_L)(\bar{q} \textbf{T}^0\sigma_{\mu\rho} q )+h.c.\,,
\notag\\
\Op^{(8)}_{89} &= F_{\nu}^\rho (\bar e_L\sigma^{\mu\nu}e_R)(\bar{q} \sigma_{\mu\rho} q )+h.c.\,,
&
\Op^{(8)}_{90} &= F_{\nu}^\rho (\nu_L^TC\sigma^{\mu\nu}\nu_L)(\bar{q} \sigma_{\mu\rho} q )+h.c.\,,
\notag\\
\Op^{(8)}_{91} &= F_{\nu}^\rho (\bar e_R\sigma^{\mu\nu}\nu_L)(\bar{q} \textbf{T}^-\sigma_{\mu\rho} q )+h.c.\,,
&
\Op^{(8)}_{92} &= F_{\nu}^\rho ( \nu_L^TC\sigma^{\mu\nu}e_L)(\bar{q} \textbf{T}^+\sigma_{\mu\rho} q )+h.c.\,,
\notag\\
\Op^{(8)}_{93} &= \partial^2(\bar\nu_L\gamma^\mu\nu_L)  (\bar{q} \textbf{T}^0\gamma_\mu q )\,,
&
\Op^{(8)}_{94} &= \partial^2(\bar e_L\gamma^\mu e_L) (\bar{q} \textbf{T}^0\gamma_\mu q )\,,
\notag\\
\Op^{(8)}_{95} &= \partial^2(\bar e_R\gamma^\mu e_R) (\bar{q} \textbf{T}^0\gamma_\mu q )\,,
&
\Op^{(8)}_{96} &= \partial^2(\bar\nu_L\gamma^\mu\nu_L) (\bar q \gamma_\mu q )\,,
\notag\\
\Op^{(8)}_{97} &= \partial^2(\bar e_L\gamma^\mu e_L) (\bar q \gamma_\mu q )\,,
&
\Op^{(8)}_{98} &= \partial^2(\bar e_R\gamma^\mu e_R)  (\bar q \gamma_\mu q )\,,
\notag\\
\Op^{(8)}_{99} &= \partial^2(\bar e_L\gamma^\mu \nu_L) (\bar{q} \textbf{T}^-\gamma_\mu q )+h.c.\,,
&
\Op^{(8)}_{100} &= \partial^2(\nu_L^TC\gamma^\mu e_R) (\bar{q} \textbf{T}^+\gamma_\mu q )+h.c.\,,
\notag\\
\Op^{(8)}_{101} &= \partial^2(\bar e_L\sigma^{\mu\nu} e_R) (\bar{q} \textbf{T}^0\sigma_{\mu\nu} q )+h.c.\,,
&
\Op^{(8)}_{102} &= \partial^2(\nu_L^TC\sigma^{\mu\nu} \nu_L) (\bar{q} \textbf{T}^0\sigma_{\mu\nu} q )+h.c.\,,
\notag\\
\Op^{(8)}_{103} &= \partial^2(\bar e_L\sigma^{\mu\nu} e_R) (\bar{q} \sigma_{\mu\nu} q )+h.c.\,,
&
\Op^{(8)}_{104} &= \partial^2(\nu_L^TC\sigma^{\mu\nu} \nu_L) (\bar{q} \sigma_{\mu\nu} q )+h.c.\,,
\notag\\
\Op^{(8)}_{105} &= \partial^2(\bar e_R\sigma^{\mu\nu} \nu_L) (\bar{q} \textbf{T}^-\sigma_{\mu\nu} q )+h.c.\,,
&
\Op^{(8)}_{106} &= \partial^2( \nu_L^TC\sigma^{\mu\nu} e_R) (\bar{q} \textbf{T}^+\sigma_{\mu\nu} q )+h.c.\,,
\notag\\
\Op^{(8)}_{107} &=  (\bar e_Li\lrpartial^\mu e_R)(\bar{q} \textbf{T}^0i\lrpartial_\mu q )+h.c.\,,
&
\Op^{(8)}_{108} &=  (\nu_L^TCi\lrpartial^\mu \nu_L)(\bar{q} \textbf{T}^0i\lrpartial_\mu q )+h.c.\,,
\notag\\
\Op^{(8)}_{109} &=  (\bar e_Li\lrpartial^\mu e_R)(\bar{q} i\lrpartial_\mu q )+h.c.\,,
&
\Op^{(8)}_{110} &= (\nu_L^TCi\lrpartial^\mu \nu_L) (\bar{q} i\lrpartial_\mu q )+h.c.\,,
\notag\\
\Op^{(8)}_{111} &= (\bar e_Ri\lrpartial^\mu\nu_L) (\bar{q} \textbf{T}^-i\lrpartial_\mu q )+h.c.\,,
&
\Op^{(8)}_{112} &= (\nu_L^TCi\lrpartial^\mu e_L) (\bar{q} \textbf{T}^+i\lrpartial_\mu q )+h.c.\,,
\end{align}
\begin{align}
C-P-:\notag\\
\Op^{(8)}_{113} &= F_{\mu\rho}F^\rho_\nu  (\bar{q} \textbf{T}^0\gamma^5\gamma^\mu\lrpartial^\nu q )\,,
&
\Op^{(8)}_{114} &= F_{\mu\rho}F^\rho_\nu  (\bar{q} \gamma^5\gamma^\mu\lrpartial^\nu q )\,,
\notag\\
\Op^{(8)}_{115} &=  (\bar e_L\gamma^\mu\lrpartial^\nu e_L) (\bar{q} \textbf{T}^0\gamma^5\gamma_\mu\lrpartial_\nu q )\,,
&
\Op^{(8)}_{116} &=  (\bar e_R\gamma^\mu\lrpartial^\nu e_R) (\bar{q} \textbf{T}^0\gamma^5\gamma_\mu\lrpartial_\nu q )\,,
\notag\\
\Op^{(8)}_{117} &=  (\bar\nu_L\gamma^\mu\lrpartial^\nu\nu_L) (\bar{q} \textbf{T}^0\gamma^5\gamma_\mu\lrpartial_\nu q )\,,
&
\Op^{(8)}_{118} &=  (\bar e_L\gamma^\mu\lrpartial^\nu e_L) (\bar{q} \gamma^5\gamma_\mu\lrpartial_\nu q )\,,
\notag\\
\Op^{(8)}_{119} &=  (\bar e_R\gamma^\mu\lrpartial^\nu e_R) (\bar{q} \gamma^5\gamma_\mu\lrpartial_\nu q )\,,
&
\Op^{(8)}_{120} &=  (\bar\nu_L\gamma^\mu\lrpartial^\nu\nu_L) (\bar{q} \gamma^5\gamma_\mu\lrpartial_\nu q )\,,
\notag\\
\Op^{(8)}_{121} &=  (\bar e_L\gamma^\mu\lrpartial^\nu\nu_L) (\bar{q} \textbf{T}^-\gamma^5\gamma_\mu\lrpartial_\nu q )+h.c.\,,
&
\Op^{(8)}_{122} &=  (\nu_L^TC\gamma^\mu\lrpartial^\nu e_R) (\bar{q} \textbf{T}^+\gamma^5\gamma_\mu\lrpartial_\nu q )+h.c.\,,
\notag\\
\Op^{(8)}_{123} &=  (\bar e_Li\lrpartial^\mu e_R) (\bar{q} \textbf{T}^0i\gamma^5i\lrpartial_\mu q )+h.c.\,,
&
\Op^{(8)}_{124} &=  (\nu_L^TCi\lrpartial^\mu \nu_L) (\bar{q} \textbf{T}^0i\gamma^5i\lrpartial_\mu q )+h.c.\,,
\notag\\
\Op^{(8)}_{125} &=  (\bar e_Li\lrpartial^\mu e_R) (\bar{q} i\gamma^5i\lrpartial_\mu q )+h.c.\,,
&
\Op^{(8)}_{126} &=  (\nu_L^TCi\lrpartial^\mu \nu_L) (\bar{q} i\gamma^5i\lrpartial_\mu q )+h.c.\,,
\notag\\
\Op^{(8)}_{127} &=  (\bar e_Ri\lrpartial^\mu\nu_L)    (\bar{q}\textbf{T}^-i\gamma^5i\lrpartial_\mu q )+h.c.\,,
&
\Op^{(8)}_{128} &= (\nu_L^TCi\lrpartial^\mu e_L)  (\bar{q}\textbf{T}^+i\gamma^5i\lrpartial_\mu q )+h.c.\,.
\end{align}

The operators with two quark bilinears are
\begin{align}
 C+P+:\notag\\
\Op^{(8)}_{129}&= (\bar q q )\partial^2( \bar q  q )\,,
&
\Op^{(8)}_{130}&= (\bar q q )\partial^2( \bar q \textbf{T}^0q )\,,
\notag\\
\Op^{(8)}_{131}&= (\bar q\textbf{T} q )\partial^2( \bar q \textbf{T} q )\,,
&
\Op^{(8)}_{132}&= (\bar qi\gamma^5 q )\partial^2( \bar q i\gamma^5 q )\,,
\notag\\
\Op^{(8)}_{133}&= (\bar qi\gamma^5 q )\partial^2( \bar q \textbf{T}^0i\gamma^5 q )\,,
&
\Op^{(8)}_{134}&= (\bar q\textbf{T}i\gamma^5 q )\partial^2( \bar q \textbf{T}i\gamma^5 q )\,,
\notag\\
\Op^{(8)}_{135}&= (\bar q\gamma^\mu q )\partial^2( \bar q \gamma_\mu q )\,,
&
\Op^{(8)}_{136}&= (\bar q\gamma^\mu q )\partial^2( \bar q \textbf{T}^0\gamma_\mu q )\,,
\notag\\
\Op^{(8)}_{137}&= (\bar q\textbf{T}\gamma^\mu q )\partial^2( \bar q \textbf{T}\gamma_\mu q )\,,
&
\Op^{(8)}_{138}&= (\bar q\gamma^5\gamma^\mu q )\partial^2( \bar q \gamma^5\gamma_\mu q )\,,
\notag\\
\Op^{(8)}_{139}&= (\bar q\gamma^5\gamma^\mu q )\partial^2( \bar q \textbf{T}^0\gamma^5\gamma_\mu q )\,,
&
\Op^{(8)}_{140}&= (\bar q\textbf{T}\gamma^5\gamma^\mu q )\partial^2( \bar q \textbf{T}\gamma^5\gamma_\mu q )\,,
\notag\\
\Op^{(8)}_{141}&= (\bar qi\lrpartial^\mu q )( \bar q i\lrpartial_\mu q )\,,
&
\Op^{(8)}_{142}&= (\bar qi\lrpartial^\mu q )( \bar q \textbf{T}^0i\lrpartial_\mu q )\,,
\notag\\
\Op^{(8)}_{143}&= (\bar q\textbf{T}i\lrpartial^\mu q )( \bar q \textbf{T}i\lrpartial_\mu q )\,,
&
\Op^{(8)}_{144}&= (\bar qi\gamma^5i\lrpartial^\mu q )( \bar q i\gamma^5i\lrpartial_\mu q )\,,
\notag\\
\Op^{(8)}_{145}&= (\bar qi\gamma^5i\lrpartial^\mu q )( \bar q \textbf{T}^0i\gamma^5i\lrpartial_\mu q )\,,
&
\Op^{(8)}_{146}&= (\bar q\textbf{T}i\gamma^5i\lrpartial^\mu q )( \bar q \textbf{T}i\gamma^5i\lrpartial_\mu q )\,,
\notag\\
\Op^{(8)}_{147}&= (\bar q\gamma^\mu\lrpartial^\nu q )( \bar q \gamma_\mu\lrpartial_\nu q )\,,
&
\Op^{(8)}_{148}&= (\bar q\gamma^\mu\lrpartial^\nu q )( \bar q \textbf{T}^0\gamma_\mu\lrpartial_\nu q )\,,
\notag\\
\Op^{(8)}_{149}&= (\bar q\textbf{T}\gamma^\mu\lrpartial^\nu q )( \bar q \textbf{T}\gamma_\mu\lrpartial_\nu q )\,,
&
\Op^{(8)}_{150}&= (\bar q\gamma^5\gamma^\mu\lrpartial^\nu q )( \bar q \gamma^5\gamma_\mu\lrpartial_\nu q )\,,
\notag\\
\Op^{(8)}_{151}&= (\bar q\gamma^5\gamma^\mu\lrpartial^\nu q )( \bar q \textbf{T}^0\gamma^5\gamma_\mu\lrpartial_\nu q )\,,
&
\Op^{(8)}_{152}&= (\bar q\textbf{T}\gamma^5\gamma^\mu\lrpartial^\nu q )( \bar q \textbf{T}\gamma^5\gamma_\mu\lrpartial_\nu q )\,,
\notag\\
\Op^{(8)}_{153}&= (\bar qT^A q )\partial^2( \bar q T^A q )\,,
&
\Op^{(8)}_{154}&= (\bar qT^A q )\partial^2( \bar qT^A \textbf{T}^0q )\,,
\notag\\
\Op^{(8)}_{155}&= (\bar qT^A\textbf{T} q )\partial^2( \bar qT^A \textbf{T} q )\,,
&
\Op^{(8)}_{156}&= (\bar qT^Ai\gamma^5 q )\partial^2( \bar q T^Ai\gamma^5 q )\,,
\notag\\
\Op^{(8)}_{157}&= (\bar qT^Ai\gamma^5 q )\partial^2( \bar q T^A\textbf{T}^0i\gamma^5 q )\,,
&
\Op^{(8)}_{158}&= (\bar qT^A\textbf{T}i\gamma^5 q )\partial^2( \bar q T^A\textbf{T}_2i\gamma^5 q )\,,
\notag\\
\Op^{(8)}_{159}&= (\bar qT^A\gamma^\mu q )\partial^2( \bar q T^A\gamma_\mu q )\,,
&
\Op^{(8)}_{160}&= (\bar qT^A\gamma^\mu q )\partial^2( \bar q T^A\textbf{T}^0\gamma_\mu q )\,,
\notag\\
\Op^{(8)}_{161}&= (\bar qT^A\textbf{T}\gamma^\mu q )\partial^2( \bar q T^A\textbf{T}\gamma_\mu q )\,,
&
\Op^{(8)}_{162}&= (\bar qT^A\gamma^5\gamma^\mu q )\partial^2( \bar q T^A\gamma^5\gamma_\mu q )\,,
\notag\\
\Op^{(8)}_{163}&= (\bar qT^A\gamma^5\gamma^\mu q )\partial^2( \bar q T^A\textbf{T}^0\gamma^5\gamma_\mu q )\,,
&
\Op^{(8)}_{164}&= (\bar qT^A\textbf{T}\gamma^5\gamma^\mu q )\partial^2( \bar q T^A\textbf{T}\gamma^5\gamma_\mu q )\,,
\notag\\
\Op^{(8)}_{165}&= (\bar qT^Ai\lrpartial^\mu q )( \bar qT^A i\lrpartial_\mu q )\,,
&
\Op^{(8)}_{166}&= (\bar qT^Ai\lrpartial^\mu q )( \bar qT^A \textbf{T}^0i\lrpartial_\mu q )\,,
\notag\\
\Op^{(8)}_{167}&= (\bar qT^A\textbf{T}i\lrpartial^\mu q )( \bar q T^A\textbf{T}i\lrpartial_\mu q )\,,
&
\Op^{(8)}_{168}&= (\bar qT^Ai\gamma^5i\lrpartial^\mu q )( \bar q T^Ai\gamma^5i\lrpartial_\mu q )\,,
\notag\\
\Op^{(8)}_{169}&= (\bar qT^Ai\gamma^5i\lrpartial^\mu q )( \bar q T^A\textbf{T}^0i\gamma^5i\lrpartial_\mu q )\,,
&
\Op^{(8)}_{170}&= (\bar qT^A\textbf{T}i\gamma^5i\lrpartial^\mu q )( \bar qT^A \textbf{T}i\gamma^5i\lrpartial_\mu q )\,,
\notag\\
\Op^{(8)}_{171}&= (\bar qT^A\gamma^\mu\lrpartial^\nu q )( \bar q T^A\gamma_\mu\lrpartial_\nu q )\,,
&
\Op^{(8)}_{172}&= (\bar qT^A\gamma^\mu\lrpartial^\nu q )( \bar q T^A\textbf{T}^0\gamma_\mu\lrpartial_\nu q )\,,
\notag\\
\Op^{(8)}_{173}&= (\bar qT^A\textbf{T}\gamma^\mu\lrpartial^\nu q )( \bar qT^A \textbf{T}\gamma_\mu\lrpartial_\nu q )\,,
&
\Op^{(8)}_{174}&= (\bar qT^A\gamma^5\gamma^\mu\lrpartial^\nu q )( \bar q T^A\gamma^5\gamma_\mu\lrpartial_\nu q )\,,
\notag\\
\Op^{(8)}_{175}&= (\bar qT^A\gamma^5\gamma^\mu\lrpartial^\nu q )( \bar qT^A \textbf{T}^0\gamma^5\gamma_\mu\lrpartial_\nu q )\,,
&
\Op^{(8)}_{176}&= (\bar qT^A\textbf{T}\gamma^5\gamma^\mu\lrpartial^\nu q )( \bar q T^A\textbf{T}\gamma^5\gamma_\mu\lrpartial_\nu q )\,,
\end{align}
\begin{align}
C+P-:\notag\\
\Op^{(8)}_{177}&= (\bar q q )\partial^2( \bar qi\gamma^5 q )\,,
&
\Op^{(8)}_{178}&= (\bar q q )\partial^2( \bar q \textbf{T}^0i\gamma^5 q )\,,
\notag\\
\Op^{(8)}_{179}&= (\bar q\textbf{T}^0 q )\partial^2( \bar q i\gamma^5 q )\,,
&
\Op^{(8)}_{180}&= (\bar q {\textbf{T}} q )\partial^2(\bar q {\textbf{T}}i\gamma^5 q )\,,
\notag\\
\Op^{(8)}_{181}&= (\bar qi\lrpartial^\mu q )( \bar qi\gamma^5i\lrpartial_\mu q )\,,
&
\Op^{(8)}_{182}&= (\bar qi\lrpartial^\mu q )( \bar q \textbf{T}^0i\gamma^5i\lrpartial_\mu q )\,,
\notag\\
\Op^{(8)}_{183}&= (\bar q\textbf{T}^0i\lrpartial^\mu q )( \bar q i\gamma^5i\lrpartial_\mu q )\,,
&
\Op^{(8)}_{184}&= (\bar q\textbf{T}i\lrpartial^\mu q )( \bar q \textbf{T}i\gamma^5i\lrpartial_\mu q )\,,
\notag\\
\Op^{(8)}_{185}&= (\bar q T^Aq )\partial^2( \bar qT^Ai\gamma^5 q )\,,
&
\Op^{(8)}_{186}&= (\bar qT^A q )\partial^2( \bar qT^A \textbf{T}^0i\gamma^5 q )\,,
\notag\\
\Op^{(8)}_{187}&= (\bar qT^A\textbf{T}^0 q )\partial^2( \bar q T^Ai\gamma^5 q )\,,
&
\Op^{(8)}_{188}&= (\bar qT^A {\textbf{T}} q )\partial^2(\bar q T^A{\textbf{T}}i\gamma^5 q )\,,
\notag\\
\Op^{(8)}_{189}&= (\bar qT^Ai\lrpartial^\mu q )( \bar qT^Ai\gamma^5i\lrpartial_\mu q )\,,
&
\Op^{(8)}_{190}&= (\bar qT^Ai\lrpartial^\mu q )( \bar qT^A \textbf{T}^0i\gamma^5i\lrpartial_\mu q )\,,
\notag\\
\Op^{(8)}_{191}&= (\bar qT^A\textbf{T}^0i\lrpartial^\mu q )( \bar q T^Ai\gamma^5i\lrpartial_\mu q )\,,
&
\Op^{(8)}_{192}&= (\bar qT^A\textbf{T}i\lrpartial^\mu q )( \bar q T^A\textbf{T}i\gamma^5i\lrpartial_\mu q )\,,
\\
C-P-:\notag\\
\Op^{(8)}_{193}&= (\bar q\gamma^\mu q )\partial^2( \bar q\gamma^5\gamma_\mu q )\,,
&
\Op^{(8)}_{194}&= (\bar q \gamma^\mu q )\partial^2( \bar q \textbf{T}^0\gamma^5\gamma_\mu q )\,,
\notag\\
\Op^{(8)}_{195}&= (\bar q\textbf{T}^0\gamma^\mu q )\partial^2( \bar q \gamma^5\gamma_\mu q )\,,
&
\Op^{(8)}_{196}&= (\bar q\gamma^\mu\textbf{T} q )\partial^2( \bar q \textbf{T}\gamma^5\gamma_\mu q )\,,
\notag\\
\Op^{(8)}_{197}&= (\bar q\gamma^\mu\lrpartial^\nu q )( \bar q\gamma^5\gamma_\mu\lrpartial_\nu q )\,,
&
\Op^{(8)}_{198}&= (\bar q \gamma^\mu\lrpartial^\nu q )( \bar q \textbf{T}^0\gamma^5\gamma_\mu\lrpartial_\nu q )\,,
\notag\\
\Op^{(8)}_{199}&= (\bar q\textbf{T}^0\gamma^\mu\lrpartial^\nu q )( \bar q \gamma^5\gamma_\mu\lrpartial_\nu q )\,,
&
\Op^{(8)}_{200}&= (\bar q\textbf{T}\gamma^\mu\lrpartial^\nu q )( \bar q \textbf{T}\gamma^5\gamma_\mu\lrpartial_\nu q )\,,
\notag\\
\Op^{(8)}_{201}&= (\bar qT^A\gamma^\mu q )\partial^2( \bar qT^A\gamma^5\gamma_\mu q )\,,
&
\Op^{(8)}_{202}&= (\bar q T^A\gamma^\mu q )\partial^2( \bar q T^A\textbf{T}^0\gamma^5\gamma_\mu q )\,,
\notag\\
\Op^{(8)}_{203}&= (\bar qT^A\textbf{T}^0\gamma^\mu q )\partial^2( \bar q T^A\gamma^5\gamma_\mu q )\,,
&
\Op^{(8)}_{204}&= (\bar qT^A\gamma^\mu\textbf{T} q )\partial^2( \bar qT^A \textbf{T}\gamma^5\gamma_\mu q )\,,
\notag\\
\Op^{(8)}_{205}&= (\bar qT^A\gamma^\mu\lrpartial^\nu q )( \bar qT^A\gamma^5\gamma_\mu\lrpartial_\nu q )\,,
&
\Op^{(8)}_{206}&= (\bar q T^A\gamma^\mu\lrpartial^\nu q )( \bar q T^A\textbf{T}^0\gamma^5\gamma_\mu\lrpartial_\nu q )\,,
\notag\\
\Op^{(8)}_{207}&= (\bar qT^A\textbf{T}^0\gamma^\mu\lrpartial^\nu q )( \bar q T^A\gamma^5\gamma_\mu\lrpartial_\nu q )\,,
&
\Op^{(8)}_{208}&= (\bar qT^A\textbf{T}\gamma^\mu\lrpartial^\nu q )( \bar qT^A \textbf{T}\gamma^5\gamma_\mu\lrpartial_\nu q )\,.
\end{align}

\subsubsection*{Dimension-9}

The dimension-9 operators could contain 3 bilinears. 
However, the Lorentz structures of the quark sector have been listed in Tab.~\ref{tab:quarkbilinear} and here we only consider the operators with two quark bilinears and one lepton bilinear which violates the lepton number, since they are of phenomenological importance.
\begin{align}
C+P+:\notag\\
\Op^{(9)}_1&=(\bar q\textbf{T} q)(\bar q\textbf{T} q)(\bar e e^c)+h.c.\,,
&
\Op^{(9)}_2&=(\bar q\gamma^\mu\textbf{T} q)(\bar q\gamma_\mu\textbf{T} q)(\bar e e^c)+h.c.\,,
\notag\\
\Op^{(9)}_3&=(\bar q\gamma^5\textbf{T} q)(\bar q\gamma^5\textbf{T} q)(\bar e e^c)+h.c.\,,
&
\Op^{(9)}_4&=(\bar q\gamma^5\gamma^\mu\textbf{T}  q)(\bar q\gamma^5\gamma_\mu\textbf{T} q)(\bar e e^c)+h.c.\,,
\notag\\
\Op^{(9)}_5&=(\bar q\gamma^5\textbf{T} q)(\bar q\gamma^5\gamma_\mu\textbf{T} q)(\bar e \gamma^\mu e^c)+h.c.\,,
&
\Op^{(9)}_6&=(\bar q\gamma^5\textbf{T} q)(\bar q\gamma^5\gamma_\mu\textbf{T} q)(\bar e \gamma^\mu e^c)+h.c.\,,
\notag\\
\Op^{(9)}_7&=(\bar qT^A\textbf{T} q)(\bar qT^A\textbf{T} q)(\bar e e^c)+h.c.\,,
&
\Op^{(9)}_8&=(\bar qT^A\gamma^\mu\textbf{T} q)(\bar qT^A\gamma_\mu\textbf{T} q)(\bar e e^c)+h.c.\,,
\notag\\
\Op^{(9)}_9&=(\bar qT^A\gamma^5\textbf{T} q)(\bar qT^A\gamma^5\textbf{T} q)(\bar e e^c)+h.c.\,,
&
\Op^{(9)}_{10}&=(\bar qT^A\gamma^5\gamma^\mu\textbf{T}  q)(\bar qT^A\gamma^5\gamma_\mu\textbf{T} q)(\bar e e^c)+h.c.\,,
\notag\\
\Op^{(9)}_{11}&=(\bar qT^A\gamma^5\textbf{T} q)(\bar qT^A\gamma^5\gamma_\mu\textbf{T} q)(\bar e \gamma^\mu e^c)+h.c.\,,
&
\Op^{(9)}_{12}&=(\bar qT^A\gamma^5\textbf{T} q)(\bar qT^A\gamma^5\gamma_\mu\textbf{T} q)(\bar e \gamma^\mu e^c)+h.c.\,,
\\
C+P-:\notag\\
\Op^{(9)}_{13}&=(\bar q\gamma^5\textbf{T} q)(\bar q\textbf{T} q)(\bar e e^c)+h.c.\,,
&
\Op^{(9)}_{14}&=(\bar q\textbf{T} q)(\bar q\gamma^5\gamma_\mu\textbf{T} q)(\bar e \gamma^\mu e^c)+h.c.\,,
\notag\\
\Op^{(9)}_{15}&=(\bar qT^A\gamma^5\textbf{T} q)(\bar qT^A\textbf{T} q)(\bar e e^c)+h.c.\,,
&
\Op^{(9)}_{16}&=(\bar qT^A\textbf{T} q)(\bar qT^A\gamma^5\gamma_\mu\textbf{T} q)(\bar e \gamma^\mu e^c)+h.c.\,,
\notag\\
\\
C-P+:\notag\\
\Op^{(9)}_{17}&=(\bar q\textbf{T} q)(\bar q\gamma_\mu\textbf{T} q)(\bar e \gamma^\mu e^c)+h.c.\,,
&
\Op^{(9)}_{18}&=(\bar qT^A\textbf{T} q)(\bar qT^A\gamma_\mu\textbf{T} q)(\bar e \gamma^\mu e^c)+h.c.\,,
\\
C-P-:\notag\\
\Op^{(9)}_{19}&=(\bar q\gamma^5\gamma^\mu\textbf{T} q)(\bar q\gamma_\mu\textbf{T} q)(\bar e e^c)+h.c.\,,
&
\Op^{(9)}_{20}&=(\bar q\gamma^5\textbf{T}_ q)(\bar q\gamma_\mu\textbf{T} q)(\bar e \gamma^\mu e^c)+h.c.\,,
\notag\\
\Op^{(9)}_{21}&=(\bar qT^A\gamma^5\gamma^\mu\textbf{T} q)(\bar qT^A\gamma_\mu\textbf{T} q)(\bar e e^c)+h.c.\,,
&
\Op^{(9)}_{22}&=(\bar qT^A\gamma^5\textbf{T} q)(\bar qT^A\gamma_\mu\textbf{T} q)(\bar e \gamma^\mu e^c)+h.c.\,.
\end{align}

\section{The Chiral Lagrangian with Spurion Field}
\label{sec:chpt}

Similar to the LEFT, in this section we reformulate the Chiral perturbation theory (ChPT) Lagrangian in terms of the spurions rather than the external sources to take the electroweak interactions into account. We present the building blocks and construct the effective operators of ChPT, which will be matched from the LEFT in Sec.~\ref{sec:matching}.

\subsection{Chiral Perturbation Theory}
The ChPT~\cite{Weinberg:1968de,Weinberg:1978kz,Gasser:1983yg,Gasser:1984gg,Gasser:1987rb} is a low-energy theory of the LEFT below $\Lambda_\chi\sim 1\text{ GeV}$, where the chiral symmetry $SU(3)_{\mathbf{L}}\times SU(3)_{\mathbf{R}}$ is spontaneously broken down to the subgroup $SU(3)_V$. According to the Goldstone theorem~\cite{Goldstone:1961eq, Goldstone:1962es}, the broken symmetries generate 8 pseudo-Nambu-Goldstone bosons (NGBs) $\phi^a$, and these NGBs are pseudoscalar bosons with negative parity which can be parameterized by the $SU(3)_V$ as
\begin{equation}
    \Pi(x)=\sum_{a=1}^8\phi^a(x)\lambda^a = \left(\begin{array}{ccc}
    \pi^0 + \frac{1}{\sqrt{3}}\eta & \sqrt{2}\pi^+ & \sqrt{2}K^+ \\
    \sqrt{2}\pi^- & -\pi^0+\frac{1}{\sqrt{3}}\eta & \sqrt{2}K^0 \\
    \sqrt{2}K^- & \sqrt{2}~\Bar{K}^0 & -\frac{2}{\sqrt{3}}\eta
    \end{array}\right)\,.
\end{equation}

The NGBs of the ChPT can be described under the CCWZ coset construction, by which they are characterized non-linearly, and are collected in a unitary matrix
\begin{equation}
    u(x) = \exp\left(\frac{\phi(x)^a \lambda^a
    }{2f}\right)\,,
\end{equation}
which transforms under the chiral symmetry as $u(x)\rightarrow \mathbf{R}u(x)V^{-1}=Vu(x)\mathbf{L}^{-1}$, where $\mathbf{L}\in SU(3)_\mathbf{L}\,,\mathbf{R}\in SU(3)_\mathbf{R}$ and $V\in SU(3)_V$. In terms of $u$ we can define the building block $u_\mu$ and the covariant derivative as
\begin{align}
    u_\mu &= i(u^\dagger\partial_\mu u-u\partial_\mu u^\dagger ) \,,\notag \\
    D_\mu&=\partial_\mu+\frac{1}{2}(u^\dagger\partial_\mu u+u\partial_\mu u^\dagger )\,. \label{eq:u_without_lr}
\end{align}
In addition, the baryons are also included in the ChPT and compose a matrix $B$ of the adjoint representation of the $SU(3)_V$, 
\begin{equation}
    B = \frac{B^a\lambda^a}{\sqrt{2}} = \left(\begin{array}{ccc}
\frac{1}{\sqrt{2}}\Sigma^0+\frac{1}{\sqrt{6}}\Lambda & \Sigma^+ & p \\
\Sigma^- & -\frac{1}{\sqrt{2}}\Sigma^0 + \frac{1}{\sqrt{6}}\Lambda & n \\
\Xi^- & \Xi^0 & -\frac{2}{\sqrt{6}}\Lambda 
    \end{array}\right)\,,
\end{equation}
which transforms as $B\rightarrow VBV^{-1}$, where $V\in SU(3)_V$. 

\subsubsection*{The External Source Method}
Traditionally, the non-hadron fields are introduced in the ChPT via the external sources. The building blocks become
\begin{align}
    u_\mu^\prime &= i(u^\dagger(\partial_\mu -ir_\mu)u-u(\partial_\mu -il_\mu)u^\dagger ) \,,\notag \\
    \chi^{\pm} &= u^\dagger \chi u^\dagger \pm u \chi^\dagger u \,, \notag \\
    f_{\pm}{}_{\mu\nu} &= u^\dagger f^R_{\mu\nu} u \pm u f^L_{\mu\nu} u^\dagger \,,\notag\\
    t_{\pm}{}_{\mu\nu}&=u^\dagger t_{\mu\nu}u^\dagger\pm ut_{\mu\nu}^\dagger u\,,\label{eq: building blocks}
\end{align}
and the baryon $B$, where
\begin{align}
    \chi &= 2B_0(s+ip) \,,\\
    f^R_{\mu\nu} &= \partial_\mu r_\nu - \partial_\nu r_\mu -i [r_\mu,r_\nu] \,,\quad r_\mu = v_\mu + a_\mu \,, \notag \\
    f^L_{\mu\nu} &= \partial_\mu l_\nu - \partial_\nu l_\mu -i [l_\mu,l_\nu] \,,\quad l_\mu = v_\mu - a_\mu \,.\label{eq: sources}
\end{align}
If we specify the external sources as the lepton bilinears or the photon field, for example, 
\begin{align}
    s\rightarrow(\bar e e)\,,\quad p\rightarrow(\bar e \gamma^5 e)\,,\quad v_\mu\rightarrow (\bar e\gamma_\mu e)\,,\quad a_\mu\rightarrow(\bar e\gamma^5\gamma_\mu e)\,,\quad\bar t_{\mu\nu}\rightarrow(\bar e\sigma_{\mu\nu}e) \,,
\end{align}
the corresponding hadronic external sources would become
\begin{align}
    \chi^+ &\sim u^\dagger \left[T_s( \bar ee)+iT_p(\bar e\gamma^5e)\right]u^\dagger + u \left[T_s( \bar ee)+iT_p(\bar e\gamma^5e)\right]^\dagger u \,,\label{eq:ext_1}\\
    \chi^- &\sim  u^\dagger \left[T_s( \bar ee)+iT_p(\bar e\gamma^5e)\right]u^\dagger - u \left[T_s( \bar ee)+iT_p(\bar e\gamma^5e)\right]^\dagger u \,, \\
    u_\mu^\prime &\sim i\{u^\dagger[\partial_\mu -i(T_v \bar e\gamma_\mu e)+(T_a\bar e\gamma^5\gamma_\mu e)]u-u[\partial_\mu -i(T_v\bar e\gamma_\mu e)+(T_a\bar e\gamma^5\gamma_\mu e)]u^\dagger \}  \,,\\
    t^{\mu\nu}_+&\sim u^\dagger [T_t(\bar e\sigma_{\mu\nu}e)]u^\dagger+u[T_t(\bar e\sigma_{\mu\nu}e)]^\dagger u\,,\\
    t^{\mu\nu}_-&\sim u^\dagger [T_t(\bar e\sigma_{\mu\nu}e)]u^\dagger-u[T_t(\bar e\sigma_{\mu\nu}e)]^\dagger u\,,\label{eq:ext_2} 
\end{align}
where $T$ are the Wilson coefficients of the LEFT Lagrangian in the flavor space. It should be noted that the types of the external sources are determined by the properties of the quark bilinears, thus the scalar external source $s$ can also become $s\rightarrow (\bar e\gamma^5 e)$ and other external sources would be similar~\cite{Bishara:2016hek}. In particular, $u'_\mu$ is different from $u_\mu$ in Eq.~\eqref{eq:u_without_lr} because of the external sources $a_\mu$ and $v_\mu$,
\begin{equation}
    u'_\mu = u_\mu + (u^\dagger a_\mu u+ua_\mu u^\dagger)+(u^\dagger v_\mu u-uv_\mu u^\dagger)\,.
\end{equation}
Thus the whole particle spectrum of the ChPT with the external sources and their transformation under the $SU(3)_V$ can be obtained,
\begin{equation}
    \left(\begin{array}{c}
         u_\mu^\prime  \\
          \chi_-\\
          \chi_+\\
          f_{-\mu\nu}\\
          f_{+\mu\nu}\\
          t_{-\mu\nu}\\
          t_{+\mu\nu}\\
          B
    \end{array}\right)\rightarrow\left(\begin{array}{c}
         Vu_\mu^\prime V^\dagger \\
          V\chi_-V^\dagger\\
          V\chi_+V^\dagger\\
          Vf_{-\mu\nu}V^\dagger\\
          Vf_{+\mu\nu}V^\dagger\\
          Vt_{-\mu\nu}V^\dagger\\
          Vt_{+\mu\nu}V^\dagger\\
          VBV^\dagger
    \end{array}\right)\,,\quad V\in SU(3)_V\,,
\end{equation}
and we list the $SU(3)_V$ representations of these building block in Tab.~\ref{Lagrangian_spurion}. 
The chiral Lagrangian with the external sources has been constructed for the pure meson sector in Ref.~\cite{Bijnens:1999sh,Ebertshauser:2001nj,Bijnens:2001bb,Bijnens:2018lez, Bijnens:2023hyv,Li:2024ghg} and the meson-baryon sector in Ref.~\cite{Ecker:1995rk,Fettes:1998ud,Fettes:2000gb,Oller:2006yh,Frink:2006hx,Jiang:2016vax,Song:2024fae}.

\subsubsection*{The Spurion Method}

Inspired by the reformulation of the LEFT in Sec.~\ref{sec:LEFT}, we can reformulate the ChPT by the spurion $\mathbf{T}$ with different dressing forms, $\Sigma_\pm$ and $Q_\pm$, that
\begin{align}
    \Sigma_{\pm} &= u^\dagger \textbf{T} u^\dagger \pm u \textbf{T}^\dagger u \,,\label{eq:chpt_sigma_1}\notag\\
    Q_{\pm}&=u^\dagger\textbf{T}u\pm u\textbf{T}^\dagger u^\dagger\,.
\end{align}
Let the leptons and the photon as building blocks of the ChPT. the building blocks of the ChPT are
\begin{equation}
\label{eq:building_blocks}
\left(u_\mu\,, \quad \Sigma_\pm,\quad Q_\pm\,,\quad B \,,\quad e_L\,,\quad e_R \,,\quad \nu_L \,,\quad F_{\mu\nu}\right)^T\,,
\end{equation}
which transform under the $SU(3)_V$ symmetry as
\begin{equation}
    \left(\begin{array}{c}
         u_\mu  \\
          \Sigma_-\\
          \Sigma_+\\
          Q_-\\
          Q_+\\
          B\\
          e_R\\
          e_L\\
          \nu_L\\
          F_{\mu\nu}
    \end{array}\right)\rightarrow\left(\begin{array}{c}
         Vu_\mu V^{\dagger} \\
          V\Sigma_-V^{\dagger}\\
          V\Sigma_+V^{\dagger}\\
          VQ_-V^{\dagger}\\
          VQ_+V^{\dagger}\\
          VBV^{\dagger}\\
          e_R\\
          e_L\\
          \nu_L\\
          F_{\mu\nu}
    \end{array}\right)\,,\quad V\in SU(3)_V\,.
\end{equation}
The $SU(3)_V$ representations of the building blocks of the chiral Lagrangian have been listed in Tab.~\ref{Lagrangian_spurion}. There is no external source, of which the leptons and the photon are decoupled as fundamental building blocks, and the coefficients are attributed to $\Sigma_\pm\,,Q_\pm$. Besides, the $CP$ properties of the building block and the baryon bilinears are obtained in Tab.~\ref{tab: building block}. In particular, the additional building blocks $\Sigma_\pm$ and $Q_\pm$ are of definite $CP$ properties. 
\begin{table}
    \centering
    \begin{tabular}{|c|c||c|c|}
    \hline
    \multicolumn{2}{|c||}{external sources method} & \multicolumn{2}{c|}{spurion method} \\
    \hline
     building blocks&$SU(3)_V$&building blocks&$SU(3)_V$\\
    \hline
         $u'_\mu$&$\mathbf{8}$&$u_\mu$&$\mathbf{8}$  \\
         \hline
         $B$&$\mathbf{8}$&$B$&$\mathbf{8}$  \\
         \hline
         $\chi_\pm$&$\mathbf{1}\oplus \mathbf{8}$&$\Sigma_\pm$&$\mathbf{8}$  \\
         $f_\pm{}_{\mu\nu}$&$\mathbf{8}$&$Q_\pm$&$\mathbf{8}$  \\
         $t_\pm{}_{\mu\nu}$&$\mathbf{1}\oplus \mathbf{8}$&$e_L$&$\mathbf{1}$  \\
         & & $e_R$&$\mathbf{1}$  \\
         & & $\nu_L$&$\mathbf{1}$  \\
         & & $F_{\mu\nu}$&$\mathbf{1}$  \\
         & & $\tilde{F}_{\mu\nu}$&$\mathbf{1}$  \\
         \hline
    \end{tabular}
    \caption{The building blocks of the ChPT and their $SU(3)_V$ representations in the external source method on the left and the spurion method on the right.}
    \label{Lagrangian_spurion}
\end{table}
In particular, the $\Sigma_\pm\,,\,Q_\pm$ are related to the external sources in Eq.~\eqref{eq:ext_1} to Eq.~\eqref{eq:ext_2} by 
\begin{align}
u_\mu^\prime &\sim i(u^\dagger\partial_\mu u-u\partial_\mu u^\dagger )+Q_+(\bar e\gamma^5\gamma_\mu e)+Q_-(\bar e\gamma^5\gamma_\mu e)\,,\label{eq:spurion_redef_1}\\
    \chi^+ &\sim \Sigma_+(\bar ee) + i \Sigma_- (\bar e\gamma^5 e)\,,\\
    \chi^- &\sim \Sigma_-(\bar ee) + i \Sigma_+ (\bar e\gamma^5 e)\,,\\
    t^{\mu\nu}_+ &\sim \Sigma_+(\bar e\sigma^{\mu\nu} e)\,,\\
    t^{\mu\nu}_- &\sim \Sigma_-(\bar e\sigma^{\mu\nu} e)\label{eq:spurion_redef_2}\,.
\end{align}
\begin{table}
    \centering
    \begin{tabular}{|l|cccccc|}
    \hline
           & $\Bar{B}B$ & $\Bar{B}\gamma^5B$ & $\Bar{B}\gamma^\mu B$ & $\Bar{B}\gamma^5\gamma^\mu B$ & $\Bar{B}\sigma^{\mu\nu}B$ & $\Bar{B}\lrD^\mu B$\\
          \hline
$C$ & $+$ & $+$ & $-$ & $+$ & $-$&$-$\\
$P$ & $+$ & $-$ & $+$ & $-$ & $+$&$+$\\
\hline
&$\Bar{B}\gamma^5\lrD^\mu B$&$\Bar{B}\gamma^\mu\lrD^\nu B$&$\Bar{B}\gamma^5\gamma^\mu\lrD^\nu B$&$u_\mu$& $\Sigma_\pm$ & $Q_\pm$\\
\hline
$C$&$-$&$+$&$-$&$+$&$+$&$\pm$\\
$P$&$-$&$+$&$-$&$-$&$\pm$&$\pm$\\
\hline
    \end{tabular}
    \caption{The $CP$ properties of the baryon bilinear and the building blocks in the hadronic part of the chiral Lagrangian. The building blocks $\Sigma_\pm$ and $Q_\pm$ of different dressing forms are of different CP properties.}
    \label{tab: building block}
\end{table}

Apart from the redundancies about the NGBs and the covariant derivatives~\cite{Li:2024ghg,Song:2024fae}, there are some new relations.
Firstly, the trace of $Q_\pm$ is zero,
\begin{equation}
    \langle Q_\pm\rangle=\langle u^\dagger\textbf{T}u\pm u\textbf{T}^\dagger u^\dagger\rangle=\langle \mathbf{T}\pm\mathbf{T}^\dagger\rangle=0\,,
\end{equation}
in which $\langle...\rangle$ stands for the trace.
Secondly, the application of derivative on the spurions is redundant,
\begin{equation}
\label{eq:redundancy}
    \partial^\mu Q_\pm\sim u^\mu Q_\mp\,,\quad\partial^\mu \Sigma_\pm\sim u^\mu\Sigma_\mp\,.
\end{equation}
Moreover, the operators with two $Q_\pm$ are
\begin{align}
    \langle Q_+Q_+\rangle=&\langle\mathbf{T}\mathbf{T}\rangle+\langle\mathbf{T}^\dagger\mathbf{T}^\dagger\rangle+2\langle \mathbf{T}^\dagger{u^2}^\dagger \mathbf{T}u^2\rangle\,,\\
    \langle Q_-Q_-\rangle=&-\langle\mathbf{T}\mathbf{T}\rangle-\langle\mathbf{T}^\dagger\mathbf{T}^\dagger\rangle+2\langle \mathbf{T}^\dagger{u^2}^\dagger \mathbf{T}u^2\rangle\,,
\end{align}
since $\langle\mathbf{T}\mathbf{T}\rangle$ and $\langle\mathbf{T}^\dagger\mathbf{T}^\dagger\rangle$ will not contribute to the hadronic process, the two operators above are actually equivalent. A similar relation exists for $\Sigma_\pm$. Thus we have redundant relations that
\begin{equation}
    \langle Q_+Q_+\rangle \sim \langle Q_-Q_- \rangle\,,\quad \langle \Sigma_+\Sigma_+\rangle \sim \langle \Sigma_-\Sigma_- \rangle\,. 
\end{equation}

We emphasize that the effective operators in terms of external sources and the spurions are equivalent because of the transformations from Eq.~\eqref{eq:spurion_redef_1} to Eq.~\eqref{eq:spurion_redef_2}. Next, we present an example. 
We consider the pure-meson operators containing at least one $u_\mu$ and a lepton bilinear $\overline{e}_Re_L$ or its complex conjugation. According to the spurion method, the building blocks are $u_\mu\,,e_L\,,e_R$ and the spurion $\Sigma_\pm$\,, and the independent next-to-leading (NLO) operators are 
\begin{align}
    &\langle u_\mu u^\mu\rangle \langle \Sigma_+\rangle (\overline{e}_Re_L+ h.c.) \,,\quad \langle u_\mu u^\mu\rangle \langle \Sigma_-\rangle (\overline{e}_Re_L+ h.c.) \,,\notag \\
    &\langle u_\mu u^\mu\Sigma_+\rangle (\overline{e}_Re_L+ h.c.) \,, \quad \langle u_\mu u^\mu\Sigma_-\rangle (\overline{e}_Re_L+ h.c.) \,,
\end{align}
On the other hand, the same operators by the external sources are also known~\cite{Li:2024ghg},
\begin{equation}
    \langle u_\mu u^\mu\rangle \langle \chi^\pm\rangle \,,\quad \langle u_\mu u^\mu \chi^\pm\rangle\,.
\end{equation}
The equivalence is obvious since the operator numbers are the same and they can be related explicitly once the $\Sigma_\pm$ takes its dressing form in Eq.~\eqref{eq:chpt_sigma_1}.

We discuss the power-counting of the ChPT. The chiral Lagrangian is organized by the expansion of $p/\Lambda_\chi$, where $\Lambda_\chi$ is the cutoff scale of the ChPT and $p$ is the momentum of the NGBs, thus the chiral power-counting corresponds to the derivative expansion in the chiral Lagrangian, and the chiral dimension of an operator is the power of $p/\Lambda_\chi$. Among the building blocks in Eq.~\eqref{eq:building_blocks}, only $u_\mu$ have chiral dimension 1, while the spurions, leptons, and the photon are of no chiral dimension. However, the leptons and the photon do have canonical dimensions, which play an important role in the LEFT and the matching procedure, and will be discussed in the next section.

Moreover, when considering the meson-baryon interaction, the heavy mass of the baryon will make the standard power counting invalid. We take the heavy baryon form~\cite{Krause:1990xc,Jenkins:1990jv,Ecker:1995rk} for the baryon field. Then the operators containing baryons can be expanded in terms of the inverse of their mass $1/m$. Thus, the baryon fields can be expressed as
\begin{equation}
    \Psi(x) = e^{-imv\cdot x}\left[\underbrace{e^{imv\cdot x} P_v^+\Psi(x)}_{\equiv \mathcal{N}_v(x)} + \underbrace{e^{imv\cdot x}P_v^- \Psi(x)}_{\equiv \mathcal{H}_v(x)}\right]\,,
\end{equation}
through the projection operators $P_v^{\pm} = (1+v\!\!\!\slash)/2$. The bilinears and derivatives of the baryon can be expanded to take the form that
\begin{align}
\label{heavy_baryon}
    \Bar{\Psi} 1 \Psi &= \Bar{\mathcal{N}_v} 1 \mathcal{N}_v +...\,, \notag \\
    \Bar{\Psi} \gamma^5 \Psi &= \frac{1}{m_N}\partial_\mu(\Bar{\mathcal{N}_v} S^\mu \mathcal{N}_v)+... \,, \notag \\
    \Bar{\Psi} \gamma^\mu \Psi &= \Bar{\mathcal{N}_v} v^\mu \mathcal{N}_v +...\,, \notag \\
    \Bar{\Psi} \gamma^5\gamma^\mu \Psi &= 2\Bar{\mathcal{N}_v} S^\mu \mathcal{N}_v +...\,, \notag \\
    \Bar{\Psi} \sigma^{\mu\nu} \Psi &= 2\epsilon^{\mu\nu\rho\lambda}\Bar{\mathcal{N}_v} v_\rho S_\lambda \mathcal{N}_v +...\,, \notag \\
    \Bar{\Psi} \gamma^5\sigma^{\mu\nu} \Psi &= 2\Bar{\mathcal{N}_v} [v^\mu ,S^\nu] \mathcal{N}_v +...\,, \notag \\
    \Bar{\Psi} \overleftrightarrow{D}^\mu \Psi &= \Bar{\mathcal{N}_v} v^\mu \mathcal{N}_v +...\,,  
\end{align}
and the $S^\mu$ in the Eq.~\eqref{heavy_baryon} is the spin-operator of the baryon, and its definition and basic properties are
\begin{equation}
    S^\mu=\frac{i}{2}\gamma^5\sigma^{\mu\nu}v^\nu\,,\quad S\cdot v=0\,,\quad\{S^\mu,S^\nu\}=\frac{1}{2}\left(v^\mu v^\nu-g^{\mu\nu}\right)\,,\quad\left[S^\mu,S^\nu\right]=i\epsilon^{\mu\nu\rho\lambda}v_\rho S_\lambda\,.
\end{equation}
Furthermore, the heavy baryon projection does not change the mesons part and we define the chiral dimension of the meson-baryon operators as the total power of $p/\Lambda_\chi$ and $p/m$ of the leading order(LO) in the heavy baryon projection. In particular, only the $\gamma^5$ of the baryon bilinear would be of chiral dimension 1, while $\lrD_\mu$ of the baryon bilinear would be chiral dimension 0.

As we will show subsequently, the redundancies such as the EOM and the IBP make the equivalent LEFT operators match to different ChPT meson-baryon operators, which may be of different chiral dimensions formally. However, they are actually equivalent, which can be realized via the IBP and the EOM of the ChPT.

\subsection{Construction of Chiral Lagrangian}

In this section, we present the effective operators of the ChPT in terms of the spurions, where the Young tensor technique~\cite{Li:2020gnx,Li:2022tec,Li:2020xlh} is used.
In the next section, we will introduce the matching rules and relate the LEFT and the ChPT operators to each other.

When the leptons and the photon are considered to be decoupled as fundamental building blocks, the chiral Lagrangian such as the type $u^2\Sigma e^2$ would be
\begin{align}
    \langle u^\mu u_\mu\rangle\langle\Sigma_\pm\rangle (\Bar{e}_Le_R)\pm h.c.\,,\quad\langle u^\mu u_\mu\Sigma_\pm\rangle (\Bar{e}_Le_R)\pm h.c.\,.
\end{align}
Similarly, the chiral Lagrangian of type $u^2F\Sigma e^2$ would be 
\begin{align}
     \langle u^\mu u_\mu\rangle\langle\Sigma_\pm\rangle (\Bar{e}_L\sigma^{\mu\nu}e_R)F_{\mu\nu}\pm h.c.\,,\langle u^\mu u_\mu\Sigma_\pm\rangle (\Bar{e}_L\sigma^{\mu\nu}e_R)F_{\mu\nu}\pm h.c.\,, 
     \langle [u^\mu,u_\nu]\Sigma_\pm\rangle (\Bar{e}_Le_R)F_{\mu\nu}\pm h.c.,
\end{align}
and these two types of operators can be combined as the eigenstates of CP for the hadronic part with the leptons and photon omitted, just as done for the LEFT operators. The chiral Lagrangian will be divided to become
\begin{align}
    C+P+:&\langle u^\mu u_\mu\rangle\langle\Sigma_+\rangle (\Bar{e}_Le_R)\pm h.c.\,,\quad\langle u^\mu u_\mu\Sigma_+\rangle (\Bar{e}_Le_R)\pm h.c.\,,\notag\\
    &\langle u^\mu u_\mu\rangle\langle\Sigma_+\rangle (\Bar{e}_L\sigma^{\mu\nu}e_R)F_{\mu\nu}\pm h.c.\,,\quad\langle u^\mu u_\mu\Sigma_+\rangle (\Bar{e}_L\sigma^{\mu\nu}e_R)F_{\mu\nu}\pm h.c.\,,\notag\\
    C+P-:&\langle u^\mu u_\mu\rangle\langle\Sigma_-\rangle (\Bar{e}_Le_R)\pm h.c.\,,\quad\langle u^\mu u_\mu\Sigma_-\rangle (\Bar{e}_Le_R)\pm h.c.\,,\notag\\
    &\langle u^\mu u_\mu\rangle\langle\Sigma_-\rangle (\Bar{e}_L\sigma^{\mu\nu}e_R)F_{\mu\nu}\pm h.c.\,,\quad\langle u^\mu u_\mu\Sigma_-\rangle (\Bar{e}_L\sigma^{\mu\nu}e_R)F_{\mu\nu}\pm h.c.,\notag\\
    C-P+:&\langle [u^\mu,u_\nu]\Sigma_+\rangle (\Bar{e}_Le_R)F_{\mu\nu}\pm h.c.,\notag\\
    C-P-:&\langle [u^\mu,u_\nu]\Sigma_-\rangle (\Bar{e}_Le_R)F_{\mu\nu}\pm h.c..
    \label{ex}
\end{align}
In addition, the Lorentz structure of the whole leptons and photon in Eq.~\eqref{ex} can be redefined as
\begin{align}
    \{(\Bar{e}_Le_R)\,,(\Bar{e}_Re_L)\,,(\Bar{e}_L\sigma^{\mu\nu}e_R)F_{\mu\nu}\,,(\Bar{e}_R\sigma^{\mu\nu}e_L)F_{\mu\nu}\}&\rightarrow j\,,\\
    \{(\Bar{e}_Le_R)F_{\mu\nu}\}&\rightarrow j^{[\mu,\nu]}\,,
\end{align}
for convenience. Then the chiral Lagrangian can be written compactly as
\begin{align}
    C+P+:\quad&\langle u^\mu u_\mu\rangle\langle\Sigma_+\rangle j\,,\quad\langle u^\mu u_\mu\Sigma_+\rangle j\,,\notag\\
    C+P-:\quad&\langle u^\mu u_\mu\rangle\langle\Sigma_+\rangle j\,,\quad\langle u^\mu u_\mu\Sigma_+\rangle j\,,\notag\\
    C-P+:\quad&\langle [u^\mu,u_\nu]\Sigma_+\rangle j^{[\mu,\nu]}\,,\notag\\
    C-P-:\quad&\langle [u^\mu,u_\nu]\Sigma_-\rangle j^{[\mu,\nu]}\,.
\end{align}

Moreover, all the Lorentz structures of the leptons and photon up to dimension 9 presented in the previous section can be collected into four types $j\,,j^\mu\,,j^{[\mu,\nu]}\,,j^{\{\mu,\nu\}}$ that
\begin{align}
    \left\{\begin{array}{c}
       \mathbf{1}\,, (\bar e_Le_R)\,,(\bar e_Re_L)\,,(\nu_L^TC\nu_L)\,,(\bar\nu_LC\bar\nu_L^T)\,,(\bar e_R\nu_L)\,,(\bar \nu_Le_R)\,,( \nu_L^TCe_L)\,,( \bar e_LC\bar \nu_L^T)\,, \\
       (\bar e_L\sigma^{\mu\nu}e_R)F_{\mu\nu}\,,(\bar e_R\sigma^{\mu\nu}e_L)F_{\mu\nu}\,,(\nu_L^TC\sigma^{\mu\nu}\nu_L)F_{\mu\nu}\,,(\bar \nu_LC\sigma^{\mu\nu}\bar\nu_L^T)F_{\mu\nu}\\
       (\bar e_R\sigma^{\mu\nu}\nu_L)F_{\mu\nu}\,,\,,(\bar\nu_L\sigma^{\mu\nu}e_R)F_{\mu\nu}\,,(\nu_L^TC\sigma^{\mu\nu}e_L)F_{\mu\nu}\,,(\bar e_LC\sigma^{\mu\nu}\bar \nu_L^T)F_{\mu\nu}\,,\\
       \partial^2(\bar e_Le_R)\,,\partial^2(\bar e_Re_L)\,,\partial^2(\nu_L^TC\nu_L)\,,\partial^2(\bar\nu_LC\bar\nu_L^T)\,,\\
       \partial^2(\bar e_R\nu_L)\,,\partial^2(\bar \nu_Le_R)\,,\partial^2( \nu_L^TCe_L)\,,\partial^2(\bar e_LC\bar\nu_L^T)\,.
    \end{array}\right\}\rightarrow j\,,
\end{align}
\begin{align}
    \left\{\begin{array}{c}
     (\bar\nu_L\gamma^\mu\nu_L)\,,(\bar e_L\gamma^\mu e_L)\,,(\bar e_R\gamma^\mu e_R)\,,(\bar e_L\gamma^\mu \nu_L)\,,(\bar\nu_L\gamma^\mu e_L)\,,(\nu_L^TC\gamma^\mu e_R)\,,(\bar e_RC\gamma^\mu \bar\nu_L^T)\,, \\ 
     (\bar e_L\lrpartial^\mu e_R)\,,(\bar e_R\lrpartial^\mu e_L)\,,(\nu_L^TC\lrpartial^\mu \nu_L)\,,(\bar\nu_LC\lrpartial^\mu \bar\nu_L^T)\,,\\
     (\bar e_R\lrpartial^\mu\nu_L)\,,(\bar \nu_L\lrpartial^\mu e_R)\,,(\nu_L^TC\lrpartial^\mu e_L)\,,(\bar e_LC\lrpartial^\mu \bar e_L^T)\,,\\
     (\bar\nu_L\gamma_\nu\nu_L)F^{\mu\nu}\,,(\bar e_L\gamma_\nu e_L)F^{\mu\nu}\,,(\bar e_R\gamma_\nu e_R)F^{\mu\nu}\,,(\bar e_L\gamma_\nu \nu_L)F^{\mu\nu}\,,\\
     (\bar\nu_L\gamma_\nu e_L)F^{\mu\nu}\,,(\nu_L^TC\gamma_\nu e_R)F^{\mu\nu}\,,(\bar e_RC\gamma_\nu \bar\nu_L^T)F^{\mu\nu}\,,\\
     \partial^2(\bar\nu_L\gamma^\mu\nu_L)\,,\partial^2(\bar e_L\gamma^\mu e_L)\,,\partial^2(\bar e_R\gamma^\mu e_R)\,,\partial^2(\bar e_L\gamma^\mu \nu_L)\,,\\
     \partial^2(\bar\nu_L\gamma^\mu e_L)\,,\partial^2(\nu_L^TC\gamma^\mu e_R)\,,\partial^2(\bar e_RC\gamma^\mu \bar\nu_L^T)
    \end{array}\right\}\rightarrow j^\mu\,,
\end{align}
\begin{align}
    \left\{\begin{array}{c}
    (\bar e_L\gamma^\mu\lrpartial^\nu e_L)\,,(\bar e_R\gamma^\mu\lrpartial^\nu e_R)\,,(\bar\nu_L\gamma^\mu\lrpartial^\nu\nu_L)\,,(\bar e_L\gamma^\mu\lrpartial^\nu\nu_L)\,,\\
    (\bar\nu_L\gamma^\mu\lrpartial^\nu e_L)\,,(\nu_L^TC\gamma^\mu\lrpartial^\nu e_R)\,,(\bar e_RC\gamma^\mu\lrpartial^\nu \bar\nu_L^T)\,,F^{\mu\rho}F_\rho^\nu 
    \end{array}\right\}\rightarrow j^{\{\mu,\nu\}}\,,
\end{align}
\begin{align}
    \left\{\begin{array}{c}
    (\bar e_L\sigma^{\mu\nu}e_R)\,,(\bar e_R\sigma^{\mu\nu}e_L)\,,(\bar\nu_L\sigma^{\mu\nu}e_R)\,,(\bar e_R\sigma^{\mu\nu}\nu_L)\,,\\
    (\nu_L^TC\sigma^{\mu\nu}\nu_L)\,,(\bar\nu_LC\sigma^{\mu\nu}\bar\nu_L^T)\,,(\nu_L^TC\sigma^{\mu\nu}e_R)\,,(\bar e_RC\sigma^{\mu\nu}\bar\nu_L^T)\,,F^{\mu\nu}\,, \\
    (\bar e_Le_R)F^{\mu\nu}\,,(\bar e_Re_L)F^{\mu\nu}\,,(\bar \nu_Le_R)F^{\mu\nu}\,,(\bar e_R\nu_L)F^{\mu\nu}\,,\\
    (\nu_L^TC\nu_L)F^{\mu\nu}\,,(\bar\nu_LC\bar\nu_L^T)F^{\mu\nu}\,,( \nu_L^TC e_L)F^{\mu\nu}\,,(\bar e_LC\bar\nu_L^T)F^{\mu\nu}\,,\\
    (\bar e_L\sigma^{\mu\rho}e_R)F_\rho^\nu\,,(\bar e_R\sigma^{\mu\rho}e_L)F_\rho^\nu\,,(\bar e_R\sigma^{\mu\rho}\nu_L)F_\rho^\nu\,,(\bar\nu_L\sigma^{\mu\rho}e_R)F_\rho^\nu\,,\\
    (\nu_L^TC\sigma^{\mu\rho}\nu_L)F_\rho^\nu\,,(\bar\nu_LC\sigma^{\mu\rho}\bar\nu_L^T)F_\rho^\nu\,,(\nu_L^TC\sigma^{\mu\rho}e_R)F_\rho^\nu\,,(\bar e_RC\sigma^{\mu\rho}\bar \nu_L^T)F_\rho^\nu
    \end{array}\right\}\rightarrow j^{[\mu,\nu]}\,,
\end{align}
in which $\{\mu,\nu\}$ and $[\mu,\nu]$ means the asymmetry and symmetry of the Lorentz index. 
In particular, we emphasize that the symbol $j$ is just a simplification for writing, and should not be confused with the conventional external sources such as the ones in Eq.~\eqref{eq:ext_1} to Eq.~\eqref{eq:ext_2}. They are distinguished from each other according to Eq.~\eqref{eq:spurion_redef_1} to Eq.~\eqref{eq:spurion_redef_2}, which imply equivalent transformations of the conventional external sources parameterization and the spurion parameterization used here. In other words, the conventional external sources are separated into two parts in our notation, the pure lepton bilinears or photon field-strength tensors, symbolized by $j$ here, and the spurions representing $SU(3)_V$ violations, $\Sigma_\pm$ and $Q_\pm$. Besides, all these structures are the ones that emerge up to dimension 9, if the LEFT operators of higher dimensions are considered, more complicated structures such as the product of the $j$'s above could be needed.

Next, we present the ChPT operators classified by the number and the kind of spurions involved, including both pure meson and meson-baryon operators. In particular, the operators without spurions can be found in Ref.~\cite{Song:2024fae, Li:2024ghg} and we only select the operators that are necessary for this work.
Subsequently, we will present the pure meson operators up to $\mathcal{O}(p^4)$ and the meson-baryon operators up to $\mathcal{O}(p^2)$. In particular, we only consider the operators with at most two spurions.

\subsubsection{Pure Meson Operators with 1 Spurion}
\begin{longtable}{|@{}c@{}|@{}c@{}|@{}c@{}|@{}c@{}|}
        \hline
   $C$+$P$+&$C$+$P$-&$C$-$P$+&$C$-$P$-\\
   \hline
    $\langle u^\mu u_\mu\rangle j$&&&\\
    \hline
    $\langle u^\mu u_\mu u^\nu u_\nu\rangle j$&&&\\
    $\langle u^\mu u^\nu u_\mu u_\nu\rangle j$&&&\\
    $\langle u^\mu u^\nu\rangle\langle u_\mu u_\nu\rangle j$&&&\\
    \hline
        $\langle D_\mu D^\lambda u^\nu u^\mu u_\nu u_\lambda\rangle j+h.c.$
        &$\langle D_\mu u^\lambda u_\nu u_\lambda u^\nu u^\mu\rangle j+h.c.$&&$\langle D_\mu u^\lambda u_\nu u_\lambda u^\nu u^\mu\rangle j-h.c.$\\
        $\langle D_\lambda u_\mu D^\lambda u^\mu u_\nu u^\nu\rangle j+h.c.$
        &$\langle D_\mu u^\lambda u_\nu u_\lambda u^\mu u^\nu\rangle j+h.c.$&&$\langle D_\mu u^\lambda u_\nu u_\lambda u^\mu u^\nu\rangle j-h.c.$\\
        $\langle D_\lambda u_\mu D_\nu u^\lambda u^\mu u^\nu\rangle j+h.c.$&$\epsilon_{\mu\nu\rho\lambda}\langle u^\mu u^\nu u^\rho u^\alpha u^\lambda u_\alpha\rangle j$&&$\epsilon_{\mu\nu\rho\lambda}\langle u^\mu u^\nu u^\rho u_\lambda u^\alpha u_\alpha\rangle j$\\
\end{longtable}

\begin{longtable}{|@{}c@{}|@{}c@{}|@{}c@{}|@{}c@{}|}
      \hline
   $C$+$P$+&$C$+$P$-&$C$-$P$+&$C$-$P$-\\
   \hline
    $\epsilon_{\mu\nu\rho\lambda}\langle u^\nu u^\rho u^\lambda\rangle j^\mu$&$\langle u^\nu u_\nu u_\mu\rangle j^\mu$&&\\
        \hline
        $\epsilon_{\mu\nu\rho\lambda}\langle u^\nu[D^\sigma u^\rho,D_\sigma u^\lambda]\rangle j^\mu$&$\langle u_\lambda\{D_\mu u_\nu,D_\lambda u_\nu\}\rangle j^\mu$&$\epsilon_{\mu\nu\rho\lambda}\langle u^\nu\{D^\sigma u^\rho,D_\sigma u^\lambda\}\rangle j^\mu$&$\langle u_\lambda[D_\mu u_\nu,D_\lambda u_\nu]\rangle j^\mu$\\
        $\langle \{D_\mu u_\nu, u^\nu\} u^\lambda u_\lambda\rangle j^\mu$&$\langle D_\mu D^\nu u^\lambda\{u_\nu,u_\lambda\}\rangle j^\mu$&$\langle [D_\mu u_\nu, u^\nu] u^\lambda u_\lambda\rangle j^\mu$&$\epsilon_{\mu\nu\rho\lambda}\langle D^\rho u^\sigma u^\nu u^\lambda u_\sigma\rangle j^\mu-h.c.$\\
        $\langle D_\mu u_\nu u^\lambda u^\nu u_\lambda\rangle j^\mu+h.c.$&$\langle u_\mu \{D_\lambda u_\nu,D^\lambda u^\nu\}\rangle j^\mu$&$\langle D_\mu u_\nu u^\lambda u^\nu u_\lambda\rangle j^\mu-h.c.$&$\epsilon_{\mu\nu\rho\lambda}\langle D^\rho u^\sigma u^\nu u_\sigma u^\lambda\rangle j^\mu$\\
        $\langle D_\mu u_\nu u^\lambda\rangle\langle u^\nu u_\lambda\rangle j^\mu$&$\epsilon_{\mu\nu\rho\lambda}\langle D^\rho u^\sigma u^\nu u^\lambda u_\sigma\rangle j^\mu+h.c.$&$\langle D_\lambda u_\nu u^\nu u_\mu u^\lambda\rangle j^\mu-h.c.$&$\epsilon_{\nu\rho\lambda\sigma}\langle D_\mu u^\nu u^\rho u^\lambda u^\sigma\rangle j^\mu$\\
        $\langle D_\lambda u_\nu u_\mu u^\nu u^\lambda\rangle j^\mu+h.c.$&&$\langle D_\lambda u_\nu u_\mu u^\nu u^\lambda\rangle j^\mu-h.c.$&\\
        $\langle D_\lambda u_\nu u^\nu u_\mu u^\lambda\rangle j^\mu+h.c.$&&$\epsilon_{\mu\nu\rho\lambda}\langle u^\sigma u_\sigma u^\nu u^\rho u^\lambda\rangle j^\mu$&\\
\end{longtable}

\begin{longtable}{|@{}c@{}|@{}c@{}|@{}c@{}|@{}c@{}|}
        \hline
   $C$+$P$+&$C$+$P$-&$C$-$P$+&$C$-$P$-\\
        \hline
            $\langle u_\mu u_\nu\rangle j^{\{\mu,\nu\}}$&&&\\
            \hline
            $\epsilon_{\mu\nu\rho\lambda}\langle D^\sigma u_\rho u_\lambda u_\sigma\rangle j^{[\mu,\nu]}$&$\langle  D^\lambda u_\nu u_\mu  u_\lambda\rangle j^{[\mu,\nu]}+h.c.$&$\epsilon_{\mu\rho\lambda\sigma}\langle  D_\nu u^\lambda u^\rho  u^\sigma\rangle j^{\{\mu,\nu\}}$&$\langle  D^\lambda u_\nu u_\mu  u_\lambda\rangle j^{\{\mu,\nu\}}-h.c.$\\
            $\langle \{u_\mu, u_\nu\} u^\lambda u_\lambda\rangle j^{\{\mu,\nu\}}$&&$\langle [u_\mu,u_\nu] u^\lambda u_\lambda\rangle j^{[\mu,\nu]}$&$\epsilon_{\mu\nu\rho\lambda}\langle u^\rho u^\lambda u^\sigma u_\sigma\rangle j^{[\mu,\nu]}$\\
            $\langle u_\mu u^\lambda u_\nu u_\lambda\rangle j^{\{\mu,\nu\}}+h.c.$&&$\langle u_\mu u^\lambda u_\nu u_\lambda\rangle j^{[\mu,\nu]}-h.c.$&$\epsilon_{\mu\nu\rho\lambda}\langle u^\rho u^\sigma u^\lambda u_\sigma\rangle j^{[\mu,\nu]}$\\
            $\langle u_\mu u_\nu\rangle\langle u^\lambda u_\lambda\rangle j^{\{\mu,\nu\}}$&&&\\
    \hline  
    \caption{The pure meson operators with $j^{\mu\nu}$ and without spurion.}
    \label{purem0jmn}
\end{longtable}

\small{\begin{longtable}{|@{}c@{}|@{}c@{}|@{}c@{}|@{}c@{}|}
\hline
   $C$+$P$+&$C$+$P$-&$C$-$P$+&$C$-$P$-\\
   \hline
    $\langle\Sigma_+\rangle j$&$\langle\Sigma_-\rangle j$&&\\
    \hline
    $\langle\Sigma_+u^\mu u_\mu\rangle j$&$\langle\Sigma_-u^\mu u_\mu\rangle j$&&\\
    $\langle \Sigma_+\rangle\langle u^\mu u_\mu\rangle j$&$\langle\Sigma_-\rangle\langle u^\mu u_\mu\rangle j$&&\\
    \hline
        $\langle\Sigma_-\rangle\langle D_\mu u^\nu u^\mu u_\nu\rangle j+h.c.$
        &$\langle\Sigma_+\rangle\langle\{D_\mu u^\nu,u^\mu\} u^\nu\rangle j$
        &$\langle D_\mu u^\nu u^\mu u_\nu\Sigma_-\rangle j-h.c.$
        &$\langle D_\mu u^\nu u^\mu u_\nu\Sigma_+\rangle j-h.c.$\\
        $\langle D_\mu u^\nu u^\mu u_\nu\Sigma_-\rangle j+h.c.$
        &$\langle D_\mu u^\nu u^\mu u_\nu\Sigma_+\rangle j+h.c.$
        &
        &\\
        $\langle D_\mu u^\nu u^\mu\Sigma_- u_\nu\rangle j+h.c.$
        &$\langle D_\mu u^\nu u^\mu\Sigma_+u_\nu\rangle j+h.c.$
        &
        &\\
        $\langle u_\mu u_\nu\rangle\langle D^\mu u^\nu\Sigma_-\rangle j$
        &$\langle u_\mu u_\nu\rangle\langle D^\mu u^\nu\Sigma_+\rangle j +h.c.$
        &&\\
\end{longtable}}

\small{\begin{longtable}{|@{}c@{}|@{}c@{}|@{}c@{}|@{}c@{}|}
        \hline
   $C$+$P$+&$C$+$P$-&$C$-$P$+&$C$-$P$-\\
      \hline
    $\langle\Sigma_-u^\mu\rangle j_{\mu}$&$\langle\Sigma_+u^\mu\rangle j_{\mu}$&&\\
    \hline
    $\langle\{\Sigma_-,u_\mu\} u^\nu u_\nu\rangle j^{\mu}$&$\langle\{\Sigma_+,u_\mu\} u^\nu u_\nu\rangle j^{\mu}$&$\langle[\Sigma_-,u_\mu] u^\nu u_\nu\rangle j^{\mu}$&$\langle[\Sigma_+,u_\mu] u^\nu u_\nu\rangle j^{\mu}$\\
    $\langle\Sigma_- u^\nu u_\mu u_\nu\rangle j^{\mu}$&$\langle\Sigma_+ u^\nu u_\mu u_\nu\rangle j^{\mu}$&$\epsilon_{\mu\nu\rho\lambda}\langle\Sigma_+\rangle\langle u^\nu u^\rho u^\lambda\rangle j^{\mu}$&$\epsilon_{\mu\nu\rho\lambda}\langle\Sigma_-\rangle\langle u^\nu u^\rho u^\lambda\rangle j^{\mu}$\\
    $\langle\Sigma_-\rangle\langle u^\nu u_\mu u_\nu\rangle j^{\mu}$&$\langle\Sigma_+\rangle\langle u^\nu u_\mu u_\nu\rangle j^{\mu}$&$\epsilon_{\mu\nu\rho\lambda}\langle\Sigma_+u^\nu u^\rho u^\lambda\rangle j^{\mu}$&$\epsilon_{\mu\nu\rho\lambda}\langle\Sigma_-u^\nu u^\rho u^\lambda\rangle j^{\mu}$\\
    $\langle D_\mu u_\nu \{u^\nu,\Sigma_+\}\rangle j^\mu$&$\langle D_\mu u_\nu \{u^\nu,\Sigma_-\}\rangle j^\mu$&$\langle D_\mu u_\nu [u^\nu,\Sigma_+]\rangle j^\mu$&$\langle D_\mu u_\nu [u^\nu,\Sigma_-]\rangle j^\mu$\\
    $\langle\Sigma_+\rangle\langle D_\mu u_\nu u^\nu\rangle j^\mu$&$\langle\Sigma_-\rangle\langle D_\mu u_\nu u^\nu\rangle j^\mu$&&\\
    \hline   
    \caption{The pure meson operators with $j^\mu$ and 1 spurion $\Sigma_\pm$.}
    \label{purem1Sjm}
\end{longtable}}

\begin{longtable}{|@{}c@{}|@{}c@{}|@{}c@{}|@{}c@{}|}
        \hline
   $C$+$P$+&$C$+$P$-&$C$-$P$+&$C$-$P$-\\
        \hline
           $\langle\Sigma_+\{u_\mu,u_\nu\}\rangle j^{\{\mu,\nu\}}$&$\langle\Sigma_-\{u_\mu,u_\nu\}\rangle j^{\{\mu,\nu\}}$&$\langle\Sigma_+[u_\mu,u_\nu]\rangle j^{[\mu,\nu]}$&$\langle\Sigma_-[u_\mu,u_\nu]\rangle j^{[\mu,\nu]}$\\
    $\langle\Sigma_+D_\mu u_\nu\rangle j^{\{\mu,\nu\}}$&$\langle\Sigma_-D_\mu u_\nu\rangle j^{\{\mu,\nu\}}$&&\\
    \hline  
    \caption{The pure meson operators with $j^{\mu\nu}$ and 1 spurion $\Sigma_\pm$.}
    \label{purem1Sjmn}
\end{longtable}

\small{
\begin{longtable}{|@{}c@{}|@{}c@{}|@{}c@{}|@{}c@{}|}
        \hline
   $C$+$P$+&$C$+$P$-&$C$-$P$+&$C$-$P$-\\
    \hline
    $\langle Q_+u^\mu u_\mu\rangle j$&$\langle Q_-u^\mu u_\mu\rangle j$&&\\
    \hline
        $\epsilon_{\mu\nu\rho\lambda}\langle u^\mu u^\nu u^\rho u^\lambda Q_-\rangle j-h.c.$&$\epsilon_{\mu\nu\rho\lambda}\langle u^\mu u^\nu u^\rho u^\lambda Q_+\rangle j+h.c.$&$\epsilon_{\mu\nu\rho\lambda}\langle u^\mu u^\nu u^\rho u^\lambda Q_-\rangle j+h.c.$
        &$\epsilon_{\mu\nu\rho\lambda}\langle u^\mu u^\nu u^\rho u^\lambda Q_+\rangle j-h.c.$\\
        $\langle u_\mu u^\mu u_\nu\rangle\langle Q_+ u^\nu\rangle j$
        &&&$\langle u_\mu u^\mu u_\nu\rangle\langle Q_- u^\nu\rangle j$
        \\
        $\langle u^\mu u_\mu u_\nu u^\nu Q_+\rangle j$
        &&&$\langle u_\mu u^\mu u_\nu u^\nu Q_-\rangle j$\\
        $\langle u^\nu u^\mu u_\mu u_\nu Q_+\rangle j$
        &&&$\langle u_\nu u_\mu u^\mu u^\nu Q_-\rangle j$\\
        $\langle u_\mu u_\nu u^\mu u^\nu Q_+\rangle j$
        &&&$\langle u_\mu u_\nu u^\mu u^\nu Q_-\rangle j$\\
    \hline
    \caption{The pure meson operators with $j$ and 1 spurion $Q_\pm$.}
    \label{purem1Qj}
\end{longtable}}

\begin{longtable}{|@{}c@{}|@{}c@{}|@{}c@{}|@{}c@{}|}
        \hline
   $C$+$P$+&$C$+$P$-&$C$-$P$+&$C$-$P$-\\
        \hline
    &$\langle Q_+u^\mu\rangle j_{\mu}$&$\langle Q_-u^\mu\rangle j_{\mu}$&\\
    \hline
    $\langle[ Q_-,u_\mu] u^\nu u_\nu\rangle j^{\mu}$&$\langle\{ Q_+,u_\mu\} u^\nu u_\nu\rangle j^{\mu}$&$\langle\{ Q_-,u_\mu\} u^\nu u_\nu\rangle j^{\mu}$&$\langle[ Q_+,u_\mu] u^\nu u_\nu\rangle j^{\mu}$\\
    $\langle D_\mu u_\nu \{u^\nu, Q_+\}\rangle j^\mu$&$\langle Q_+ u^\nu u_\mu u_\nu\rangle j^{\mu}$&$\langle Q_- u^\nu u_\mu u_\nu\rangle j^{\mu}$&$\langle D_\mu u_\nu \{u^\nu, Q_-\}\rangle j^\mu$\\
    &$\epsilon_{\mu\nu\rho\lambda}\langle Q_-u^\nu u^\rho u^\lambda\rangle j^{\mu}$&$\epsilon_{\mu\nu\rho\lambda}\langle Q_+u^\nu u^\rho u^\lambda\rangle j^{\mu}$&\\
    &$\langle D_\mu u_\nu [u^\nu, Q_-]\rangle j^\mu$&$\langle D_\mu u_\nu [u^\nu, Q_+]\rangle j^\mu$&\\
    \hline   
    \caption{The pure meson operators with $j^{\mu}$ and 1 spurion $Q_\pm$.}
    \label{purem1Qjm}
\end{longtable}

\begin{longtable}{|@{}c@{}|@{}c@{}|@{}c@{}|@{}c@{}|}
        \hline
   $C$+$P$+&$C$+$P$-&$C$-$P$+&$C$-$P$-\\
        \hline
           $\langle Q_+\{u_\mu,u_\nu\}\rangle j^{\{\mu,\nu\}}$&$\langle Q_-[u_\mu,u_\nu]\rangle j^{[\mu,\nu]}$&$\langle Q_+[u_\mu,u_\nu]\rangle j^{[\mu,\nu]}$&$\langle Q_-\{u_\mu,u_\nu\}\rangle j^{\{\mu,\nu\}}$\\
    \hline  
    \caption{The pure meson operators with $j^{\mu\nu}$ and 1 spurion $Q_\pm$.}
    \label{purem1Qjmn}
\end{longtable}

\subsubsection{Meson-Baryon Operators with 1 Spurion}

\begin{longtable}{|@{}c@{}|@{}c@{}|@{}c@{}|@{}c@{}|}
        \hline
   $C$+$P$+&$C$+$P$-&$C$-$P$+&$C$-$P$-\\
       \hline
       $\langle\Bar{B}B\rangle j$&&&\\
       \hline
       $\langle\Bar{B}\gamma^5\gamma_\mu u^\mu B\rangle j$&$\langle\Bar{B}\gamma^5B\rangle j$&&$\langle\Bar{B}\gamma_\mu u^\mu B\rangle j$\\
    $\langle\Bar{B}\gamma^5\gamma_\mu Bu^\mu \rangle j$&&&$\langle\Bar{B}\gamma_\mu Bu^\mu \rangle j$\\
    \hline
        $\langle\Bar{B} B\rangle\langle u_\mu u^\mu\rangle j$&$\langle\Bar{B}\gamma^5\sigma^{\mu\nu} u_\mu\rangle\langle u_\nu B\rangle j$&$\langle\Bar{B}\sigma^{\mu\nu} u_\mu B u_\nu\rangle j$&$\langle\Bar{B}\gamma^5\sigma^{\mu\nu} u^\mu B u^\nu\rangle j$\\
        $\langle\Bar{B} u_\mu\rangle\langle u^\mu B\rangle j$&$\langle\Bar{B}\gamma^5\sigma^{\mu\nu} B[u^\rho,u^\lambda]\rangle j$&&$\langle\Bar{B} \gamma^5\gamma^\mu\lrD^\nu B\rangle\langle u_\mu u_\nu\rangle j$\\
        $\langle\Bar{B}\sigma^{\mu\nu} u_\mu\rangle\langle u_\nu B\rangle j$&$\langle\Bar{B}\gamma^5\sigma^{\mu\nu}[u^\rho,u^\lambda] B\rangle j$&&$\langle\Bar{B}\gamma^5 \gamma^\mu\lrD^\nu u_\mu\rangle\langle  u_\nu B\rangle j$\\
        $\langle\Bar{B} B u_\mu u^\mu\rangle j$&&&$\langle\Bar{B} \gamma^5\gamma^\mu\lrD^\nu B u_\mu u_\nu\rangle j$\\
        $\langle\Bar{B} u_\mu u^\mu B\rangle j$&&&$\langle\Bar{B} \gamma^5\gamma^\mu\lrD^\nu  u_\mu u_\nu B\rangle j$\\
        $\langle\Bar{B}\sigma^{\mu\nu} B[u_\mu,u_\nu]\rangle j$&&&\\
        $\langle\Bar{B}\sigma^{\mu\nu}[u_\mu,u_\nu] B\rangle j$&&&\\
        $\langle\Bar{B} \gamma^\mu\lrD^\nu B\rangle\langle u_\mu u_\nu\rangle j$&&&\\
        $\langle\Bar{B} \gamma^\mu\lrD^\nu u_\mu\rangle\langle  u_\nu B\rangle j$&&&\\
        $\langle\Bar{B} \gamma^\mu\lrD^\nu B u_\mu u_\nu\rangle j$&&&\\
        $\langle\Bar{B} \gamma^\mu\lrD^\nu  u_\mu u_\nu B\rangle j$&&&\\
        \hline 
    \caption{The meson-baryon operators with $j$ and without spurion.}
    \label{mb0j}
\end{longtable}

\begin{longtable}{|@{}c@{}|@{}c@{}|@{}c@{}|@{}c@{}|}
        \hline
   $C$+$P$+&$C$+$P$-&$C$-$P$+&$C$-$P$-\\
   \hline
       &$\langle\Bar{B}\gamma^5\gamma_\mu B\rangle j^\mu$&$\langle\Bar{B}\gamma_\mu B\rangle j^\mu$&\\
\hline
$\langle\Bar{B}\gamma^5 u_\mu B\rangle j^\mu$&$\langle\Bar{B}u_\mu B\rangle j^\mu$&&\\
    $\langle\Bar{B}\gamma^5B u_\mu \rangle j^\mu$&$\langle\Bar{B}Bu_\mu \rangle j^\mu$&&\\
    \hline
$\langle\Bar{B}\gamma^\nu [u_\nu,u_\mu] B\rangle j^\mu $&$\langle\Bar{B}\gamma^5\gamma_\mu B\rangle\langle u_\nu u^\nu\rangle j^\mu $&$\langle\Bar{B}\gamma_\mu B\rangle\langle u_\nu u^\nu\rangle j^\mu $&$\langle\Bar{B}\gamma^5\gamma^\nu [u_\nu,u_\mu] B\rangle j^\mu $\\
        $\langle\Bar{B}\gamma^\nu B[u_\nu,u_\mu] \rangle j^\mu $&$\langle\Bar{B}\gamma^5\gamma_\mu u_\nu\rangle\langle u^\nu B\rangle j^\mu $&$\langle\Bar{B}\gamma_\mu u_\nu\rangle\langle u^\nu B\rangle j^\mu $&$\langle\Bar{B}\gamma^5\gamma^\nu B[u_\nu,u_\mu] \rangle j^\mu $\\
        $\langle\Bar{B}\gamma^\nu u_\mu\rangle\langle Bu_\nu\rangle j^\mu-h.c. $&$\langle\Bar{B}\gamma^5\gamma_\mu B u_\nu u^\nu\rangle j^\mu $& $\langle\Bar{B}\gamma_\mu B u_\nu u^\nu\rangle j^\mu $&$\langle\Bar{B}\gamma^5\gamma^\nu u_\mu\rangle\langle Bu_\nu\rangle j^\mu-h.c. $\\
        $\langle\Bar{B}\sigma_{\mu\nu}\lrD_\rho B\{u^\nu,u^\rho\} \rangle j^\mu$&$\langle\Bar{B}\gamma^5\gamma_\mu u_\nu u^\nu B\rangle j^\mu $&$\langle\Bar{B}\gamma_\mu u_\nu u^\nu B\rangle j^\mu $&$\langle\Bar{B}\gamma^5\sigma_{\mu\nu}\lrD_\rho B[u^\nu,u^\rho] \rangle j^\mu$\\
\end{longtable}

\begin{longtable}{|@{}c@{}|@{}c@{}|@{}c@{}|@{}c@{}|}
        \hline
   $C$+$P$+&$C$+$P$-&$C$-$P$+&$C$-$P$-\\
        \hline
            $\langle\Bar{B}\gamma_\mu \lrD_\nu B\rangle j^{\{\mu,\nu\}}$&&$\langle\Bar{B}\sigma_{\mu\nu} B\rangle j^{[\mu,\nu]}$&$\langle\Bar{B}\gamma^5\sigma_{\mu\nu} B\rangle j^{[\mu,\nu]}$\\  
            &&&$\langle\Bar{B}\gamma^5\gamma_\mu \lrD_\nu B\rangle j^{\{\mu,\nu\}}$\\
    \hline  
    $\langle\Bar{B}\gamma^5\gamma_\mu u_\nu B\rangle j^{\{\mu,\nu\}}$&&&$\langle\Bar{B}\gamma_\mu u_\nu B\rangle j^{\{\mu,\nu\}}$\\
    $\langle\Bar{B}\gamma^5\gamma_\mu u_\nu B\rangle j^{[\mu,\nu]}$&&&$\langle\Bar{B}\gamma_\mu u_\nu B\rangle j^{[\mu,\nu]}$\\
    $\langle\Bar{B}\gamma^5\gamma_\mu Bu_\nu \rangle j^{\{\mu,\nu\}}$&&&$\langle\Bar{B}\gamma_\mu Bu_\nu \rangle j^{\{\mu,\nu\}}$\\
    $\langle\Bar{B}\gamma^5\gamma_\mu Bu_\nu \rangle j^{[\mu,\nu]}$&&&$\langle\Bar{B}\gamma_\mu Bu_\nu \rangle j^{[\mu,\nu]}$\\
    \hline
    \caption{The meson-baryon operators with $j^{\mu\nu}$ and without spurion.}
    \label{mb0jmn}
\end{longtable}

\begin{longtable}{|@{}c@{}|@{}c@{}|@{}c@{}|@{}c@{}|}
        \hline
   $C$+$P$+&$C$+$P$-&$C$-$P$+&$C$-$P$-\\
        \hline
        $\langle\Bar{B}\Sigma_+B\rangle j$&$\langle\Bar{B}\Sigma_-B\rangle j$&&\\
   $\langle\Bar{B}B\Sigma_+\rangle j$&$\langle\Bar{B}B\Sigma_-\rangle j$&&\\
   $\langle\Bar{B}B\rangle\langle\Sigma_+\rangle j$&$\langle\Bar{B}B\rangle\langle\Sigma_-\rangle j$&&\\
   \hline
   $\langle\Bar{B}\gamma^5\Sigma_-B\rangle j$&$\langle\Bar{B}\gamma^5\Sigma_+B\rangle j$&&\\
   $\langle\Bar{B}\gamma^5B\Sigma_-\rangle j$&$\langle\Bar{B}\gamma^5B\Sigma_+\rangle j$&&\\
   $\langle\Bar{B}\gamma^5B\rangle\langle\Sigma_-\rangle j$&$\langle\Bar{B}\gamma^5B\rangle\langle\Sigma_+\rangle j$&&\\
   $\langle\Bar{B}\gamma^\mu[\Sigma_-,u_\mu] B\rangle j$&$\langle\Bar{B}\gamma^\mu[\Sigma_+,u_\mu] B\rangle j$&$\langle\Bar{B}\gamma^\mu\{\Sigma_-,u_\mu\} B\rangle j$&$\langle\Bar{B}\gamma^\mu\{\Sigma_+,u_\mu\} B\rangle j$\\
\end{longtable}

{\small

\begin{longtable}{|@{}c@{}|@{}c@{}|@{}c@{}|@{}c@{}|}
        \hline
   $C$+$P$+&$C$+$P$-&$C$-$P$+&$C$-$P$-\\
        \hline
        $\langle\Bar{B}\gamma^5\gamma^\mu\Sigma_-B\rangle j^{\mu}$&$\langle\Bar{B}\gamma^5\gamma^\mu\Sigma_+B\rangle j^{\mu}$&$\langle\Bar{B}\gamma^\mu\Sigma_+B\rangle j^{\mu}$&$\langle\Bar{B}\gamma^\mu\Sigma_-B\rangle j^{\mu}$\\
   $\langle\Bar{B}\gamma^5\gamma^\mu B\Sigma_-\rangle j^{\mu}$&$\langle\Bar{B}\gamma^5\gamma^\mu B\Sigma_+\rangle j^{\mu}$&$\langle\Bar{B}\gamma^\mu B\Sigma_+\rangle j^{\mu}$&$\langle\Bar{B}\gamma^\mu B\Sigma_-\rangle j^{\mu}$\\
   $\langle\Bar{B}\gamma^5\gamma^\mu B\rangle\langle\Sigma_-\rangle j^{\mu}$&$\langle\Bar{B}\gamma^5\gamma^\mu B\rangle\langle\Sigma_+\rangle j^{\mu}$&$\langle\Bar{B}\gamma^\mu B\rangle\langle\Sigma_+\rangle j^{\mu}$&$\langle\Bar{B}\gamma^\mu B\rangle\langle\Sigma_-\rangle j^{\mu}$\\
   \hline
   $\langle\Bar{B}\{\Sigma_-,u^\mu\} B\rangle j^\mu$&$\langle\Bar{B}\{\Sigma_+,u^\mu\} B\rangle j^\mu$&$\langle\Bar{B}[\Sigma_-,u^\mu] B\rangle j^\mu$&$\langle\Bar{B}[\Sigma_+,u^\mu] B\rangle j^\mu$\\
   $\langle\Bar{B}B\{\Sigma_-,u^\mu\} \rangle j^\mu$&$\langle\Bar{B}B\{\Sigma_+,u^\mu\} \rangle j^\mu$&$\langle\Bar{B}B[\Sigma_-,u^\mu] \rangle j^\mu$&$\langle\Bar{B}B[\Sigma_+,u^\mu] \rangle j^\mu$\\
   $\langle\Bar{B}\Sigma_-Bu^\mu \rangle j^\mu$&$\langle\Bar{B}\Sigma_+Bu^\mu \rangle j^\mu$&$\langle\Bar{B} u^\mu\rangle\langle B\Sigma_-\rangle j^\mu-h.c.$&$\langle\Bar{B} u^\mu\rangle\langle B\Sigma_+\rangle j^\mu-h.c.$\\
   $\langle\Bar{B}u^\mu B\Sigma_-\rangle j^\mu$&$\langle\Bar{B}u^\mu B\Sigma_+\rangle j^\mu$&$\langle\Bar{B}\gamma^5\sigma^{\mu\nu}u_\nu B\rangle\langle \Sigma_+\rangle j^\mu$&$\langle\Bar{B}\gamma^5\sigma^{\mu\nu}u_\nu B\rangle\langle \Sigma_-\rangle j^\mu$\\
\end{longtable}

}
\begin{longtable}{|@{}c@{}|@{}c@{}|@{}c@{}|@{}c@{}|}
        \hline
   $C$+$P$+&$C$+$P$-&$C$-$P$+&$C$-$P$-\\
         \hline
         $\langle \Bar{B}\gamma_\mu\lrD_\nu\Sigma_+B\rangle j^{\{\mu,\nu\}}$&$\langle \Bar{B}\gamma_\mu\lrD_\nu\Sigma_-B\rangle j^{\{\mu,\nu\}}$&$\langle \Bar{B}\sigma_{\mu\nu}\Sigma_+B\rangle j^{[\mu,\nu]}$&$\langle \Bar{B}\sigma_{\mu\nu}\Sigma_-B\rangle j^{[\mu,\nu]}$\\
   $\langle \Bar{B}\gamma_\mu\lrD_\nu B\Sigma_+\rangle j^{\{\mu,\nu\}}$&$\langle \Bar{B}\gamma_\mu\lrD_\nu B\Sigma_-\rangle j^{\{\mu,\nu\}}$&$\langle \Bar{B}\gamma^5\sigma_{\mu\nu}\Sigma_-B\rangle j^{[\mu,\nu]}$&$\langle \Bar{B}\gamma^5\sigma_{\mu\nu}\Sigma_+B\rangle j^{[\mu,\nu]}$\\
   $\langle \Bar{B}\gamma_\mu\lrD_\nu B\rangle\langle\Sigma_+\rangle j^{\{\mu,\nu\}}$&$\langle \Bar{B}\gamma_\mu\lrD_\nu B\rangle\langle\Sigma_-\rangle j^{\{\mu,\nu\}}$&$\langle \Bar{B}\sigma_{\mu\nu}B\Sigma_+\rangle j^{[\mu,\nu]}$&$\langle \Bar{B}\sigma_{\mu\nu}B\Sigma_-\rangle j^{[\mu,\nu]}$\\
   &&$\langle \Bar{B}\gamma^5\sigma_{\mu\nu}B\Sigma_-\rangle j^{[\mu,\nu]}$&$\langle \Bar{B}\gamma^5\sigma_{\mu\nu}B\Sigma_+\rangle j^{[\mu,\nu]}$\\
   &&$\langle \Bar{B}\sigma_{\mu\nu}B\rangle\langle\Sigma_+\rangle j^{[\mu,\nu]}$&$\langle \Bar{B}\sigma_{\mu\nu}B\rangle\langle\Sigma_-\rangle j^{[\mu,\nu]}$\\
   &&$\langle \Bar{B}\gamma^5\sigma_{\mu\nu}B\rangle\langle\Sigma_-\rangle j^{[\mu,\nu]}$&$\langle \Bar{B}\gamma^5\sigma_{\mu\nu}B\rangle\langle\Sigma_+\rangle j^{[\mu,\nu]}$\\
   &&$\langle \Bar{B}\gamma^5\gamma_\mu\lrD_\nu\Sigma_-B\rangle j^{\{\mu,\nu\}}$&$\langle \Bar{B}\gamma^5\gamma_\mu\lrD_\nu\Sigma_+B\rangle j^{\{\mu,\nu\}}$\\
\end{longtable}

\begin{longtable}{|@{}c@{}|@{}c@{}|@{}c@{}|@{}c@{}|}
        \hline
   $C$+$P$+&$C$+$P$-&$C$-$P$+&$C$-$P$-\\
       \hline
   $\langle\Bar{B}Q_+B\rangle j$&&&$\langle\Bar{B}Q_-B\rangle j$\\
   $\langle\Bar{B}BQ_+\rangle j$&&&$\langle\Bar{B}BQ_-\rangle j$\\
   \hline  
   &$\langle\Bar{B}\gamma^5Q_+B\rangle j$&$\langle\Bar{B}\gamma^5Q_-B\rangle j$&\\
   &$\langle\Bar{B}\gamma^5BQ_+\rangle j$&$\langle\Bar{B}\gamma^5BQ_-\rangle j$&\\
   \hline
   $\langle\Bar{B}\gamma^5\gamma^\mu\{Q_+,u_\mu\} B\rangle j$&$\langle\Bar{B}\gamma^\mu[Q_+,u_\mu] B\rangle j$&$\langle\Bar{B}\gamma^5\gamma^\mu[Q_+,u_\mu] B\rangle j$&$\langle\Bar{B}\gamma^\mu\{Q_+,u_\mu\} B\rangle j$\\
   $\langle\Bar{B}\gamma^5\gamma^\mu B[Q_+,u_\mu] \rangle j$&$\langle\Bar{B}\gamma^\mu B\{Q_+,u_\mu\} \rangle j$&$\langle\Bar{B}\gamma^5\gamma^\mu B\{Q_+,u_\mu\} \rangle j$&$\langle\Bar{B}\gamma^\mu B[Q_+,u_\mu] \rangle j$\\
   $\langle\Bar{B}\gamma^5\gamma^\mu u_\mu BQ_+ \rangle j$&$\langle\Bar{B}\gamma^\mu Q_+\rangle\langle Bu_\mu \rangle j-h.c.$&$\langle\Bar{B}\gamma^5\gamma^\mu Q_+\rangle\langle Bu_\mu \rangle j-h.c.$&$\langle\Bar{B}\gamma^\mu u_\mu BQ_+ \rangle j$\\
\end{longtable}

\begin{longtable}{|@{}c@{}|@{}c@{}|@{}c@{}|@{}c@{}|}
        \hline
   $C$+$P$+&$C$+$P$-&$C$-$P$+&$C$-$P$-\\
   \hline
   &$\langle\Bar{B}\gamma^5\gamma_\mu Q_+B\rangle j^{\mu}$&$\langle\Bar{B}\gamma_\mu Q_+B\rangle j^{\mu}$&\\
   &$\langle\Bar{B}\gamma_\mu Q_-B\rangle j^{\mu}$&$\langle\Bar{B}\gamma^5\gamma_\mu Q_-B\rangle j^{\mu}$&\\
   &$\langle\Bar{B}\gamma^5\gamma_\mu BQ_+\rangle j^{\mu}$&$\langle\Bar{B}\gamma_\mu BQ_+\rangle j^{\mu}$&\\
   &$\langle\Bar{B}\gamma_\mu BQ_-\rangle j^{\mu}$&$\langle\Bar{B}\gamma^5\gamma_\mu BQ_-\rangle j^{\mu}$&\\
   \hline
   $\langle\Bar{B}[Q_-,u^\mu] B\rangle j^{\mu}$&$\langle\Bar{B}\{Q_+,u^\mu\} B\rangle j^{\mu}$&$\langle\Bar{B}\{Q_-,u^\mu\} B\rangle j^{\mu}$&$\langle\Bar{B}[Q_+,u^\mu] B\rangle j^{\mu}$\\
   $\langle\Bar{B}B[Q_-,u^\mu] \rangle j^{\mu}$&$\langle\Bar{B}B\{Q_+,u^\mu\} \rangle j^{\mu}$&$\langle\Bar{B}B\{Q_-,u^\mu\} \rangle j^{\mu}$&$\langle\Bar{B}B[Q_+,u^\mu] \rangle j^{\mu}$\\
   $\langle\Bar{B} u^\mu\rangle\langle BQ_-\rangle j^{\mu}-h.c.$&$\langle\Bar{B}Q_+Bu^\mu \rangle j^{\mu}$&$\langle\Bar{B}Q_-Bu^\mu \rangle j^{\mu}$&$\langle\Bar{B} u^\mu\rangle\langle BQ_+\rangle j^{\mu}-h.c.$\\
\end{longtable}

\begin{longtable}{|@{}c@{}|@{}c@{}|@{}c@{}|@{}c@{}|}
        \hline
   $C$+$P$+&$C$+$P$-&$C$-$P$+&$C$-$P$-\\
   \hline
   $\langle \Bar{B}\gamma^5\sigma_{\mu\nu}Q_-B\rangle j^{[\mu,\nu]}$&$\langle \Bar{B}\sigma_{\mu\nu}Q_-B\rangle j^{[\mu,\nu]}$&$\langle \Bar{B}\sigma_{\mu\nu}Q_+B\rangle j^{[\mu,\nu]}$&$\langle \Bar{B}\gamma^5\sigma_{\mu\nu}Q_+B\rangle j^{[\mu,\nu]}$\\
   $\langle \Bar{B}\gamma^5\sigma_{\mu\nu}BQ_-\rangle j^{[\mu,\nu]}$&$\langle \Bar{B}\sigma_{\mu\nu}BQ_-\rangle j^{[\mu,\nu]}$&$\langle \Bar{B}\sigma_{\mu\nu}BQ_+\rangle j^{[\mu,\nu]}$&$\langle \Bar{B}\gamma^5\sigma_{\mu\nu}BQ_+\rangle j^{[\mu,\nu]}$\\
   $\langle \Bar{B}\gamma^5\gamma_\mu\lrD_\nu Q_-B\rangle j^{\{\mu,\nu\}}$&&&$\langle \Bar{B}\gamma_\mu\lrD_\nu Q_-B\rangle j^{\{\mu,\nu\}}$\\
   $\langle \Bar{B}\gamma^5\gamma_\mu\lrD_\nu BQ_-\rangle j^{\{\mu,\nu\}}$&&&$\langle \Bar{B}\gamma_\mu\lrD_\nu BQ_-\rangle j^{\{\mu,\nu\}}$\\
   $\langle \Bar{B}\gamma_\mu\lrD_\nu Q_+B\rangle j^{\{\mu,\nu\}}$&&&$\langle \Bar{B}\gamma^5\gamma_\mu\lrD_\nu Q_+B\rangle j^{\{\mu,\nu\}}$\\
   $\langle \Bar{B}\gamma_\mu\lrD_\nu BQ_+\rangle j^{\{\mu,\nu\}}$&&&$\langle \Bar{B}\gamma^5\gamma_\mu\lrD_\nu BQ_+\rangle j^{\{\mu,\nu\}}$\\
   \hline     
    \caption{The meson-baryon operators with $j^{\mu\nu}$ and 1 spurion $Q_\pm$.}
    \label{mb1Qjmn}
\end{longtable}


\subsubsection{Pure Meson Operators with 2 Spurions}

\begin{center}
\begin{longtable}{|c|c|c|c|}
\hline
   $C$+$P$+&$C$+$P$-&$C$-$P$+&$C$-$P$-\\
   \hline
    $\langle Q_{+}Q_{+}\rangle$&&&$\langle Q_{-}Q_{+}\rangle$\\
    \hline
    $\langle\{ Q_{+}, Q_{+}\}u^\mu u_\mu\rangle$&$\langle[ Q_{-}, Q_{+}]u^\mu u_\mu\rangle$&&$\langle\{ Q_{-}, Q_{+}\}u^\mu u_\mu\rangle$\\
    $\langle\{ Q_{-}, Q_{-}\}u^\mu u_\mu\rangle$&&&$\langle Q_{+}u_\mu\rangle\langle Q_{-} u^\mu\rangle$\\
    $\langle Q_{+}u_\mu\rangle\langle Q_{+} u^\mu\rangle$&&&\\
    $\langle Q_{-}u_\mu\rangle\langle Q_{-} u^\mu\rangle$&&&\\
    \hline
    \caption{The pure meson operators with 2 spurions $Q_\pm$.}
    \label{pm2Q}
\end{longtable}    
\end{center}

\begin{center}
\begin{longtable}{|c|c|c|c|}
\hline
   $C$+$P$+&$C$+$P$-&$C$-$P$+&$C$-$P$-\\
   \hline
    $\langle\Sigma_{+}\Sigma_{+}\rangle$&$\langle\Sigma_{-}\Sigma_{+}\rangle$&&\\
    $\langle\Sigma_{+}\rangle\langle\Sigma_{+}\rangle$&$\langle\Sigma_{-}\rangle\langle\Sigma_{+}\rangle$&&\\
    $\langle\Sigma_{-}\rangle\langle\Sigma_{-}\rangle$&&&\\
    \hline
    $\langle\{\Sigma_{+},\Sigma_{+}\}u^\mu u_\mu\rangle$&$\langle\{\Sigma_{-},\Sigma_{+}\}u^\mu u_\mu\rangle$&&$\langle[\Sigma_{-},\Sigma_{+}]u^\mu u_\mu\rangle$\\
    $\langle\{\Sigma_{-},\Sigma_{-}\}u^\mu u_\mu\rangle$&&&\\
    $\langle\Sigma_{+}\rangle\langle\Sigma_{+}u_\mu u^\mu\rangle$&$\langle\Sigma_{-}\rangle\langle\Sigma_{+}u_\mu u^\mu\rangle$&&\\
    $\langle\Sigma_{-}\rangle\langle\Sigma_{-}u_\mu u^\mu\rangle$&$\langle\Sigma_{+}\rangle\langle\Sigma_{-}u_\mu u^\mu\rangle$&&\\
    $\langle\Sigma_{+}\rangle\langle\Sigma_{+}\rangle\langle u_\mu u^\mu\rangle$&&&\\
    $\langle\Sigma_{-}\rangle\langle\Sigma_{-}\rangle\langle u_\mu u^\mu\rangle$&$\langle\Sigma_{+}\rangle\langle\Sigma_{-}\rangle\langle u_\mu u^\mu\rangle$&&\\
    $\langle\Sigma_{+}\Sigma_{+}\rangle\langle u_\mu u^\mu\rangle$&$\langle\Sigma_{-}\Sigma_{+}\rangle\langle u_\mu u^\mu\rangle$&&\\
    $\langle\Sigma_{-}\Sigma_{-}\rangle\langle u_\mu u^\mu\rangle$&&&\\
    $\langle\Sigma_{+}u_\mu\rangle\langle\Sigma_{+} u^\mu\rangle$&$\langle\Sigma_{-}u_\mu\rangle\langle\Sigma_{+} u^\mu\rangle$&&\\
    $\langle\Sigma_{-}u_\mu\rangle\langle\Sigma_{-} u^\mu\rangle$&&&\\
    \hline
    \caption{The pure meson operators with 2 spurions $\Sigma_\pm$.}
    \label{pm2S}
\end{longtable}    
\end{center}

\begin{center}
\begin{longtable}{|c|c|c|c|}
\hline
   $C$+$P$+&$C$+$P$-&$C$-$P$+&$C$-$P$-\\
   \hline
    $\langle \Sigma_{+}Q_{+}\rangle$&$\langle \Sigma_{-}Q_{+}\rangle$&$\langle \Sigma_{-}Q_{-}\rangle$&$\langle \Sigma_{+}Q_{-}\rangle$\\
    \hline
    $\langle[ Q_{-},\Sigma_{-}]u^\mu u_\mu\rangle$&$\langle[ Q_{-},\Sigma_{+}]u^\mu u_\mu\rangle$&$\langle[ Q_{+},\Sigma_{+}]u^\mu u_\mu\rangle$&$\langle[ Q_{+},\Sigma_{-}]u^\mu u_\mu\rangle$\\
    $\langle\{ Q_{+},\Sigma_{+}\}u^\mu u_\mu\rangle$&$\langle\{ Q_{+},\Sigma_{-}\}u^\mu u_\mu\rangle$&$\langle\{ Q_{-},\Sigma_{-}\}u^\mu u_\mu\rangle$&$\langle\{ Q_{-},\Sigma_{+}\}u^\mu u_\mu\rangle$\\
    $\langle\Sigma_{+}\rangle\langle Q_{+}u_\mu u^\mu\rangle$&$\langle\Sigma_{-}\rangle\langle Q_{+}u_\mu u^\mu\rangle$&$\langle\Sigma_{-}\rangle\langle Q_{-}u_\mu u^\mu\rangle$&$\langle\Sigma_{+}\rangle\langle Q_{-}u_\mu u^\mu\rangle$\\
    $\langle Q_{+}\Sigma_{+}\rangle\langle u_\mu u^\mu\rangle$&$\langle Q_{+}\Sigma_{-}\rangle\langle u_\mu u^\mu\rangle$&$\langle Q_{-}\Sigma_{-}\rangle\langle u_\mu u^\mu\rangle$&$\langle Q_{-}\Sigma_{+}\rangle\langle u_\mu u^\mu\rangle$\\
    $\langle Q_{+}u_\mu\rangle\langle\Sigma_{+} u^\mu\rangle$&$\langle Q_{+}u_\mu\rangle\langle\Sigma_{-} u^\mu\rangle$&$\langle Q_{-}u_\mu\rangle\langle\Sigma_{-} u^\mu\rangle$&$\langle Q_{-}u_\mu\rangle\langle\Sigma_{+} u^\mu\rangle$\\
    \hline
    \caption{The pure meson operators with spurions $\Sigma_\pm$ and $Q_\pm$.}
    \label{pmSQ}
\end{longtable}    
\end{center}

\begin{center}
\begin{longtable}{|c|c|c|c|}
\hline
   $C$+$P$+&$C$+$P$-&$C$-$P$+&$C$-$P$-\\
   \hline
    $\langle[ Q_{-},\Sigma_{+}] u_\mu\rangle\bar e^c\gamma^\mu e$&$\langle[ Q_{-},\Sigma_{-}] u_\mu\rangle\bar e^c\gamma^\mu e$&$\langle[ Q_{+},\Sigma_{-}] u_\mu\rangle\bar e^c\gamma^\mu e$&$\langle[ Q_{+},\Sigma_{+}] u_\mu\rangle\bar e^c\gamma^\mu e$\\
    $\langle\{ Q_{+},\Sigma_{-}\} u_\mu\rangle\bar e^c\gamma^\mu e$&$\langle\{ Q_{+},\Sigma_{+}\} u_\mu\rangle\bar e^c\gamma^\mu e$&$\langle\{ Q_{-},\Sigma_{+}\} u_\mu\rangle\bar e^c\gamma^\mu e$&$\langle\{ Q_{-},\Sigma_{-}\} u_\mu\rangle\bar e^c\gamma^\mu e$\\
    \hline
    \caption{The pure meson operators for dimenson-9 with $\bar e^c\gamma^\mu e$.}
    \label{md9}
\end{longtable}    
\end{center}

\subsubsection{Meson-Baryon Operators with 2 Spurions}

\begin{center}
\small
    \begin{longtable}{|@{}c@{}|@{}c@{}|@{}c@{}|@{}c@{}|}
\hline
   $C$+$P$+&$C$+$P$-&$C$-$P$+&$C$-$P$-\\
   \hline 
   $\langle \Bar{B}\Sigma_{+}\Sigma_{+}B\rangle$&$\langle \Bar{B}\{\Sigma_{+},\Sigma_{-}\}B\rangle$&&$\langle \Bar{B}[\Sigma_{+},\Sigma_{-}]B\rangle$\\
   $\langle \Bar{B}B\Sigma_{+}\Sigma_{+}\rangle$&$\langle \Bar{B}B\{\Sigma_{+},\Sigma_{-}\}\rangle$&&$\langle \Bar{B}B[\Sigma_{+},\Sigma_{-}]\rangle$\\
   $\langle \Bar{B}\Sigma_{+}B\Sigma_{+}\rangle$&$\langle \Bar{B}\Sigma_{+}B\Sigma_{-}\rangle$&&$\langle \Bar{B}\Sigma_{+}\rangle\langle B\Sigma_{-}\rangle-h.c.$\\
   $\langle \Bar{B}B\rangle\langle\Sigma_{+}\Sigma_{+}\rangle$&$\langle \Bar{B}\Sigma_{-}B\Sigma_{+}\rangle$&&$\langle \Bar{B}\Sigma_{-}\rangle\langle B\Sigma_{+}\rangle-h.c.$\\
   $\langle \Bar{B}B\Sigma_{+}\rangle\langle\Sigma_{+}\rangle$&$\langle \Bar{B}B\rangle\langle\Sigma_{+}\Sigma_{-}\rangle$&&\\
   $\langle \Bar{B}\Sigma_{+}B\rangle\langle\Sigma_{+}\rangle$&$\langle \Bar{B}B\Sigma_{+}\rangle\langle\Sigma_{-}\rangle$&&\\
   $\langle \Bar{B}B\rangle\langle\Sigma_{+}\rangle\langle\Sigma_{+}\rangle$&$\langle \Bar{B}B\Sigma_{-}\rangle\langle\Sigma_{+}\rangle$&&\\
\end{longtable}
\end{center}
\begin{center}
\small
    \begin{longtable}{|@{}c@{}|@{}c@{}|@{}c@{}|@{}c@{}|}
\hline
        $C$+$P$+&$C$+$P$-&$C$-$P$+&$C$-$P$-\\
   \hline 
   $\langle \Bar{B}\{ Q_{+}, Q_{+}\}B\rangle$&$\langle \Bar{B}[ Q_{+}, Q_{-}]B\rangle$&&$\langle \Bar{B}B\{ Q_{-}, Q_{+}\}\rangle$\\
   $\langle \Bar{B}B\{ Q_{+}, Q_{+}\}\rangle$&$\langle \Bar{B}B[ Q_{+}, Q_{-}]\rangle$&&$\langle \Bar{B}\{ Q_{-}, Q_{+}\}B\rangle$\\
   $\langle \Bar{B} Q_{+}B Q_{+}\rangle$&$\langle \Bar{B}[ Q_{-}, Q_{+}]B\rangle$&&$\langle \Bar{B} Q_{+}B Q_{-}\rangle$\\
   $\langle \Bar{B}B\rangle\langle Q_{+} Q_{+}\rangle$&$\langle \Bar{B}B[ Q_{-} Q_{+}]\rangle$&&$\langle \Bar{B} Q_{-}B Q_{+}\rangle$\\
   $\langle \Bar{B} Q_{-}B Q_{-}\rangle$&$\langle \Bar{B} Q_{+}\rangle\langle B Q_{-}\rangle-h.c.$&&$\langle \Bar{B}B\rangle\langle Q_{+} Q_{-}\rangle$\\
   $\langle \Bar{B}B\rangle\langle Q_{-} Q_{-}\rangle$&$\langle \Bar{B} Q_{-}\rangle\langle B Q_{+}\rangle-h.c.$&&\\
   $\langle \Bar{B}\{ Q_{-}, Q_{-}\}B\rangle$&&&\\
\end{longtable}
\end{center}
\begin{center}
\small
    \begin{longtable}{|@{}c@{}|@{}c@{}|@{}c@{}|@{}c@{}|}
\hline
   $C$+$P$+&$C$+$P$-&$C$-$P$+&$C$-$P$-\\
   \hline 
   $\langle \Bar{B}\{ Q_{+},\Sigma_{+}\}B\rangle$&$\langle \Bar{B}\{ Q_{+},\Sigma_{-}\}B\rangle$&$\langle \Bar{B}[ Q_{+},\Sigma_{+}]B\rangle$&$\langle \Bar{B}[ Q_{+},\Sigma_{-}]B\rangle$\\
   $\langle \Bar{B}B\{ Q_{+},\Sigma_{+}\}\rangle$&$\langle \Bar{B}B\{ Q_{+},\Sigma_{-}\}\rangle$&$\langle \Bar{B}B[ Q_{+},\Sigma_{+}]\rangle$&$\langle \Bar{B}B[ Q_{+},\Sigma_{-}]\rangle$\\
   $\langle \Bar{B} Q_{+}B\Sigma_{+}\rangle$&$\langle \Bar{B} Q_{+}B\Sigma_{-}\rangle$&$\langle \Bar{B}\{ Q_{-},\Sigma_{-}\}B\rangle$&$\langle \Bar{B}\{ Q_{-},\Sigma_{+}\}B\rangle$\\
   $\langle \Bar{B}\Sigma_{+}B Q_{+}\rangle$&$\langle \Bar{B}\Sigma_{-}B Q_{+}\rangle$&$\langle \Bar{B}B\{ Q_{-},\Sigma_{-}\}\rangle$&$\langle \Bar{B}B\{ Q_{-}\Sigma_{+}\}\rangle$\\
   $\langle \Bar{B}B\rangle\langle Q_{+}\Sigma_{+}\rangle$&$\langle \Bar{B}B\rangle\langle Q_{+}\Sigma_{-}\rangle$&$\langle \Bar{B} Q_{+}\rangle\langle B\Sigma_{+}\rangle-h.c.$&$\langle \Bar{B} Q_{+}\rangle\langle B\Sigma_{-}\rangle-h.c.$\\
   $\langle \Bar{B}B Q_{+}\rangle\langle\Sigma_{+}\rangle$&$\langle \Bar{B}B Q_{+}\rangle\langle\Sigma_{-}\rangle$&$\langle \Bar{B}\Sigma_{-}B Q_{-}\rangle$&$\langle \Bar{B}\Sigma_{+}B Q_{-}\rangle$\\
   $\langle \Bar{B} Q_{+}B\rangle\langle\Sigma_{+}\rangle$&$\langle \Bar{B} Q_{+}B\rangle\langle\Sigma_{-}\rangle$&$\langle \Bar{B} Q_{-}B\Sigma_{-}\rangle$&$\langle \Bar{B} Q_{-}B\Sigma_{+}\rangle$\\
\end{longtable}
\end{center}

\begin{center}
\small
    \begin{longtable}{|@{}c@{}|@{}c@{}|@{}c@{}|@{}c@{}|}
\hline
   $C$+$P$+&$C$+$P$-&$C$-$P$+&$C$-$P$-\\
   \hline 
   $\langle \Bar{B}\gamma_\mu [ Q_{+},\Sigma_{+}]B\rangle{j^\mu}$&$\langle \Bar{B}\gamma_\mu [ Q_{+},\Sigma_{-}]B\rangle{j^\mu}$&$\langle \Bar{B}\gamma_\mu \{ Q_{+},\Sigma_{+}\}B\rangle{j^\mu}$&$\langle \Bar{B}\gamma_\mu \{ Q_{+},\Sigma_{-}\}B\rangle{j^\mu}$\\
   $\langle \Bar{B}\gamma_\mu B[ Q_{+},\Sigma_{+}]\rangle{j^\mu}$&$\langle \Bar{B}\gamma_\mu B[ Q_{+},\Sigma_{-}]\rangle{j^\mu}$&$\langle \Bar{B}\gamma_\mu B\{ Q_{+},\Sigma_{+}\}\rangle{j^\mu}$&$\langle \Bar{B}\gamma_\mu B\{ Q_{+},\Sigma_{-}\}\rangle{j^\mu}$\\
   $\langle \Bar{B}\gamma_\mu \{ Q_{-},\Sigma_{-}\}B\rangle{j^\mu}$&$\langle \Bar{B}\gamma_\mu \{ Q_{-},\Sigma_{+}\}B\rangle{j^\mu}$&$\langle \Bar{B}\gamma_\mu  Q_{+}B\Sigma_{+}\rangle{j^\mu}$&$\langle \Bar{B}\gamma_\mu  Q_{+}B\Sigma_{-}\rangle{j^\mu}$\\
   $\langle \Bar{B}\gamma_\mu B\{ Q_{-},\Sigma_{-}\}\rangle{j^\mu}$&$\langle \Bar{B}\gamma_\mu B\{ Q_{-}\Sigma_{+}\}\rangle{j^\mu}$&$\langle \Bar{B}\gamma_\mu \Sigma_{+}B Q_{+}\rangle{j^\mu}$&$\langle \Bar{B}\gamma_\mu \Sigma_{-}B Q_{+}\rangle{j^\mu}$\\
   $\langle \Bar{B}\gamma_\mu  Q_{+}\rangle\langle B\Sigma_{+}\rangle{j^\mu}-h.c.$&$\langle \Bar{B}\gamma_\mu  Q_{+}\rangle\langle B\Sigma_{-}\rangle{j^\mu}-h.c.$&$\langle \Bar{B}\gamma_\mu B\rangle\langle Q_{+}\Sigma_{+}\rangle{j^\mu}$&$\langle \Bar{B}\gamma_\mu B\rangle\langle Q_{+}\Sigma_{-}\rangle{j^\mu}$\\
   $\langle \Bar{B}\gamma_\mu \Sigma_{-}B Q_{-}\rangle{j^\mu}$&$\langle \Bar{B}\gamma_\mu \Sigma_{+}B Q_{-}\rangle{j^\mu}$&$\langle \Bar{B}\gamma_\mu B Q_{+}\rangle\langle\Sigma_{+}\rangle{j^\mu}$&$\langle \Bar{B}\gamma_\mu B Q_{+}\rangle\langle\Sigma_{-}\rangle{j^\mu}$\\
   $\langle \Bar{B}\gamma_\mu  Q_{-}B\Sigma_{-}\rangle{j^\mu}$&$\langle \Bar{B}\gamma_\mu  Q_{-}B\Sigma_{+}\rangle{j^\mu}$&$\langle \Bar{B}\gamma_\mu  Q_{+}B\rangle\langle\Sigma_{+}\rangle{j^\mu}$&$\langle \Bar{B}\gamma_\mu  Q_{+}B\rangle\langle\Sigma_{-}\rangle{j^\mu}$\\
\end{longtable}

\end{center}
\section{Spurion Matching between LEFT and ChPT}
\label{sec:matching}
\linespread{1.0}\selectfont

In the previous sections, we outlined the LEFT and the ChPT effective operators using the spurions. In this section, we will present the matching between them. Firstly we present the matching rules, which utilize the properties kept at both the quark and the hadronic levels. The naive dimension analysis (NDA)~\cite{Manohar:1983md,Gavela:2016bzc,Jenkins:2013sda,Panico:2015jxa,Buchalla:2013eza} is used to assess and organize the matching operators, thus the spurion method can be used for the matching beyond the leading order systematically. Then we will illustrate the matching procedure with some examples and finally present the LO and NLO matching results obtained from the LEFT operators up to dimension 9.

\subsection{The Matching Rules}

From the quark condensation, the quark interactions in the LEFT are described by the effective operators composed of the mesons and the baryons in the ChPT~\cite{Weinberg:1968de,Weinberg:1978kz,Gasser:1983yg,Gasser:1984gg,Gasser:1987rb}. Thus the matching of the operators between these two EFTs is important. Because the degrees of freedom change from the quarks to the hadrons, it is helpful to look at the properties that are kept at both the two EFTs. The commonly used external source method does well when the LEFT operators considered are simple, otherwise, there are several shortages mentioned in the introduction. 

Utilizing the spurion technique, we have shown that both the LEFT and the ChPT operators can be reformulated, which also implies the new matching procedure using these operators of the new form without the external sources. The quantities kept in both two EFTs are the spurion structures, the CP properties of the quark bilinears, and the non-quark fields including the leptons and the photon. Next, we discuss them in detail.
\begin{itemize}
    \item The spurion structures should be reserved in the matching procedure. As discussed before, for the spurion introduced in the LEFT, there are two kinds of dressing forms in the ChPT with different CP eigenvalues,  $\Sigma_\pm, Q_\pm$. When matching, the chiral correspondences of the spurion $\mathbf{T}$ are dependent on the quark bilinears of which the LEFT spurions are. The spurions of the quark bilinears changing chirality are matched to $Q_\pm$, while the ones reserving chirality are matched to $\Sigma_\pm$.
    Up to dimension 9, all possible quark bilinears and their corresponding chiral spurions are shown in Tab.~\ref{tab:corresponding}.
    \begin{table}[b]
    \centering
    \begin{tabular}{|c|ccccccccc|}
    \hline
          $\Gamma$ & $1$ & $\gamma^5$ & $\gamma^\mu$ & $\gamma^5\gamma^\mu$ & $\sigma^{\mu\nu}$ & $\lrpartial^\mu$&$\gamma^5\lrpartial^\mu$&$\gamma^\mu\lrpartial^\nu$&$\gamma^5\gamma^\mu\lrpartial^\nu$\\
          \hline
spurion & $\Sigma_\pm$ &$\Sigma_\pm$ & $Q_\pm$ &$Q_\pm$ & $\Sigma_\pm$&$\Sigma_\pm$&$\Sigma_\pm$&$Q_\pm$&$Q_\pm$\\
\hline
    \end{tabular}
    \caption{The corresponding between the spurion contracted with the quark bilinear and the building blocks in the ChPT.}
    \label{tab:corresponding}
\end{table}
    \item The CP properties of the quark parts are non-perturbative, and should be kept by the hadronic counterparts. The ChPT operators violating CP are important and have been discussed extensively for the matching, for example, the kaon decay~\cite{Bernard:1985wf,Grinstein:1985ut,Cheng:1987dk,Kambor:1989tz,Pich:1990mw,Kambor:1992he,Cirigliano:2003gt,Akdag:2022sbn,Pich:2021yll,Cornella:2023kjq}, the $0\nu\beta\beta$~\cite{Savage:1998yh,Prezeau:2003xn,Graesser:2016bpz,Cirigliano:2017ymo,Cirigliano:2017djv,Cirigliano:2017tvr,Pastore:2017ofx,Cirigliano:2018yza,Cirigliano:2019vdj}, and so on.
    
    \item Because the leptons and the photon are degrees of freedom in both the LEFT and ChPT, and they are independent of the quark condensation, the ChPT operators should have the same leptons or photon structure as the ones of the LEFT.
\end{itemize}
In particular, the relations such as the EOM and the IBP make the forms of the independent operators not unique. The matching results should not depend on this ambiguity, especially the power-counting should not change when matching to the meson-baryon operators, which is transparent in this spurion method and will be illustrated in the example next.

\subsubsection{Naive Dimension Analysis}

The matching between the LEFT and the ChPT is not a one-to-one procedure, which means that every LEFT operator can be matched to several ChPT operators, and every ChPT operator can be contributed from several LEFT operators (will be clarified in the subsequent examples). Thus we need a method to assess all these contributions systematically. In detail, we want to distinguish which one is more dominant among the LEFT operators matching to the same ChPT operator, and which one is more dominant among the ChPT operators matched from the same LEFT operator. In this section, we use the naive dimension analysis (NDA)~\cite{Manohar:1983md,Gavela:2016bzc,Jenkins:2013sda,Panico:2015jxa,Buchalla:2013eza} of the effective operators to organize them.

The LEFT is described to be expanded in powers of the electroweak scale inverse $1/\Lambda_{\text{EW}}$. The NDA master formula~\cite{Manohar:1983md,Gavela:2016bzc} in the 4-dimension spacetime is that the operator in the Lagrangian is normalized according to
\begin{equation}
\label{eq:NDA_1}
\frac{\Lambda_{\text{EW}}^4}{16\pi^2} {\left[\frac{\partial}{\Lambda_{\text{EW}}}\right]}^{N_p}{\left[\frac{4\pi\phi}{\Lambda_{\text{EW}}}\right]}^{N_\phi}{\left[\frac{4\pi A}{\Lambda_{\text{EW}}}\right]}^{N_A}{\left[\frac{4\pi\psi}{\Lambda_{\text{EW}}^{3/2}}\right]}^{N_\psi}{\left[\frac{g}{4\pi}\right]}^{N_g}{\left[\frac{y}{4\pi}\right]}^{N_y}{\left[\frac{\lambda}{16\pi^2}\right]}^{N_\lambda}\,.
\end{equation}
For convenience, we express the master formula by replacing the vector field $A$ by the field-strength tensor $F\sim \partial A$ and omitting the renormalizable coupling constants $g,y,\lambda$,
\begin{equation}
\text{LEFT:\quad }
    \frac{\Lambda_{\text{EW}}^4}{16\pi^2}\left[\frac{\partial}{\Lambda_{\text{EW}}}\right]^{N_p} \left[\frac{4\pi F}{\Lambda_{\text{EW}}^2}\right]^{N_F} \left[\frac{4\pi \psi}{\Lambda_{\text{EW}}^{3/2}}\right]^{N_\psi}\,,
\end{equation}
according to which the 4-fermion operators of dimension-6, 7, and 8 are normalized as
\begin{equation}
    \frac{{(4\pi)}^2}{\Lambda_{\text{EW}}^2}\psi^4\,,\quad\frac{{(4\pi)}^2}{\Lambda_{\text{EW}}^2}\frac{\partial}{\Lambda_{\text{EW}}}\psi^4\,,\quad\frac{{(4\pi)}^2}{\Lambda_{\text{EW}}^2}\frac{\partial^2}{\Lambda_{\text{EW}}^2}\psi^4\,.
\end{equation}
In particular, the spurions are dimensionless in the NDA.
On the other hand, the NDA master formula of the ChPT is 
\begin{equation}
    \text{ChPT:\quad }f^2\Lambda_\chi^2 \left[\frac{\partial}{\Lambda_\chi}\right]^{N_p} \left[\frac{\psi}{f\sqrt{\Lambda_\chi}}\right]^{N_\psi} \left[\frac{F}{\Lambda_\chi f}\right]^{N_A} \,,  \label{eq:nda_chiral}
\end{equation}
where $4\pi f\sim \Lambda_\chi$. To relate these two NDA formulae we first replace all the scale $\Lambda_{\text{EW}}$ in Eq.~\eqref{eq:NDA_1} by the scale $\Lambda_\chi$ that
\begin{align}
    & \frac{\Lambda_{\text{EW}}^4}{16\pi^2} \left[\frac{\Lambda_\chi}{\Lambda_{\text{EW}}}\right]^{N_p+2N_F + \frac{3}{2}N_\psi} \left[\frac{\partial}{\Lambda_\chi}\right]^{N_p} \left[\frac{F}{\Lambda_\chi^2}\right]^{N_F} \left[\frac{\psi}{\sqrt{\Lambda_\chi}f}\right]^{N_\psi} \notag \\
    =& \left[\frac{\Lambda_\chi}{\Lambda_{\text{EW}}}\right]^{\mathcal{D}} \left(f^2\Lambda_\chi^2 \left[\frac{\partial}{\Lambda_\chi}\right]^{N_p} \left[\frac{F}{\Lambda_\chi^2}\right]^{N_F} \left[\frac{\psi}{\sqrt{\Lambda_\chi}f}\right]^{N_\psi}\right)\,,
\end{align}
where $\mathcal{D}=N_p+2N_F + \frac{3}{2}N_\psi-4$. The expression inside the circle bracket is similar to the NDA formula of the ChPT, so we replace it with the ChPT NDA formula in Eq.~\eqref{eq:nda_chiral} and obtain the NDA formula of the matching that
\begin{align}
    \text{matching:\quad }& \frac{\Lambda_{\text{EW}}^4}{16\pi^2}\left[\frac{\partial}{\Lambda_{\text{EW}}}\right]^{N_p} \left[\frac{4\pi F}{\Lambda_{\text{EW}}^2}\right]^{N_F} \left[\frac{4\pi \psi}{\Lambda_{\text{EW}}^{3/2}}\right]^{N_\psi} \notag \\
    \sim & \left[\frac{\Lambda_\chi}{\Lambda_{\text{EW}}}\right]^{\mathcal{D}} \left(f^2\Lambda_\chi^2 \left[\frac{\partial}{\Lambda_\chi}\right]^{N_p} \left[\frac{\psi}{f\sqrt{\Lambda_\chi}}\right]^{N_\psi} \left[\frac{F}{\Lambda_\chi f}\right]^{N_A} \right)\,, \label{eq:NDA_2}
\end{align}
which is the main result of our argument. 

According to Eq.~\eqref{eq:NDA_2}, we can conclude that
\begin{itemize}
    \item For a specific ChPT operator, the contributions from different LEFT operators are dominated by the factor $[\Lambda_\chi/\Lambda_{\text{EW}}]^\mathcal{D}$, thus the contribution from the higher-dimension LEFT operators are less important.
    \item For a specific LEFT operator, all the matching ChPT operators are organized by the number of derivatives and fields. In the matching procedure, the photon and the leptons are fixed, and the matched operators are expanded in terms of the numbers of derivatives and baryons, which is consistent with the chiral power-counting scheme in Sec.~\ref{sec:chpt}.
\end{itemize}

\subsubsection*{Matching Example for Dim-7 Operators with One-derivative}
Here is a simple example to illustrate the matching procedure. We consider the dimension-7 four-fermion operators with one derivative
\begin{align}
    \Op_1=&(\bar q\lrpartial^\mu\textbf{T} q)(\bar e_L\gamma_\mu e_L)\,,\\
    \Op_2=&(\bar q\lrpartial^\mu \textbf{T}q)(\bar e_R\gamma_\mu e_R)\,,\\
    \Op_3=&(\bar q\gamma^5\lrpartial^\mu\textbf{T} q)(\bar e_L\gamma_\mu e_L)\,,\\
    \Op_4=&(\bar q\gamma^5\lrpartial^\mu\textbf{T} q)(\bar e_R\gamma_\mu e_R)\,,
\end{align}
which are from the operators $\mathcal{O}^{(7)}_{16}\,,\mathcal{O}^{(7)}_{17}$ in Eq.~\eqref{eq:ops1} and $\mathcal{O}^{(7)}_{30}\,,\mathcal{O}^{(7)}_{31}$ in Eq.~\eqref{eq:ops2}, and has been relabeled for convenience.
There is only 1 spurion, and the corresponding chiral spurion is $\Sigma_\pm$ according to Tab.~\ref{tab:corresponding}. The CP properties of the quarks can be obtained according to Tab.~\ref{tab:quarkbilinear}.
The leptons of all these operators are $e_L$ or $e_R$. All these properties are listed in Tab.~\ref{tab:example_1}.
\begin{table}[]
    \centering
    \begin{tabular}{|c|c|c|c|c|}
\hline
operators & $\mathcal{O}_1$& $\mathcal{O}_2$ & $\mathcal{O}_3$ & $\mathcal{O}_4$ \\
\hline
spurion & \multicolumn{4}{|c|}{$\Sigma_\pm$} \\
\hline
CP & \multicolumn{2}{|c|}{C-P+} & \multicolumn{2}{|c|}{C-P-} \\
\hline
leptons & $e_L$ & $e_R$ & $e_L$ & $e_R$ \\
\hline
    \end{tabular}
    \caption{The properties of the operators $\mathcal{O}_1\sim \mathcal{O}_4$.}
    \label{tab:example_1}
\end{table}

According to the matching rules, the relevant building blocks of the ChPT are 
\begin{equation}
    \left(\begin{array}{c}
         u_\mu  \\
          \Sigma_-\\
          \Sigma_+\\
          e_R\\
          e_L\\
          B
    \end{array}\right)\,,
\end{equation}
and the ChPT operators after matching are the ones composed of these fields invariant under both the Lorentz symmetry and the $SU(3)_V$ symmetry with CP properties C-P+ and C-P- respectively. In particular, the lepton current should not change during the matching. 

Let us consider $\op_1$ as an example. It is C-P+, and the first two pure meson operators in the matching are 
\begin{equation}
    \langle \Sigma_+ [u_\nu ,D^\nu u_\mu]\rangle(\bar e_L\gamma^\mu e_L)\,,\quad\langle[\Sigma_-,u_\mu] u^\nu u_\nu\rangle(\bar e_L\gamma^\mu e_L)\,,
\end{equation}
where at least 2 $u_\mu$ are needed to form the C-P+ operators. The nonvanishing operators with minimal derivatives are called the leading-order (LO) operators in the matching. 
On the other hand, the LO meson-baryon operators are 
\begin{equation}
    \langle\bar B\gamma^\mu \Sigma_+B\rangle(\bar e_L\gamma_\mu e_L)\,,\quad \langle\bar B\gamma^\mu B\Sigma_+\rangle(\bar e_L\gamma_\mu e_L)\,,\quad \langle\bar B\gamma^\mu B\rangle\langle\Sigma_+\rangle(\bar e_L\gamma_\mu e_L)\,.
\end{equation}
According to the NDA formula in Eq.~\eqref{eq:NDA_2}, the normalization tells
\begin{equation}
\label{eq:nor_1}
    \left(\frac{\Lambda_\chi}{\Lambda_{\text{EW}}}\right)^2 f^2\Lambda_\chi^2 \left(\frac{\psi}{f\sqrt{\Lambda_\chi}}\right)^4 \sim \frac{(4\pi)^2}{\Lambda_{\text{EW}}^2}\psi^4\,.
\end{equation}
With more $u_\mu$ added, there are more operators such as 
\begin{equation}
    \langle\bar B[u^\mu, \Sigma_-]B\rangle(\bar e_L\gamma_\mu e_L)\,,\quad \langle\bar BB[u^\mu, \Sigma_-]\rangle(\bar e_L\gamma_\mu e_L)\,,\quad \langle\bar B u^\mu\rangle\langle\Sigma_-B\rangle(\bar e_L\gamma_\mu e_L)-h.c.\dots
\end{equation}
The normalization of them has one more factor $\frac{1}{\Lambda_\chi}$, thus are sub-leading. 

Here we present the matching result of the operators $\op_1\sim\op_4$ with at most 3 $u_\mu$ for the pure meson operators and at most 1 $u_\mu$ for the meson-baryon operators.
{\small
\begin{equation}
    \begin{aligned}
C-P+:\,\Op_1,\Op_2\rightarrow&\langle \Sigma_+ [u_\nu ,D^\nu u_\mu]\rangle(\bar e_L\gamma^\mu e_L)\,,\quad\langle[\Sigma_-,u_\mu] u^\nu u_\nu\rangle(\bar e_L\gamma^\mu e_L)\,,\\
    &\langle\bar B\gamma^\mu \Sigma_+B\rangle(\bar e_L\gamma_\mu e_L)\,,\langle\bar B\gamma^\mu B\Sigma_+\rangle(\bar e_L\gamma_\mu e_L)\,,\langle\bar B\gamma^\mu B\rangle\langle\Sigma_+\rangle(\bar e_L\gamma_\mu e_L)\,,\\
    &\langle\bar B[u^\mu, \Sigma_-]B\rangle(\bar e_L\gamma_\mu e_L)\,,\langle\bar BB[u^\mu, \Sigma_-]\rangle(\bar e_L\gamma_\mu e_L)\,,\langle\bar B u^\mu\rangle\langle\Sigma_-B\rangle(\bar e_L\gamma_\mu e_L)-h.c.,\\
    &\langle\bar B\gamma^5\sigma^{\mu\nu} u_\nu B\Sigma_+\rangle(\bar e_L\gamma_\mu e_L)\,,\langle\bar B\gamma^5\sigma^{\mu\nu}\Sigma_+ Bu_\nu\rangle(\bar e_L\gamma_\mu e_L)\,,\langle\bar B\gamma^5\sigma^{\mu\nu}B\{\Sigma_+, u_\nu\}\rangle(\bar e_L\gamma_\mu e_L)\,,\\
    &\langle\bar B\gamma^5\sigma^{\mu\nu}\{\Sigma_+, u_\nu\}B\rangle(\bar e_L\gamma_\mu e_L)\,,\langle\bar B\gamma^5\sigma^{\mu\nu}B\rangle\langle\Sigma_+ u_\nu\rangle(\bar e_L\gamma_\mu e_L)\,,\langle\bar B\gamma^5\sigma^{\mu\nu}Bu_\nu\rangle\langle\Sigma_+ \rangle(\bar e_L\gamma_\mu e_L)\,,\\
    &\langle\bar B\gamma^5\sigma^{\mu\nu}u_\nu B\rangle\langle\Sigma_+ \rangle(\bar e_L\gamma_\mu e_L)\,,\langle\bar B\sigma^{\mu\nu}u_\nu B\rangle\langle\Sigma_- \rangle(\bar e_L\gamma_\mu e_L)\,,\\
    &\langle\bar B\sigma^{\mu\nu} u_\nu B\Sigma_-\rangle(\bar e_L\gamma_\mu e_L)\,,\langle\bar B\sigma^{\mu\nu}\Sigma_- Bu_\nu\rangle(\bar e_L\gamma_\mu e_L)\,,\langle\bar B\sigma^{\mu\nu}B\{\Sigma_-, u_\nu\}\rangle(\bar e_L\gamma_\mu e_L)\,,\\
    &\langle\bar B\sigma^{\mu\nu}\{\Sigma_-, u_\nu\}B\rangle(\bar e_L\gamma_\mu e_L)\,,\langle\bar B\sigma^{\mu\nu}B\rangle\langle\Sigma_- u_\nu\rangle(\bar e_L\gamma_\mu e_L)\,,\langle\bar B\sigma^{\mu\nu}Bu_\nu\rangle\langle\Sigma_- \rangle(\bar e_L\gamma_\mu e_L)\,,\\
    &\langle\bar B\gamma^5\gamma^\mu\lrD^\nu u_\nu B\Sigma_+\rangle(\bar e_L\gamma_\mu e_L)\,,\langle\bar B\gamma^5\gamma^\mu\lrD^\nu\Sigma_+ Bu_\nu\rangle(\bar e_L\gamma_\mu e_L)\,,\langle\bar B\gamma^5\gamma^\mu\lrD^\nu B\{\Sigma_+, u_\nu\}\rangle(\bar e_L\gamma_\mu e_L)\,,\\
    &\langle\bar B\gamma^5\gamma^\mu\lrD^\nu\{\Sigma_+, u_\nu\}B\rangle(\bar e_L\gamma_\mu e_L)\,,\langle\bar B\gamma^5\gamma^\mu\lrD^\nu B\rangle\langle\Sigma_+ u_\nu\rangle(\bar e_L\gamma_\mu e_L)\,,\\
    &\langle\bar B\gamma^5\gamma^\mu\lrD^\nu u_\nu B\rangle\langle\Sigma_+ \rangle(\bar e_L\gamma_\mu e_L)\,,\langle\bar B\gamma^5\gamma^\mu\lrD^\nu Bu_\nu\rangle\langle\Sigma_+ \rangle(\bar e_L\gamma_\mu e_L)\,,\\
    &\langle\bar B\gamma^\mu\lrD^\nu  B[u_\nu,\Sigma_-]\rangle(\bar e_L\gamma_\mu e_L)\,,\langle\bar B\gamma^\mu\lrD^\nu  [u_\nu,\Sigma_-]B\rangle(\bar e_L\gamma_\mu e_L)\,,\langle\bar B\gamma^\mu\lrD^\nu  u_\nu\rangle\langle\Sigma_-B\rangle(\bar e_L\gamma_\mu e_L)-h.c.\,,\\
    &(L\leftrightarrow R)\,,\\
C-P-:\,\Op_3,\Op_4\rightarrow&\langle \Sigma_- [u_\nu ,D^\nu u_\mu]\rangle(\bar e_L\gamma^\mu e_L)\,,\quad\langle[\Sigma_+,u_\mu] u^\nu u_\nu\rangle(\bar e_L\gamma^\mu e_L)\,,\\
    &\langle\bar B\gamma^\mu \Sigma_-B\rangle(\bar e_L\gamma_\mu e_L)\,,\langle\bar B\gamma^\mu B\Sigma_-\rangle(\bar e_L\gamma_\mu e_L)\,,\langle\bar B\gamma^\mu B\rangle\langle\Sigma_-\rangle(\bar e_L\gamma_\mu e_L)\,,\\
    &\langle\bar B[u^\mu, \Sigma_+]B\rangle(\bar e_L\gamma_\mu e_L)\,,\langle\bar BB[u^\mu, \Sigma_+]\rangle(\bar e_L\gamma_\mu e_L)\,,\langle\bar B u^\mu\rangle\langle\Sigma_+B\rangle(\bar e_L\gamma_\mu e_L)-h.c.\,,\\
    &\langle\bar B\gamma^5\sigma^{\mu\nu} u_\nu B\Sigma_-\rangle(\bar e_L\gamma_\mu e_L)\,,\langle\bar B\gamma^5\sigma^{\mu\nu}\Sigma_- Bu_\nu\rangle(\bar e_L\gamma_\mu e_L)\,,\langle\bar B\gamma^5\sigma^{\mu\nu}B\{\Sigma_-, u_\nu\}\rangle(\bar e_L\gamma_\mu e_L)\,,\\
    &\langle\bar B\gamma^5\sigma^{\mu\nu}\{\Sigma_-, u_\nu\}B\rangle(\bar e_L\gamma_\mu e_L)\,,\langle\bar B\gamma^5\sigma^{\mu\nu}B\rangle\langle\Sigma_- u_\nu\rangle(\bar e_L\gamma_\mu e_L)\,,\langle\bar B\gamma^5\sigma^{\mu\nu}Bu_\nu\rangle\langle\Sigma_- \rangle(\bar e_L\gamma_\mu e_L)\,,\\
    &\langle\bar B\gamma^5\sigma^{\mu\nu}u_\nu B\rangle\langle\Sigma_- \rangle(\bar e_L\gamma_\mu e_L)\,,\langle\bar B\sigma^{\mu\nu}u_\nu B\rangle\langle\Sigma_+ \rangle(\bar e_L\gamma_\mu e_L)\,,\\
    &\langle\bar B\sigma^{\mu\nu} u_\nu B\Sigma_+\rangle(\bar e_L\gamma_\mu e_L)\,,\langle\bar B\sigma^{\mu\nu}\Sigma_+ Bu_\nu\rangle(\bar e_L\gamma_\mu e_L)\,,\langle\bar B\sigma^{\mu\nu}B\{\Sigma_+, u_\nu\}\rangle(\bar e_L\gamma_\mu e_L)\,,\\
    &\langle\bar B\sigma^{\mu\nu}\{\Sigma_+, u_\nu\}B\rangle(\bar e_L\gamma_\mu e_L)\,,\langle\bar B\sigma^{\mu\nu}B\rangle\langle\Sigma_+ u_\nu\rangle(\bar e_L\gamma_\mu e_L)\,,\langle\bar B\sigma^{\mu\nu}Bu_\nu\rangle\langle\Sigma_+ \rangle(\bar e_L\gamma_\mu e_L)\,,\\
    &\langle\bar B\gamma^5\gamma^\mu\lrD^\nu u_\nu B\Sigma_-\rangle(\bar e_L\gamma_\mu e_L)\,,\langle\bar B\gamma^5\gamma^\mu\lrD^\nu\Sigma_- Bu_\nu\rangle(\bar e_L\gamma_\mu e_L)\,,\langle\bar B\gamma^5\gamma^\mu\lrD^\nu B\{\Sigma_-, u_\nu\}\rangle(\bar e_L\gamma_\mu e_L)\,,\\
    &\langle\bar B\gamma^5\gamma^\mu\lrD^\nu\{\Sigma_-, u_\nu\}B\rangle(\bar e_L\gamma_\mu e_L)\,,\langle\bar B\gamma^5\gamma^\mu\lrD^\nu B\rangle\langle\Sigma_- u_\nu\rangle(\bar e_L\gamma_\mu e_L)\,,\\
    &\langle\bar B\gamma^5\gamma^\mu\lrD^\nu u_\nu B\rangle\langle\Sigma_- \rangle(\bar e_L\gamma_\mu e_L)\,,\langle\bar B\gamma^5\gamma^\mu\lrD^\nu  Bu_\nu\rangle\langle\Sigma_- \rangle(\bar e_L\gamma_\mu e_L)\,,\\
    &\langle\bar B\gamma^\mu\lrD^\nu  B[u_\nu,\Sigma_+]\rangle(\bar e_L\gamma_\mu e_L)\,,\langle\bar B\gamma^\mu\lrD^\nu  [u_\nu,\Sigma_+]B\rangle(\bar e_L\gamma_\mu e_L)\,,\langle\bar B\gamma^\mu\lrD^\nu  u_\nu\rangle\langle\Sigma_+B\rangle(\bar e_L\gamma_\mu e_L)-h.c.\,,\\
    &(L\leftrightarrow R)\,.
    \end{aligned}
    \label{ope}
\end{equation}
}
In addition, the EOM and the IBP for the operators will not affect the matching between the LEFT operators and the chiral Lagrangian. For example, one of the above operator 
\begin{equation}
  \Op_1= (\bar q\lrpartial_\mu\textbf{T} q)(\bar e_L\gamma^\mu e_L)\,,  
\end{equation}
can translate to the operator
\begin{equation}
   \Op'_1= (\bar q\sigma_{\mu\nu}\textbf{T} q)\partial^\nu(\bar e_L\gamma^\mu e_L)\,,
\end{equation}
through the EOM and IBP, while the $CP$ properties and the spurion of the operators are the same, thus the particle spectrum in the hadronic level will be the same. When the lepton part is fixed, the two operators should have the same matching results.

Here we will show the equivalence of these two operators after the matching explicitly. According to the matching rules, the non-vanish leading chiral Lagrangian of the operator $\Op_1$ are
\begin{equation}
    \begin{aligned}
&\langle \Sigma_+ [u_\nu ,D^\nu u_\mu]\rangle(\bar e_L\gamma^\mu e_L)\,,\quad \langle[\Sigma_-,u_\mu] u^\nu u_\nu\rangle(\bar e_L\gamma^\mu e_L)\,,\\
&\langle\bar B\gamma^\mu \Sigma_+B\rangle(\bar e_L\gamma_\mu e_L)\,,\quad \langle\bar B\gamma^\mu B\Sigma_+\rangle(\bar e_L\gamma_\mu e_L)\,,\quad \langle\bar B\gamma^\mu B\rangle\langle\Sigma_+\rangle(\bar e_L\gamma_\mu e_L)\,.
\end{aligned}
\label{re}
\end{equation}

In addition, the non-vanish LO chiral Lagrangian for $\Op'_1$ seem to be different with $\Op_1$ due to the difference of the lepton part and become
\begin{equation}
\begin{aligned}
    &\langle \Sigma_+ [u_\nu , u_\mu]\rangle D^\nu(\bar e_L\gamma^\mu e_L)\,,\\
    &\langle\bar B\gamma^5\sigma_{\mu\nu} \Sigma_-B\rangle D^\nu(\bar e_L\gamma^\mu e_L)\,,\quad \langle\bar B\gamma^5\sigma_{\mu\nu} B\Sigma_-\rangle D^\nu(\bar e_L\gamma^\mu e_L)\,,\quad \langle\bar B\gamma^5\sigma_{\mu\nu} B\rangle\langle\Sigma_-\rangle D^\nu(\bar e_L\gamma^\mu e_L)\,,\\
    &\langle\bar B\gamma^5\gamma_\mu\lrD_\nu \Sigma_-B\rangle D^\nu(\bar e_L\gamma^\mu e_L)\,,\quad \langle\bar B\gamma^5\gamma_\mu\lrD_\nu B\Sigma_-\rangle D^\nu(\bar e_L\gamma^\mu e_L)\,,\quad \langle\bar B\gamma^5\gamma_\mu\lrD_\nu B\rangle\langle\Sigma_-\rangle D^\nu(\bar e_L\gamma^\mu e_L)\,,\\
    &\langle\bar B\sigma_{\mu\nu} \Sigma_+B\rangle D^\nu(\bar e_L\gamma^\mu e_L)\,,\quad \langle\bar B\sigma_{\mu\nu} B\Sigma_+\rangle D^\nu(\bar e_L\gamma^\mu e_L)\,,\quad \langle\bar B\sigma_{\mu\nu} B\rangle\langle\Sigma_+\rangle D^\nu(\bar e_L\gamma^\mu e_L)\,.
\end{aligned}
\end{equation}
According to Eq.~\eqref{eq:redundancy}, the IBP, and the EOM of the hadrons, we can obtain 
\begin{equation}
 \begin{aligned}
\langle \Sigma_+ [u_\nu , u_\mu]\rangle D^\nu(\bar e_L\gamma^\mu e_L)&\sim\langle D^\nu\Sigma_+ [u_\nu , u_\mu]\rangle (\bar e_L\gamma^\mu e_L)+\langle \Sigma_+ [D^\nu u_\nu , u_\mu]\rangle (\bar e_L\gamma^\mu e_L)\\
    &+\langle \Sigma_+ [u_\nu , D^\nu u_\mu]\rangle (\bar e_L\gamma^\mu e_L)\\
    &\sim\langle[\Sigma_-,u_\mu]u^\nu u_\nu\rangle (\bar e_L\gamma^\mu e_L)+\langle \Sigma_+ [ u_\nu , D^\nu u_\mu]\rangle (\bar e_L\gamma^\mu e_L)\,,
 \end{aligned}   
\end{equation}
where the first equivalence uses the IBP, the second equivalence uses the EOM of $u_\mu$ and Eq.~\eqref{eq:redundancy}. This means the pure meson operators matched from $\op_1$ are equivalent to the ones matched from $\op'_1$ up to some operators of higher chiral dimension. Similarly, the meson-baryon part can also translate that 
\begin{equation}
    \begin{aligned}
        \langle\bar B\gamma^5\sigma_{\mu\nu}B \Sigma_-\rangle D^\nu(\bar e_L\gamma^\mu e_L)&\sim\langle D^\nu(\bar B\gamma^5\sigma_{\mu\nu}B) \Sigma_-\rangle (\bar e_L\gamma^\mu e_L)+\langle \bar B\gamma^5\sigma_{\mu\nu}B D^\nu\Sigma_-\rangle (\bar e_L\gamma^\mu e_L)\\
        &\sim\langle\bar B\gamma^5\sigma_{\mu\nu}B \{\Sigma_+,u^\nu\} \rangle (\bar e_L\gamma^\mu e_L)\,,\\
        \langle\bar B\gamma^5\gamma_\mu\lrD_\nu B\Sigma_-\rangle D^\nu(\bar e_L\gamma^\mu e_L)&\sim\langle D^\nu(\bar B\gamma^5\gamma_\mu\lrD_\nu B)\Sigma_-\rangle (\bar e_L\gamma^\mu e_L)+\langle \bar B\gamma^5\gamma_\mu\lrD_\nu BD^\nu\Sigma_-\rangle (\bar e_L\gamma^\mu e_L)\\
        &\sim\langle \bar B\gamma^5\gamma_\mu\lrD_\nu B\{u^\nu,\Sigma_-\}\rangle (\bar e_L\gamma^\mu e_L)\,,\\
        \langle\bar B\sigma_{\mu\nu} B\Sigma_+\rangle D^\nu(\bar e_L\gamma^\mu e_L)&\sim\langle D^\nu(\bar B\sigma_{\mu\nu} B)\Sigma_+\rangle (\bar e_L\gamma^\mu e_L)+\langle\bar B\sigma_{\mu\nu} BD^\nu\Sigma_+\rangle (\bar e_L\gamma^\mu e_L)\\
        &\sim\langle\Bar{B}\lrD_\mu B\Sigma_+\rangle(\bar e_L\gamma^\mu e_L)+\langle\bar B\sigma_{\mu\nu} B\{u^\nu,\Sigma_+\}\rangle (\bar e_L\gamma^\mu e_L)\,,
    \end{aligned}
    \label{trans}
\end{equation}
where the first equivalence uses the IBP, and the second equivalence uses the EOM of $B$ and Eq.~\eqref{eq:redundancy}. The equivalence relations of other operators can also be obtained similarly. According to the heavy baryon projection Eq.~\eqref{heavy_baryon}
\begin{align}
    \langle\Bar{B}\lrD_\mu B\Sigma_+\rangle(\bar e_L\gamma^\mu e_L)=\langle\Bar{B}_v v_\mu B_v\Sigma_+\rangle(\bar e_L\gamma^\mu e_L)+...\,,\\
    \langle\bar B\gamma^\mu B\Sigma_+\rangle(\bar e_L\gamma_\mu e_L)=\langle\bar B_vv^\mu B_v\Sigma_+\rangle(\bar e_L\gamma_\mu e_L)+...\,,
\end{align}
which means the meson-baryon part is also consistent at LO. The other operators in Eq.~\eqref{trans} with more $u_\mu$ which can also be found in Eq.~\eqref{ope} are of higher chiral dimension. Thus the EOM and the IBP relations do not affect the matching result. 
Although this example is about pure meson operators and meson-baryon operators, the same argument applies when matching to other operators such as the ones with more baryons.

\subsection{Matching Example for Dim-6 and 8 LEFT Operators}

In this subsection, we present 2 more examples and clarify more details of our matching procedure.

Firstly, we consider the dimension-8 four-fermion operators with two derivatives 
\begin{align}
    \mathcal{O}_5=&(\bar q\gamma^\mu\lrpartial^\nu\textbf{T} q)(\bar e_L\gamma_\mu\lrpartial_\nu e_L)\,,\\   
    \mathcal{O}_6=&(\bar q\gamma^\mu\lrpartial^\nu\textbf{T} q)(\bar e_R\gamma_\mu\lrpartial_\nu e_R)\,,\\
    \mathcal{O}_7=&(\bar q\gamma^5\gamma^\mu\lrpartial^\nu\textbf{T} q)(\bar e_L\gamma_\mu\lrpartial_\nu e_L)\,,\\
    \mathcal{O}_8=&(\bar q\gamma^5\gamma^\mu\lrpartial^\nu\textbf{T} q)(\bar e_R\gamma_\mu\lrpartial_\nu e_R)\,,
\end{align}
where the Lorentz indices are symmetric. Here $\textbf{T}$ corresponds to the building block $Q_\pm$. The CP properties of the quark part of the operators $\mathcal{O}_5\,,\mathcal{O}_6$ are $C$+$P$+, while the CP properties of the operators $\mathcal{O}_7\,,\mathcal{O}_8$ are $C$-$P$-. Thus the relevant particle spectrum of the chiral Lagrangian is
\begin{equation}
    \left(\begin{array}{c}
         u_\mu  \\
          Q_-\\
          Q_+\\
          e_R\\
          e_L\\
          B
    \end{array}\right)\,,
\end{equation}
and the Lorentz structures of the lepton part should be fixed. Thus the chiral Lagrangian for the CP eigenstates of the non-vanish leading term becomes
\begin{equation}
\begin{aligned}
    C+P+:\quad&\langle Q_+ u^\mu u^\nu\rangle(\bar e_L\gamma_\mu\lrpartial_\nu e_L)\,,\\
    &\langle \bar B\gamma^\mu\lrD^\nu Q_+B\rangle(\bar e_L\gamma_\mu\lrpartial_\nu e_L)\,,\quad\langle \bar B\gamma^5\gamma^\mu\lrD^\nu Q_-B\rangle(\bar e_L\gamma_\mu\lrpartial_\nu e_L)\,,\\
    &\langle \bar B\gamma^\mu\lrD^\nu BQ_+\rangle(\bar e_L\gamma_\mu\lrpartial_\nu e_L)\,,\quad\langle \bar B\gamma^5\gamma^\mu\lrD^\nu BQ_-\rangle(\bar e_L\gamma_\mu\lrpartial_\nu e_L)\,,\\
    &(L\leftrightarrow R)\,,\\
    C-P-:\quad&\langle Q_- u^\mu u^\nu\rangle(\bar e_L\gamma_\mu\lrpartial_\nu e_L)\,,\\
    &\langle \bar B\gamma^5\gamma^\mu\lrD^\nu Q_+B\rangle(\bar e_L\gamma_\mu\lrpartial_\nu e_L)\,,\quad\langle \bar B\gamma^\mu\lrD^\nu Q_-B\rangle(\bar e_L\gamma_\mu\lrpartial_\nu e_L)\,,\\
    &\langle \bar B\gamma^5\gamma^\mu\lrD^\nu BQ_+\rangle(\bar e_L\gamma_\mu\lrpartial_\nu e_L)\,,\quad\langle \bar B\gamma^\mu\lrD^\nu BQ_-\rangle(\bar e_L\gamma_\mu\lrpartial_\nu e_L)\,,\\
    &(L\leftrightarrow R)\,,
\end{aligned}
\end{equation}
thus the matching will be obtained as
\begin{equation}
\begin{aligned}
    \Op_5\,,\Op_6\rightarrow&\langle Q_+ u^\mu u^\nu\rangle(\bar e_L\gamma_\mu\lrpartial_\nu e_L)\,,\\
    &\langle \bar B\gamma^\mu\lrD^\nu Q_+B\rangle(\bar e_L\gamma_\mu\lrpartial_\nu e_L)\,,\quad\langle \bar B\gamma^5\gamma^\mu\lrD^\nu Q_-B\rangle(\bar e_L\gamma_\mu\lrpartial_\nu e_L)\,,\\
    &\langle \bar B\gamma^\mu\lrD^\nu BQ_+\rangle(\bar e_L\gamma_\mu\lrpartial_\nu e_L)\,,\quad\langle \bar B\gamma^5\gamma^\mu\lrD^\nu BQ_-\rangle(\bar e_L\gamma_\mu\lrpartial_\nu e_L)\,,\\
    &(L\leftrightarrow R)\,,\\
    \Op_7\,,\Op_8\rightarrow&\langle Q_- u^\mu u^\nu\rangle(\bar e_L\gamma_\mu\lrpartial_\nu e_L)\,,\\
    &\langle \bar B\gamma^5\gamma^\mu\lrD^\nu Q_+B\rangle(\bar e_L\gamma_\mu\lrpartial_\nu e_L)\,,\quad\langle \bar B\gamma^\mu\lrD^\nu Q_-B\rangle(\bar e_L\gamma_\mu\lrpartial_\nu e_L)\,,\\
    &\langle \bar B\gamma^5\gamma^\mu\lrD^\nu BQ_+\rangle(\bar e_L\gamma_\mu\lrpartial_\nu e_L)\,,\quad\langle \bar B\gamma^\mu\lrD^\nu BQ_-\rangle(\bar e_L\gamma_\mu\lrpartial_\nu e_L)\,,\\
    &(L\leftrightarrow R)\,,
\end{aligned}
\end{equation}
and the higher order of the chiral Lagrangian can be also obtained similarly, which can be found in the tables in the next section. The Wilson coefficients of mass dimension according to the NDA formula in Eq.~\eqref{eq:NDA_2} are
\begin{equation}
    \langle Q_+ u^\mu u^\nu\rangle(\bar e_L\gamma_\mu\lrpartial_\nu e_L)\sim \frac{\Lambda_\chi^2}{\Lambda_{\text{EW}}^4}\partial^2\psi^2 \,,\quad 
    \langle \bar B\gamma^\mu\lrD^\nu Q_+B\rangle(\bar e_L\gamma_\mu\lrpartial_\nu e_L) \sim \frac{\Lambda_\chi}{\Lambda_{\text{EW}}^2} \frac{(4\pi)^2}{\Lambda_{\text{EW}}^2} \partial\psi^4\,,
\end{equation}
where the derivatives on the baryons are of no chiral dimension. The normalization of the meson-baryon operators here are suppressed by a factor $\frac{\Lambda_\chi}{\Lambda_{\text{EW}}^2} $ compared to the ones of the previous example in Eq.~\eqref{eq:nor_1}.

Secondly, we consider LEFT operators involving four quarks
\begin{align}
   C+P+: \mathcal{O}_9=&(\Bar{q}\textbf{T} q)(\Bar{q}\textbf{T} q)\,,\\
    \mathcal{O}_{10}=&(\Bar{q}\textbf{T}\gamma^5 q)(\Bar{q}\textbf{T}\gamma^5 q)\,,\\
   C+P-: \mathcal{O}_{11}=&(\Bar{q}\textbf{T} q)(\Bar{q}\textbf{T}\gamma^5 q)\,,\\
   \mathcal{O}_{12}=&(\Bar{q}\textbf{T}\gamma^5 q)(\Bar{q}\textbf{T} q)\,,
\end{align}
which may contain a lepton bilinear either free of electric charge or not. These operators are useful, for example, by adding a lepton bilinear $(\bar e e^c)+h.c.$ they describe the $0\nu\beta\beta$ process. Regardless of the leptons, the corresponding chiral spurions are $\Sigma_\pm$ and the chiral Lagrangian of the pure meson operators are
\begin{align}
C+P+:\,&\langle\Sigma_{+}\Sigma_{+}\rangle\,,\quad\langle\Sigma_{+}\rangle\langle\Sigma_{+}\rangle\,,\quad\langle\Sigma_{-}\rangle\langle\Sigma_{-}\rangle\,,\\
   &\langle\{\Sigma_{+},\Sigma_{+}\}u^\mu u_\mu\rangle\,,\quad\langle\{\Sigma_{-},\Sigma_{-}\}u^\mu u_\mu\rangle\,,\quad\langle\Sigma_{+}\rangle\langle\Sigma_{+}u_\mu u^\mu\rangle\,,\\
   &\langle\Sigma_{-}\rangle\langle\Sigma_{-}u_\mu u^\mu\rangle\,,\quad\langle\Sigma_{+}\rangle\langle\Sigma_{+}u_\mu u^\mu\rangle\,,\quad\langle\Sigma_{-}\rangle\langle\Sigma_{-}u_\mu u^\mu\rangle\,,\\
   &\langle\Sigma_{+}\rangle\langle\Sigma_{+}\rangle\langle u_\mu u^\mu\rangle\,,\quad\langle\Sigma_{-}\rangle\langle\Sigma_{-}\rangle\langle u_\mu u^\mu\rangle\,,\\
   &\langle\Sigma_{-}\Sigma_{-}\rangle\langle u_\mu u^\mu\rangle\,,\langle\Sigma_{+}u_\mu\rangle\langle\Sigma_{+} u^\mu\rangle\,,\langle\Sigma_{-}u_\mu\rangle\langle\Sigma_{-} u^\mu\rangle\,,\\
   C+P-:\, &\langle\Sigma_{+}\Sigma_{-}\rangle\,,\quad\langle\Sigma_{+}\rangle\langle\Sigma_{-}\rangle\,,\quad\langle\Sigma_{-}\rangle\langle\Sigma_{+}\rangle\,,\\
   &\langle\{\Sigma_{+},\Sigma_{-}\}u^\mu u_\mu\rangle\,,\quad\langle\{\Sigma_{-},\Sigma_{+}\}u^\mu u_\mu\rangle\,,\quad\langle\Sigma_{+}\rangle\langle\Sigma_{-}u_\mu u^\mu\rangle\,,\\
   &\langle\Sigma_{-}\rangle\langle\Sigma_{+}u_\mu u^\mu\rangle\,,\quad\langle\Sigma_{-}\rangle\langle\Sigma_{+}u_\mu u^\mu\rangle\,,\quad\langle\Sigma_{+}\rangle\langle\Sigma_{-}u_\mu u^\mu\rangle\,,\\
   &\langle\Sigma_{+}\rangle\langle\Sigma_{-}\rangle\langle u_\mu u^\mu\rangle\,,\quad\langle\Sigma_{-}\rangle\langle\Sigma_{+}\rangle\langle u_\mu u^\mu\rangle\,,\\
   &\langle\Sigma_{-}\Sigma_{+}\rangle\langle u_\mu u^\mu\rangle\,,\langle\Sigma_{+}u_\mu\rangle\langle\Sigma_{-} u^\mu\rangle\,,\langle\Sigma_{-}u_\mu\rangle\langle\Sigma_{+} u^\mu\rangle\,,
\end{align}
where only the LO and NLO operators are presented.
On the other hand, the LEFT operators can be combined as
\begin{align}
 (\bar q_R\textbf{T} q_L)(\bar q_R\textbf{T} q_L)\sim \mathcal{O}_9+\mathcal{O}_{10}-\mathcal{O}_{11}-\mathcal{O}_{12}\,,\\
    (\bar q_L\textbf{T} q_R)(\bar q_L\textbf{T} q_R)\sim \mathcal{O}_9+\mathcal{O}_{10}+\mathcal{O}_{11}+\mathcal{O}_{12}\,,\\
    (\bar q_R\textbf{T} q_L)(\bar q_L\textbf{T} q_R)\sim \mathcal{O}_9-\mathcal{O}_{10}-\mathcal{O}_{11}+\mathcal{O}_{12}\,,\\
    (\bar q_L\textbf{T} q_R)(\bar q_R\textbf{T} q_L)\sim \mathcal{O}_9-\mathcal{O}_{10}+\mathcal{O}_{11}-\mathcal{O}_{12}\,,   
\end{align}
so that we recover the LEFT operators in terms of the symmetry $SU(3)_{\mathbf{L}}\times SU(3)_{\mathbf{R}}$.
The LO ChPT operators after matching would be 
\begin{align}
    (\bar q_R\textbf{T} q_L)(\bar q_R\textbf{T} q_L):\quad&2\langle\Sigma_{+}\Sigma_{-}\rangle+\langle\Sigma_{+}\rangle\langle\Sigma_{+}\rangle+\langle\Sigma_{-}\rangle\langle\Sigma_{-}\rangle-2\langle\Sigma_{+}\Sigma_{-}\rangle\\
    &-\langle\Sigma_{+}\rangle\langle\Sigma_{-}\rangle-\langle\Sigma_{-}\rangle\langle\Sigma_{+}\rangle+...\,,\\
    (\bar q_L\textbf{T} q_R)(\bar q_L\textbf{T} q_R):\quad&2\langle\Sigma_{+}\Sigma_{-}\rangle+\langle\Sigma_{+}\rangle\langle\Sigma_{+}\rangle+\langle\Sigma_{-}\rangle\langle\Sigma_{-}\rangle+2\langle\Sigma_{+}\Sigma_{-}\rangle\\
    &+\langle\Sigma_{+}\rangle\langle\Sigma_{-}\rangle+\langle\Sigma_{-}\rangle\langle\Sigma_{+}\rangle+...\,,\\
    (\bar q_R\textbf{T} q_L)(\bar q_L\textbf{T} q_R):\quad&\langle\Sigma_{+}\rangle\langle\Sigma_{+}\rangle-\langle\Sigma_{-}\rangle\langle\Sigma_{-}\rangle-\langle\Sigma_{+}\rangle\langle\Sigma_{-}\rangle+\langle\Sigma_{-}\rangle\langle\Sigma_{+}\rangle+...\,,\\
    (\bar q_L\textbf{T} q_R)(\bar q_R\textbf{T} q_L):\quad&\langle\Sigma_{+}\rangle\langle\Sigma_{+}\rangle-\langle\Sigma_{-}\rangle\langle\Sigma_{-}\rangle+\langle\Sigma_{+}\rangle\langle\Sigma_{-}\rangle-\langle\Sigma_{-}\rangle\langle\Sigma_{+}\rangle+...\,,
\end{align}
which can be compared to the previous result about $0\nu\beta\beta$~\cite{Cirigliano:2017ymo}.

Similarly, we have another 4-quark structure composed of two vector (axial vector) bilinears,
\begin{align}
   C+P+:\,& \mathcal{O}_{13}=(\Bar{q}\textbf{T}\gamma^\mu q)(\Bar{q}\textbf{T}\gamma_\mu q)\,,\\
    &\mathcal{O}_{14}=(\Bar{q}\textbf{T}\gamma^5\gamma^\mu q)(\Bar{q}\textbf{T}\gamma^5\gamma_\mu q)\,,\\
   C-P-:\,&\mathcal{O}_{15}=(\Bar{q}\textbf{T}\gamma^5\gamma^\mu q)(\Bar{q}\textbf{T}\gamma_\mu q)\,,  \\
    &\mathcal{O}_{16}=(\Bar{q}\textbf{T}\gamma^\mu q)(\Bar{q}\textbf{T}\gamma^5\gamma_\mu q)\,.
\end{align}
The chiral spurions are $Q_\pm$ and they would match to the chiral Lagrangian
\begin{align}
   C+P+\,:&\langle Q_{+}Q_{+}\rangle\,,\\
   &\langle\{ Q_{+}, Q_{+}\}u^\mu u_\mu\rangle\,,\quad\langle\{ Q_{-}, Q_{-}\}u^\mu u_\mu\rangle\,,\\
   &\langle Q_{-} Q_{-}\rangle\langle u_\mu u^\mu\rangle\,,\langle Q_{+}u_\mu\rangle\langle Q_{+} u^\mu\rangle\,,\langle Q_{-}u_\mu\rangle\langle Q_{-} u^\mu\rangle\,,\\
   C-P-\,:&\langle Q_{+}Q_{-}\rangle\,,\\
   &\langle\{ Q_{+}, Q_{-}\}u^\mu u_\mu\rangle\,,\quad\langle\{ Q_{-}, Q_{+}\}u^\mu u_\mu\rangle\,,\\
   &\langle Q_{-} Q_{+}\rangle\langle u_\mu u^\mu\rangle\,,\langle Q_{+}u_\mu\rangle\langle Q_{-} u^\mu\rangle\,,\langle Q_{-}u_\mu\rangle\langle Q_{+} u^\mu\rangle\,.
\end{align}
As argued before, we can combine the LEFT operators as
\begin{align}
    (\bar q_L\gamma^\mu\textbf{T} q_L)(\bar q_L\gamma_\mu\textbf{T} q_L)\sim \mathcal{O}_{13}+\mathcal{O}_{14}-\mathcal{O}_{15}-\mathcal{O}_{16}\,,\\
    (\bar q_R\gamma^\mu\textbf{T} q_R)(\bar q_R\gamma_\mu\textbf{T} q_R)\sim \mathcal{O}_{13}+\mathcal{O}_{14}+\mathcal{O}_{15}+\mathcal{O}_{16}\,,\\
    (\bar q_L\gamma^\mu\textbf{T} q_L)(\bar q_R\gamma_\mu\textbf{T} q_R)\sim \mathcal{O}_{13}-\mathcal{O}_{14}-\mathcal{O}_{15}+\mathcal{O}_{16}\,,\\
    (\bar q_R\gamma^\mu\textbf{T} q_R)(\bar q_L\gamma_\mu\textbf{T} q_L)\sim \mathcal{O}_{13}-\mathcal{O}_{14}+\mathcal{O}_{15}-\mathcal{O}_{16}\,,
\end{align}
and their LO ChPT operators after matching would be
\begin{align}
   (\bar q_L\gamma^\mu\textbf{T} q_L)(\bar q_L\gamma_\mu\textbf{T} q_L):&\quad\langle Q_{+}Q_{+}\rangle-\langle Q_{+}Q_{-}\rangle\,,\\
    (\bar q_R\gamma^\mu\textbf{T} q_R)(\bar q_R\gamma_\mu\textbf{T} q_R):&\quad\langle Q_{+}Q_{+}\rangle+\langle Q_{+}Q_{-}\rangle\,,\\
    (\bar q_L\gamma^\mu\textbf{T} q_L)(\bar q_R\gamma_\mu\textbf{T} q_R):&\quad0\,,\\
    (\bar q_R\gamma^\mu\textbf{T} q_R)(\bar q_L\gamma_\mu\textbf{T} q_L):&\quad0\,,
\end{align}
which is consistent with Ref.~\cite{Cirigliano:2017ymo}. 
We have mentioned that a ChPT operator can be matched from more than one LEFT operator. The 4-quark operators here offer an appropriate example. If we consider the operators with 2 more derivatives
\begin{align}
   C+P+: &\mathcal{O}'_9=(\Bar{q}\textbf{T} q)D^2(\Bar{q}\textbf{T} q)\,,\\
    &\mathcal{O}'_{10}=(\Bar{q}\textbf{T}\gamma^5 q)D^2(\Bar{q}\textbf{T}\gamma^5 q)\,,\\
   C+P-: &\mathcal{O}'_{11}=(\Bar{q}\textbf{T}q)D^2(\Bar{q}\textbf{T}\gamma^5 q)\,,\\
   &\mathcal{O}'_{12}=(\Bar{q}\textbf{T}\gamma^5 q)D^2(\Bar{q}\textbf{T} q)\,,
\end{align}
the spurions and the CP properties are the same, thus the matching operators should not change. However, since of the two derivatives, these LEFT operators are of higher dimension, and thus their matching contributions to the ChPT operators are sub-dominant according to the NDA formula in Eq.~\eqref{eq:NDA_2}.

\subsection{The Matching Results Up to NLO}

The spurion method is systematic and can be applied to any effective operators of the two EFTs.
In this section, we present the matching result of the LEFT operators listed in Sec.~\ref{sec:LEFT}. And the matching result can be classified by the number of spurions.
Because we only consider the LO and NLO operators of the pure meson and the meson-baryon sectors, some LEFT operators can match to nothing, whose nonvanish matching results are of higher order. 
In particular, we match the dimension 9 LEFT operators about the $0\nu\beta\beta$ to the ChPT ones, but we do not present the correspondences of the operators with scalar lepton bilinear such as $\mathcal{O}^{(9)}_1,\mathcal{O}^{(9)}_2,\mathcal{O}^{(9)}_7,\mathcal{O}^{(9)}_8$, since they can be obtained from some ChPT operators by adding the bilinear, as shown in Tab.~\ref{mm3_1}.

\begin{center}
\begin{longtable}{|c|c|c|}
\hline
    LEFT operators&pure meson&meson-baryon\\
\hline
$\Op^{(6)}_2\,,\Op^{(6)}_4\,,\Op^{(6)}_{35}\,,\Op^{(6)}_{38}\,,\Op^{(6)}_{41}\,,\Op^{(6)}_{44}\,,\Op^{(6)}_{47}\,,\Op^{(6)}_{50}\,,$&{\multirow{4}{*}{$C$+$P$+ in Tab.~\ref{purem0j}}}&{\multirow{4}{*}{$C$+$P$+ in Tab.~\ref{mb0j}}}\\
$\Op^{(6)}_{53}\,,\Op^{(6)}_{56}\,,\Op^{(7)}_{2}\,,\Op^{(7)}_{4}\,,\Op^{(7)}_{37}\,,\Op^{(7)}_{42}\,,\Op^{(8)}_{5}\,,\Op^{(8)}_{6}\,,$&&\\
$\Op^{(8)}_{11}\,,\Op^{(8)}_{12}\,,\Op^{(8)}_{129}\,,\Op^{(8)}_{132}\,,\Op^{(8)}_{135}\,,\Op^{(8)}_{138}\,,\Op^{(8)}_{141}\,,\Op^{(8)}_{144}\,,\Op^{(8)}_{147}\,,$&&\\
$\Op^{(8)}_{150}\,,\Op^{(8)}_{153}\,,\Op^{(8)}_{156}\,,\Op^{(8)}_{159}\,,\Op^{(8)}_{162}\,,\Op^{(8)}_{165}\,,\Op^{(8)}_{168}\,,\Op^{(8)}_{171}\,,\Op^{(8)}_{174}\,.$&&\\
\hline
$\Op^{(8)}_{2}\,,\Op^{(8)}_{18}\,,\Op^{(8)}_{19}\,,\Op^{(8)}_{20}\,.$&$C$+$P$+ in Tab.~\ref{purem0jmn}&$C$+$P$+ in Tab.~\ref{mb0jmn}\\
\hline
$\Op^{(6)}_8\,,\Op^{(6)}_{10}\,,\Op^{(6)}_{59}\,,\Op^{(6)}_{63}\,,\Op^{(7)}_{6}\,,\Op^{(7)}_{8}\,,\Op^{(7)}_{45}\,,\Op^{(7)}_{50}\,,$&{\multirow{2}{*}{$C$+$P$- in Tab.~\ref{purem0j}}}&{\multirow{3}{*}{$C$+$P$- in Tab.~\ref{mb0j}}}\\
$\Op^{(8)}_{41}\,,\Op^{(8)}_{42}\,,\Op^{(8)}_{47}\,,\Op^{(8)}_{48}\,,\Op^{(8)}_{177},,\Op^{(8)}_{181}\,,\Op^{(8)}_{185}\,,\Op^{(8)}_{189}\,.$&&\\
\hline
$\Op^{(6)}_{16}\,,\Op^{(6)}_{17}\,,\Op^{(6)}_{18}\,,\Op^{(7)}_{11}\,,\Op^{(7)}_{12}\,,\Op^{(8)}_{26}\,,$&{\multirow{2}{*}{$C$+$P$- in Tab.~\ref{purem0jm}}}&{\multirow{2}{*}{$C$+$P$- in Tab.~\ref{mb0jm}}}\\
$\Op^{(8)}_{27}\,,\Op^{(8)}_{28}\,,\Op^{(8)}_{34}\,,\Op^{(8)}_{35}\,,\Op^{(8)}_{36}\,,\Op^{(8)}_{54}\,,\Op^{(8)}_{55}\,,\Op^{(8)}_{56}\,.$&&\\
\hline
$\Op^{(6)}_{24}\,,\Op^{(6)}_{25}\,,\Op^{(6)}_{26}\,,\Op^{(7)}_{18}\,,\Op^{(7)}_{19}\,,\Op^{(7)}_{20}\,,\Op^{(7)}_{25}\,,\Op^{(7)}_{26}\,,\Op^{(8)}_{62}\,,$&{\multirow{2}{*}{$C$-$P$+ in Tab.~\ref{purem0jm}}}&{\multirow{2}{*}{$C$-$P$+ in Tab.~\ref{mb0jm}}}\\
$\Op^{(8)}_{63}\,,\Op^{(8)}_{64}\,,\Op^{(8)}_{70}\,,\Op^{(8)}_{71}\,,\Op^{(8)}_{72}\,,\Op^{(8)}_{96}\,,\Op^{(8)}_{97}\,,\Op^{(8)}_{98}\,,\Op^{(8)}_{109}\,,\Op^{(8)}_{110}\,.$&&\\
\hline
$\Op^{(5)}_{1}\,,\Op^{(5)}_{3}\,,\Op^{(6)}_{31}\,,\Op^{(6)}_{32}\,,\Op^{(8)}_{77}\,,\Op^{(8)}_{78}\,,$&{\multirow{2}{*}{$C$-$P$+ in Tab.~\ref{purem0jmn}}}&{\multirow{2}{*}{$C$-$P$+ in Tab.~\ref{mb0jmn}}}\\
$\Op^{(8)}_{83}\,,\Op^{(8)}_{84}\,,\Op^{(8)}_{89}\,,\Op^{(8)}_{90}\,,\Op^{(8)}_{103}\,,\Op^{(8)}_{104}\,.$&&\\
\hline
$\Op^{(7)}_{53}$\,,$\Op^{(7)}_{58}\,.$&$C$-$P$+ in Tab.~\ref{purem0j}&$C$-$P$+ in Tab.~\ref{mb0j}\\
\hline
$\Op^{(6)}_{67}\,,\Op^{(6)}_{71}\,,\Op^{(7)}_{61}\,,\Op^{(7)}_{66}\,,\Op^{(8)}_{193}\,,\Op^{(8)}_{197}\,,\Op^{(8)}_{201}\,,\Op^{(8)}_{205}\,.$&$C$-$P$- in Tab.~\ref{purem0j}&$C$-$P$- in Tab.~\ref{mb0j}\\
\hline
$\Op^{(7)}_{32}\,,\Op^{(7)}_{33}\,,\Op^{(7)}_{34}\,,\Op^{(8)}_{125}\,,\Op^{(8)}_{126}\,.$&$C$-$P$- in Tab.~\ref{purem0jm}&$C$-$P$- in Tab.~\ref{mb0jm}\\
\hline
$\Op^{(8)}_{114}\,,\Op^{(8)}_{118}\,,\Op^{(8)}_{119}\,,\Op^{(8)}_{120}\,.$&$C$-$P$- in Tab.~\ref{purem0jmn}&$C$-$P$- in Tab.~\ref{mb0jmn}\\
\hline
\caption{The matching for operators without spurion.}
\end{longtable}
\end{center}

\begin{center}
    \begin{longtable}{|c|c|c|c|}
    \hline
    LEFT operators&pure meson&meson-baryon\\
    \hline
    $\Op_{2}^{(5)}\,,\Op_{4}^{(5)}\,,\Op^{(6)}_{29}\,,\Op^{(6)}_{30}\,,\Op^{(6)}_{33}\,,\Op^{(6)}_{34}\,,\Op^{(8)}_{75}\,,$&&\\
    $\Op^{(8)}_{76}\,,\Op^{(8)}_{79}\,,\Op^{(8)}_{80}\,,\Op^{(8)}_{81}\,,\Op^{(8)}_{82}\,,\Op^{(8)}_{85}\,,\Op^{(8)}_{86}\,,$&$C$-$P$+ in Tab.~\ref{purem1Sjmn}&$C$-$P$+ in Tab.~\ref{mb1Sjmn}\\
    $\Op^{(8)}_{87}\,,\Op^{(8)}_{88}\,,\Op^{(8)}_{91}\,,\Op^{(8)}_{92}\,,\Op^{(8)}_{101}\,,\Op^{(8)}_{102}\,,\Op^{(8)}_{105}\,,\Op^{(8)}_{106}\,.$&&\\
    \hline
    $\mathcal{O}_1^{(6)}\,,\Op^{(6)}_3\,,\mathcal{O}_5^{(6)}\,,\Op^{(6)}_6\,,\Op^{(6)}_{36}\,,\Op^{(6)}_{39}\,,\Op^{(6)}_{48}\,,\Op^{(6)}_{51}\,,\Op^{(7)}_1\,,$&&\\
    $\Op^{(7)}_3\,,\Op^{(7)}_{38}\,,\Op^{(7)}_{43}\,,\Op^{(8)}_3\,,\Op^{(8)}_4\,,\Op^{(8)}_7\,,\Op^{(8)}_8\,,\Op^{(8)}_{9}\,,\Op^{(8)}_{10}\,,$  &$C$+$P$+ in Tab.~\ref{purem1Sj}&$C$+$P$+ in Tab.~\ref{mb1Sj}\\
    $\Op^{(8)}_{13}\,,\Op^{(8)}_{14}\,,\Op^{(8)}_{130}\,,\Op^{(8)}_{133}\,,\Op^{(8)}_{142}\,,\Op^{(8)}_{145}\,,\Op^{(8)}_{154}\,,\Op^{(8)}_{157}\,,\Op^{(8)}_{166}\,,\Op^{(8)}_{169}\,.$&&\\
    \hline
    $\mathcal{O}_7^{(6)}\,,\Op^{(6)}_9\,,\mathcal{O}_{11}^{(6)}\,,\Op^{(6)}_{12}\,,\Op^{(6)}_{60}\,,\Op^{(6)}_{61}\,,\Op^{(6)}_{64}\,,\Op^{(6)}_{65}\,,\Op^{(7)}_5\,,$&&\\
    $\Op^{(7)}_7\,,\Op^{(7)}_{46}\,,\Op^{(7)}_{51}\,,\Op^{(8)}_{39}\,,\Op^{(8)}_{40}\,,\Op^{(8)}_{43}\,,\Op^{(8)}_{44}\,,\Op^{(8)}_{45}\,,\Op^{(8)}_{46}\,,$  &$C$+$P$- in Tab.~\ref{purem1Sj}&$C$+$P$- in Tab.~\ref{mb1Sj}\\
    $\Op^{(8)}_{49}\,,\Op^{(8)}_{50}\,,\Op^{(8)}_{178}\,,\Op^{(8)}_{179}\,,\Op^{(8)}_{182}\,,\Op^{(8)}_{183}\,,\Op^{(8)}_{186}\,,\Op^{(8)}_{187}\,,\Op^{(8)}_{190}\,,\Op^{(8)}_{191}\,.$&&\\
    \hline
    $\mathcal{O}_{54}^{(7)}\,,\mathcal{O}_{59}^{(7)}\,.$  &$C$-$P$+ in Tab.~\ref{purem1Sj}&$C$+$P$- in Tab.~\ref{mb1Sj}\\
    \hline
    $\mathcal{O}_{62}^{(7)}\,,\mathcal{O}_{67}^{(7)}\,.$  &$C$-$P$- in Tab.~\ref{purem1Sj}&$C$+$P$- in Tab.~\ref{mb1Sj}\\
    \hline
    $\Op^{(6)}_{13}\,,\Op^{(6)}_{14}\,,\Op^{(6)}_{15}\,,\Op^{(6)}_{19}\,,\Op^{(6)}_{20}\,,\Op^{(7)}_9\,,\Op^{(7)}_{10}\,,\Op^{(7)}_{13}\,,$&&\\
    $\Op^{(7)}_{14}\,,\Op^{(8)}_{23}\,,\Op^{(8)}_{24}\,,\Op^{(8)}_{25}\,,\Op^{(8)}_{29}\,,\Op^{(8)}_{30}\,,\Op^{(8)}_{31}\,,\Op^{(8)}_{32}\,,$&$C$+$P$- in Tab.~\ref{purem1Qjm}&$C$+$P$- in Tab.~\ref{mb1Qjm}\\
    $\Op^{(8)}_{33}\,,\Op^{(8)}_{37}\,,\Op^{(8)}_{38}\,,\Op^{(8)}_{51}\,,\Op^{(8)}_{52}\,,\Op^{(8)}_{53}\,,\Op^{(8)}_{57}\,,\Op^{(8)}_{58}\,.$&&\\
    \hline
    $\Op^{(6)}_{21}\,,\Op^{(6)}_{22}\,,\Op^{(6)}_{23}\,,\Op^{(6)}_{27}\,,\Op^{(6)}_{28}\,,\Op^{(7)}_{23}\,,\Op^{(7)}_{24}\,,\Op^{(7)}_{27}\,,$&&\\
    $\Op^{(7)}_{28}\,,\Op^{(8)}_{59}\,,\Op^{(8)}_{60}\,,\Op^{(8)}_{61}\,,\Op^{(8)}_{65}\,,\Op^{(8)}_{66}\,,\Op^{(8)}_{67}\,,\Op^{(8)}_{68}\,,$&$C$-$P$+ in Tab.~\ref{purem1Qjm}&$C$-$P$+ in Tab.~\ref{mb1Qjm}\\
    $\Op^{(8)}_{69}\,,\Op^{(8)}_{73}\Op^{(8)}_{74}\,,\Op^{(8)}_{93}\,,\Op^{(8)}_{94}\,,\Op^{(8)}_{95}\,,\Op^{(8)}_{99}\,,\Op^{(8)}_{100}\,.$&&\\
    \hline
    $\Op^{(6)}_{42}\,,\Op^{(6)}_{45}\,,\Op^{(6)}_{54}\,,\Op^{(6)}_{57}\,,\Op^{(7)}_{39}\,,\Op^{(7)}_{44}\,,\Op^{(8)}_{136}\,,$&{\multirow{2}{*}{$C$+$P$+ in Tab.~\ref{purem1Qj}}}&{\multirow{2}{*}{$C$+$P$+ in Tab.~\ref{mb1Qj}}}\\
    $\Op^{(8)}_{139}\,,\Op^{(8)}_{148}\,,\Op^{(8)}_{151}\,,\Op^{(8)}_{160}\,,\Op^{(8)}_{163}\,,\Op^{(8)}_{172}\,,\Op^{(8)}_{175}\,.$&&\\
    \hline
    $\Op^{(6)}_{68}\,,\Op^{(6)}_{69}\,,\Op^{(6)}_{72}\,,\Op^{(6)}_{73}\,,\Op^{(7)}_{63}\,,\Op^{(7)}_{68}\,,\Op^{(8)}_{194}\,,$&{\multirow{2}{*}{$C$-$P$- in Tab.~\ref{purem1Qj}}}&{\multirow{2}{*}{$C$-$P$- in Tab.~\ref{mb1Qj}}}\\ $\Op^{(8)}_{195}\,,\Op^{(8)}_{198}\,,\Op^{(8)}_{199}\,,\Op^{(8)}_{202}\,,\Op^{(8)}_{203}\,,\Op^{(8)}_{206}\,,\Op^{(8)}_{207}\,.$&&\\
    \hline
    $\Op^{(7)}_{47}\,,\Op^{(7)}_{52}\,.$&$C$+$P$- in Tab.~\ref{purem1Qj}&$C$+$P$- in Tab.~\ref{mb1Qj}\\
    \hline
    $\Op^{(7)}_{55}\,,\Op^{(7)}_{60}\,.$&$C$-$P$+ in Tab.~\ref{purem1Qj}&$C$-$P$+ in Tab.~\ref{mb1Qj}\\
    \hline
    $\Op^{(7)}_{15}\,,\Op^{(7)}_{16}\,,\Op^{(7)}_{17}\,,\Op^{(7)}_{21}\,,\Op^{(7)}_{22}\,,\Op^{(8)}_{107}\,,\Op^{(8)}_{108}\,,\Op^{(8)}_{111}\,,\Op^{(8)}_{112}\,.$&$C$-$P$+ in Tab.~\ref{purem1Sjm}&$C$-$P$+ in Tab.~\ref{mb1Sjm}\\
    \hline 
    $\Op^{(7)}_{29}\,,\Op^{(7)}_{30}\,,\Op^{(7)}_{31}\,,\Op^{(7)}_{35}\,,\Op^{(7)}_{36}\,,\Op^{(8)}_{123}\,,\Op^{(8)}_{124}\,,\Op^{(8)}_{127}\,,\Op^{(8)}_{128}\,.$&$C$-$P$- in Tab.~\ref{purem1Sjm}&$C$-$P$- in Tab.~\ref{mb1Sjm}\\
    \hline
    $\Op^{(8)}_{1}\,,\Op^{(8)}_{15}\,,\Op^{(8)}_{16}\,,\Op^{(8)}_{17}\,,\Op^{(8)}_{21}\,,\Op^{(8)}_{22}\,.$&$C$+$P$+ in Tab.~\ref{purem1Qjmn}&$C$+$P$+ in Tab.~\ref{mb1Qjmn}\\
    \hline
    $\Op^{(8)}_{113}\,,\Op^{(8)}_{115}\,,\Op^{(8)}_{116}\,,\Op^{(8)}_{117}\,,\Op^{(8)}_{121}\,,\Op^{(8)}_{122}\,.$&$C$-$P$- in Tab.~\ref{purem1Qjmn}&$C$-$P$- in Tab.~\ref{mb1Qjmn}\\
    \hline
    \caption{The matching for operators with 1 spurion.}
    \label{tab:my_label}
\end{longtable}
\end{center}

\begin{center}
    \begin{longtable}{|c|c|c|c|}
    \hline
       LEFT operators&pure meson&meson-baryon\\
    \hline
    $\Op^{(6)}_{43}\,,\Op^{(6)}_{46}\,,\Op^{(6)}_{55}\,,\Op^{(6)}_{58}\,,\Op^{(8)}_{137}\,,\Op^{(8)}_{140}\,,$&{\multirow{2}{*}{$C$+$P$+ in Tab.~\ref{pm2Q}}}&{\multirow{2}{*}{$C$+$P$+ in Tab.~\ref{twoQ}}}\\
    $\Op^{(8)}_{149}\,,\Op^{(8)}_{152}\,,\Op^{(8)}_{161}\,,\Op^{(8)}_{164}\,,\Op^{(8)}_{173}\,,\Op^{(8)}_{176}\,.$&&\\
       \hline
       $\Op^{(6)}_{70}\,,\Op^{(6)}_{74}\,,\Op^{(8)}_{196}\,,$&{\multirow{2}{*}{$C$-$P$- in Tab.~\ref{pm2Q}}}&{\multirow{2}{*}{$C$-$P$- in Tab.~\ref{twoQ}}}\\
       $\Op^{(8)}_{200}\,,\Op^{(8)}_{204}\,,\Op^{(8)}_{208}\,.$&&\\
       \hline
       $\Op^{(6)}_{37}\,,\Op^{(6)}_{40}\,,\Op^{(6)}_{49}\,,\Op^{(6)}_{52}\,,\Op^{(8)}_{131}\,,\Op^{(8)}_{134}\,,$&{\multirow{2}{*}{$C$+$P$+ in Tab.~\ref{pm2S}}}&{\multirow{2}{*}{$C$+$P$+ in Tab.~\ref{twosigma}}}\\
       $\Op^{(8)}_{143}\,,\Op^{(8)}_{146}\,,\Op^{(8)}_{155}\,,\Op^{(8)}_{158}\,,\Op^{(8)}_{167}\,,\Op^{(8)}_{170}\,.$&&\\
       \hline
       $\Op^{(6)}_{62}\,,\Op^{(6)}_{66}\,,\Op^{(8)}_{180}\,,$&{\multirow{2}{*}{$C$+$P$- in Tab.~\ref{pm2S}}}&{\multirow{2}{*}{$C$+$P$- in Tab.~\ref{twosigma}}}\\
       $\Op^{(8)}_{184}\,,\Op^{(8)}_{188}\,,\Op^{(8)}_{192}\,.$&&\\
       \hline
       $\Op^{(7)}_{40}\,,\Op^{(7)}_{41}\,.$&$C$+$P$+ in Tab.~\ref{pmSQ}&$C$+$P$+ in Tab.~\ref{sigmaQ}\\
       \hline
       $\Op^{(7)}_{48}\,,\Op^{(7)}_{49}\,.$&$C$+$P$- in Tab.~\ref{pmSQ}&$C$+$P$- in Tab.~\ref{sigmaQ}\\
       \hline
       $\Op^{(7)}_{56}\,,\Op^{(7)}_{57}\,.$&$C$-$P$+ in Tab.~\ref{pmSQ}&$C$-$P$+ in Tab.~\ref{sigmaQ}\\
       \hline
       $\Op^{(7)}_{64}\,,\Op^{(7)}_{65}\,.$&$C$-$P$- in Tab.~\ref{pmSQ}&$C$-$P$- in Tab.~\ref{sigmaQ}\\
       \hline
       \caption{The matching for operators with 2 spurion.}
    \label{mm2}
    \end{longtable}
\end{center}

\begin{center}
    \begin{longtable}{|c|c|c|c|}
    \hline
       LEFT operators&pure meson&meson-baryon\\
    \hline
    $\Op_1^{(9)}\,,\Op_3^{(9)}\,,\Op_7^{(9)}\,,\Op_9^{(9)}\,.$&$C$+$P$+ in Tab.~\ref{pm2S} with $(\bar e^c e)$&$C$+$P$+ in Tab.~\ref{twosigma} with $(\bar e^c e)$\\
    \hline
    $\Op_{13}^{(9)}\,,\Op_{15}^{(9)}\,.$&$C$+$P$- in Tab.~\ref{pm2S} with $(\bar e^c e)$&$C$+$P$- in Tab.~\ref{twosigma} with $(\bar e^c e)$\\
    \hline
    $\Op_2^{(9)}\,,\Op_4^{(9)}\,,\Op_8^{(9)}\,,\Op_{10}^{(9)}\,.$&$C$+$P$+ in Tab.~\ref{pm2Q} with $(\bar e^c e)$&$C$+$P$+ in Tab.~\ref{twoQ} with $(\bar e^c e)$\\
    \hline
    $\Op_{19}^{(9)}\,,\Op_{21}^{(9)}\,.$&$C$-$P$- in Tab.~\ref{pm2Q} with $(\bar e^c e)$&$C$-$P$- in Tab.~\ref{twoQ} with $(\bar e^c e)$\\
    \hline
    $\Op_5^{(9)}\,,\Op_6^{(9)}\,,\Op_{11}^{(9)}\,,\Op_{12}^{(9)}\,.$&$C$+$P$+ in Tab.~\ref{md9}&$C$+$P$+ in Tab.~\ref{d9}\\
    \hline
    $\Op_{14}^{(9)}\,,\Op_{16}^{(9)}\,.$&$C$+$P$- in Tab.~\ref{md9}&$C$+$P$+ in Tab.~\ref{d9}\\
    \hline
    $\Op_{17}^{(9)}\,,\Op_{18}^{(9)}\,.$&$C$-$P$+ in Tab.~\ref{md9}&$C$+$P$+ in Tab.~\ref{d9}\\
    \hline
    $\Op_{20}^{(9)}\,,\Op_{22}^{(9)}\,.$&$C$-$P$- in Tab.~\ref{md9}&$C$+$P$+ in Tab.~\ref{d9}\\
       \hline
       \caption{The matching for dimension-9 operators.}
    \label{mm3_1}
    \end{longtable}
\end{center}

\section{Conclusion}
\label{sec:con}


The LEFT and the ChPT describe the quark interactions above and below the scale $\Lambda_{\chi}=1\text{ GeV}$ respectively, 
thus the matching between the operators of these two theories is important. Such matching has been considered for different procedures such as the kaon decay, the $0\nu\beta\beta$, 
and so on, of which the external source method is commonly used. As the precision measurements higher and higher, the higher-dimension effective operators of both the LEFT and the ChPT are demanded. However, the external source method and the old-fashioned spurion method are not convenient to do the matching as order of Lagrangian becomes higher and higher. On the one hand, the higher-dimension LEFT operators could contain complicated lepton and quark bilinears, which means the external sources introduced in Eq.~\eqref{eq:external_sources_1} or the spurions under the $SU(3)_\mathbf{L} \times SU(3)_\mathbf{R}$ symmetry in Eq.~\eqref{eq:spurion_lr_method} are not adequate. Besides, the old-fashioned spurion method parametrizes the LEFT Lagrangian by the $SU(3)_\mathbf{L}\times SU(3)_\mathbf{R}$ symmetry, and maps the quarks to the hadronic degrees of freedom one by one as shown in Eq.~\eqref{eq:correspondence_1}, which is complicated for the matching beyond the LO.

In this paper, we propose a systematic matching of any orders of LEFT operators to the ChPT operators using a single $SU(3)_V$ spurions. We present an new formulation of the LEFT operators via only one spurion covariant under the $SU(3)_V$,
\begin{equation}
    \mathbf{T}\rightarrow V\mathbf{T}V^\dagger \,,\quad V\in SU(3)_V\,.
\end{equation}
Similarly, we construct the ChPT operators for the $SU(3)_V$ symmetry using $\Sigma_\pm$ and $Q_\pm$, which is related to the quark level spurion by the following dressing
\begin{align}
    \Sigma_{\pm} &= u^\dagger \textbf{T} u^\dagger \pm u \textbf{T}^\dagger u \,,\notag\\
    Q_{\pm}&=u^\dagger\textbf{T}u\pm u\textbf{T}^\dagger u^\dagger\,.
\end{align}
At the same time, the leptons and the photon are regarded as building blocks of the LEFT and the ChPT.

The matching between the LEFT and the ChPT utilizes the properties kept in both the quark level and the hadronic level including the spurion structures, the CP properties, and the non-quark fields. In the spurion method, any LEFT operators have ChPT counterparts in principle. In particular, the redundancies such as the EOM and the IBP do not affect the matching results. To assess the importance of the operators after the matching we use the NDA to normalize them. 
Such an assessment is consistent with the power-counting of both the LEFT and the ChPT.
This spurion method is systematic and convenient when matching the high-dimension operators of both LEFT and ChPT.

Using the spurion matching method described above, we present the matching results of the LEFT operators of dimension 5, dimension 6, dimension 7, and part of dimension 8 and dimension 9. For the dimension 8 LEFT operators, we exclude the operators with two quark bilinears and one photon field, since the matching of them is similar with the pure quark operators. 
For the dimension 9 LEFT operators, we only consider the ones with two quark bilinears and one lepton bilinear, which can be matched to the ChPT operators about the $0\nu\beta\beta$.
Because every LEFT operator can be matched to more than one ChPT operator, we only list the LO and NLO ones according to the NDA.

Finally, The matching procedure presented here for the LEFT and ChPT is quite general. Although we only list the matching operators in the pure meson sector and the meson-baryon sector, this spurion method is general and can be used in other sectors such as the 4-baryon sector. This matching procedure can also be applied to other theories, such as the dark matter direct detection, the axion searches, etc. These low energy processes probes high energy scale physics, which involve in several different scales.



\section*{Acknowledgments}

We would like to thank Gang Li, Xiao-Dong Ma, and Yi Liao for their valuable discussions and comments. This work is supported by the National Science Foundation of China under Grants No. 12347105, No. 12375099 and No. 12047503, and the National Key Research and Development Program of China Grant No. 2020YFC2201501, No. 2021YFA0718304.

\appendix
\section{Comparison to the $SU(3)_\mathbf{L}\times SU(3)_\mathbf{R}$ Parameterization }
\label{app:comparison}

With $SU(3)_V$ spurions in Eq.~\eqref{spurion1}, the relevant higher-dimension operators of the LEFT have been constructed in Sec.~\ref{sec:LEFT}. In this appendix, we present some examples to compare the operators here to the ones parameterized by the $SU(3)_\mathbf{L}\times SU(3)_\mathbf{R}$ symmetry and illustrate that these two different symmetries are equivalent. We repeat here the equivalence means the independent coefficients are the same after the spurions get their VEVs.

Firstly, consider the dimension-5 operators, all the operators containing quarks are composed of a quark bilinear and a field-strength tensor of photon field. Due to charge conservation, there are 10 operators,
\begin{align}
    \mathcal{L}^{(5)}_{LEFT} \supset \frac{1}{\Lambda_{\text{EW}}}\sum_{a=1}^{5} c_a^{(5)}\mathcal{O}_a^{(5)} + h.c.\,, \label{eq:left_lagrangian_3}
\end{align}
where $c_a^{(5)}$ are dimensionless coefficients and the five operators are
\begin{align}
    \mathcal{O}^{(5)}_1 = F^{\mu\nu}(\overline{u}_\mathbf{L}\sigma_{\mu\nu}u_\mathbf{R})\,, \\
    \mathcal{O}^{(5)}_2 = F^{\mu\nu}(\overline{d}_\mathbf{L}\sigma_{\mu\nu}d_\mathbf{R})\,, \\
    \mathcal{O}^{(5)}_3 = F^{\mu\nu}(\overline{s}_\mathbf{L}\sigma_{\mu\nu}s_\mathbf{R})\,, \\
    \mathcal{O}^{(5)}_4 = F^{\mu\nu}(\overline{d}_\mathbf{L}\sigma_{\mu\nu}s_\mathbf{R})\,, \\
    \mathcal{O}^{(5)}_5 = F^{\mu\nu}(\overline{s}_\mathbf{L}\sigma_{\mu\nu}d_\mathbf{R})\,.
\end{align}
Firstly we organize the Lagrangian in terms of $q_\mathbf{L}$ and $q_\mathbf{R}$ as 
\begin{equation}
\label{eq:lr_d5}
    \frac{1}{\Lambda_{\text{EW}}}F^{\mu\nu}(\overline{q}_\mathbf{L}\mathbf{C}\sigma_{\mu\nu}q_\mathbf{R}) + h.c.\,,
\end{equation}
where $\mathbf{C}$ is a matrix composed by the Wilson coefficients
\begin{equation}
\label{eq:vev_c}
    \mathbf{C} = \left(\begin{array}{ccc}
c^{(5)}_1 & 0 & 0 \\
0 & c^{(5)}_2 & c^{(5)}_5 \\
0 & c^{(5)}_4 & c^{(5)}_3 
    \end{array}\right)\,.
\end{equation}
The Lagrangian is invariant under $SU(3)_\mathbf{L}\times SU(3)_\mathbf{R}$ if $\mathbf{C}$ is covariant $\mathbf{C}\rightarrow L\mathbf{C}R^\dagger$. 
Furthermore we combine $q_\mathbf{L}$ and $q_\mathbf{R}$ to form $q$ utilizing parity symmetry,
\begin{equation}
    q = (u,d,s)^T = \mathcal{R} \cdot (q_\mathbf{L}\oplus q_\mathbf{R})\,,
\end{equation}
the Lagrangian becomes 
\begin{equation}
    \mathcal{L}^{(5)}_{LEFT} \supset \frac{1}{\Lambda_{\text{EW}}}\left(F^{\mu\nu}\overline{q} \mathbf{C}_1 \sigma_{\mu\nu} q + \tilde{F}^{\mu\nu}\overline{q} \mathbf{C}_2 \sigma_{\mu\nu} q\right)\,,
\end{equation}
where the first operator is of positive parity and the second on is of negative parity. The coefficients are
\begin{align}
    \mathbf{C}_1 &= \mathcal{R}\,2\text{Re}\mathbf{C}\,\mathcal{R}^{-1} = \frac{\text{Tr}\mathbf{C}_1}{3}\mathbf{I} + \mathbf{t}_1\,, \\
    \mathbf{C}_2 &= \mathcal{R}\, 2i\text{Im}\mathbf{C}\,\mathcal{R}^{-1} = \frac{\text{Tr}\mathbf{C}_2}{3}\mathbf{I} + \mathbf{t}_2 \,,
\end{align}
where the matrices $\mathbf{t}_1$ and $\mathbf{t}_2$ can be prompted to spurion field $\mathbf{T}$. 
Thus we obtain the final expression of the Lagrangian
\begin{equation}
\label{eq:left_lagrangian_4}
    \mathcal{L}^{(5)}_{LEFT} \supset \frac{1}{\Lambda_{\text{EW}}}\left[ \frac{\text{Tr}\mathbf{C}_1}{3} F^{\mu\nu}(\overline{q}\sigma_{\mu\nu}q) + \frac{\text{Tr}\mathbf{C}_2}{3} \tilde{F}^{\mu\nu}(\overline{q}\sigma_{\mu\nu}q) + a_1F^{\mu\nu}(\overline{q}\mathbf{T}\sigma_{\mu\nu}q) + a_2\tilde{F}^{\mu\nu}(\overline{q}\mathbf{T}\sigma_{\mu\nu}q)\right]\,,
\end{equation}
where $a_1\,,a_2$ are two independent Wilson coefficients.
This expression is equivalent to Eq.~\eqref{eq:left_lagrangian_3} since the numbers of the independent operators are the same once the spurion $\mathbf{T}$ gets VEV. The Lagrangian of $SU(3)_\textbf{L}\times SU(3)_\textbf{R}$ symmetry in Eq.~\eqref{eq:lr_d5} looks different though, the expression of  $\mathbf{C}$ in Eq.~\eqref{eq:vev_c} implies that the number of independent operators is the same. 

Such an argument illustrates the correspondence of the LEFT operators between the $SU(3)_\mathbf{L}\times SU(3)_\mathbf{R}$ symmetry and the $SU(3)_V$ symmetry,
\begin{align}
 (\overline{\mathbf{3}}_\mathbf{L},\mathbf{3}_\mathbf{R}) \oplus (\mathbf{3}_R,\overline{\mathbf{3}}_L) &\rightarrow \mathbf{1}^+ \oplus \mathbf{1}^-\oplus \mathbf{8}^+ \oplus \mathbf{8}^-\,,\notag \\
   A_i^j (\bar q_\mathbf{L}^i\Gamma q_{\mathbf{R}j})+ B_i^j(\bar q_\mathbf{R}^i\Gamma q_{\mathbf{L}j})
   &\sim c_0(\bar q\Gamma q)+c_0^\prime(\bar q\Gamma\gamma^5 q)+ c_a(\bar q\Gamma\mathbf{t}^a q)+ c_a^\prime(\bar q\Gamma\gamma^5\mathbf{t}^a q)\,,
   \label{tranc}
\end{align}
in which $A, B$ are coefficients in the $SU(3)_\mathbf{L}\times SU(3)_\mathbf{R}$ symmetry, $\Gamma$ is a Dirac matrix, the superscripts on the right are the parity eigenvalues corresponding to the $SU(3)_V$ irreducible representations and $a=1,2,...,8$. In addition, the translation of the coefficients for these two representations become
\begin{align}
    c_0\sim&A^1_1+A^2_2+A^3_3+B^1_1+B^2_2+B^3_3\,,\quad c_0^\prime\sim A^1_1+A^2_2+A^3_3-B^1_1+B^2_2+B^3_3\,,\notag\\
    c_1\sim &A^1_2+B^1_2\,,\quad c_1^\prime\sim A^1_2-B^1_2\,,\quad c_2\sim A^2_1+B^2_1\,,\quad c_2^\prime\sim A^2_1-B^2_1\,,\notag\\
    c_3\sim &A^1_3+B^1_3\,,\quad c_3^\prime\sim A^1_3-B^1_3\,,\quad c_4\sim A^3_1+B^3_1\,,\quad c_4^\prime\sim A^3_1-B^3_1\,,\notag\\
    c_5\sim &A^2_3+B^2_3\,,\quad c_5^\prime\sim A^2_3-B^2_3\,,\quad c_6\sim A^3_2+B^3_2\,,\quad c_6^\prime\sim A^3_2-B^3_2\,,\notag\\
    c_7\sim &A^1_1-A^2_2+B^1_1-B^2_2\,,\quad c_7^\prime\sim A^1_1-A^2_2-B^1_1+B^2_2\,,\notag\\
    c_8\sim &\frac{1}{\sqrt{3}}(A^1_1+A^2_2-2A^3_3+B^1_1+B^2_2-2B^3_3)\,,\quad c_8^\prime\sim \frac{1}{\sqrt{3}}(A^1_1+A^2_2-2A^3_3-B^1_1-B^2_2+2B^3_3)\,,
\label{coff}
\end{align}
where $\sim$ means the similar transformation in Eq.~\eqref{eq:R_tras} is used.
Similarly, we can get the correspondences of the other representations, for example
\begin{align}
  (\overline{\mathbf{3}}_\mathbf{L},\mathbf{3}_\mathbf{L}) \oplus (\mathbf{3}_\mathbf{R},\overline{\mathbf{3}}_\mathbf{R}) & \rightarrow \mathbf{1}^+ + \mathbf{1}^- + \mathbf{8}^+ + \mathbf{8}^-: \notag \\
  (\overline{q}_\mathbf{L}A\Gamma q_\mathbf{L}) + (\overline{q}_\mathbf{R}B \Gamma q_\mathbf{R}) &=c_0(\bar q\Gamma q)+c_0'(\bar q\Gamma\gamma^5 q)+c_a(\bar q\Gamma\mathbf{t}^a q)+c_a'(\bar q\Gamma\gamma^5\mathbf{t}^a q)\,,
\end{align}
where the translation of the coefficients is the same as Eq.~\eqref{tranc}. 


Considering the dimension-6 Lagrangian and beyond, the relevant operators can contain more than one quark bilinear. For the operators with single quark bilinears, the reformulations are similar, for example, the dimension-6 operators
\begin{align}
    \mathcal{O}^{(6)}_1 = (\nu_L^T C\nu_L)(\overline{u}_\mathbf Lu_\mathbf R)\,, \\
    \mathcal{O}^{(6)}_2 = (\nu_L^T C\nu_L)(\overline{d}_\mathbf Ld_\mathbf R)\,, \\
    \mathcal{O}^{(6)}_3 = (\nu_L^T C\nu_L)(\overline{s}_\mathbf Ls_\mathbf R)\,, \\
    \mathcal{O}^{(6)}_4 = (\nu_L^T C\nu_L)(\overline{d}_\mathbf Ls_\mathbf R)\,, \\
    \mathcal{O}^{(6)}_5 = (\nu_L^T C\nu_L)(\overline{s}_\mathbf Ld_\mathbf R)\,,
\end{align}
can be reformulated as 
\begin{equation}
    \mathcal{L}^{(6)}_{LEFT} \supset \frac{1}{\Lambda_{\text{EW}}^2}\left[a_1(\nu_L^T C\nu_L)(\overline{q}q)+b_1(\nu_L^T C\nu_L)(\overline{q}\mathbf{T}q) + a_2(\nu_L^T C\nu_L)(\overline{q}\gamma^5q)+b_2(\nu_L^T C\nu_L)(\overline{q}\mathbf{T}\gamma^5q)\right]\,.
\end{equation}
However, the operators with 2 quark bilinears such as the 4-quark operators of the dimension-6 Lagrangian can be complicated. For example, an operator is composed of two quark bilinears
\begin{equation}
    A^{jl}_{ik}(\overline{q}_\mathbf{L}^i\Gamma_1 q_{\mathbf{R}j}) (\overline{q}_\mathbf{L}^k\Gamma_2 q_{\mathbf{R}l}) \oplus B^{jl}_{ik}(\overline{q}_\mathbf{R}^i\Gamma_1 q_{\mathbf{L}j}) (\overline{q}_\mathbf{L}^k\Gamma_2 q_{\mathbf{R}l}) \oplus C^{jl}_{ik}(\overline{q}_\mathbf{L}^i\Gamma_1 q_{\mathbf{R}j}) (\overline{q}_\mathbf{R}^k\Gamma_2 q_{\mathbf{L}l}) \oplus D^{jl}_{ik}(\overline{q}_\mathbf{R}^i\Gamma_1 q_{\mathbf{L}j}) (\overline{q}_\mathbf{R}^k\Gamma_2 q_{\mathbf{L}l})\,,
\end{equation}
the correspondence can be obtained by the tensor product that
\begin{align}
    & \left[(\overline{\mathbf{3}}_\mathbf{L},\mathbf{3}_\mathbf{R}) \oplus (\mathbf{3}_\mathbf{L},\overline{\mathbf{3}}_\mathbf{R})\right]\otimes \left[(\overline{\mathbf{3}}_\mathbf{L},\mathbf{3}_\mathbf{R}) \oplus (\mathbf{3}_\mathbf{L},\overline{\mathbf{3}}_\mathbf{R})\right] \notag \\
    &\rightarrow \left(\mathbf{1}^+ + \mathbf{1}^- + \mathbf{8}^+ + \mathbf{8}^-\right) \otimes \left(\mathbf{1}^+ + \mathbf{1}^- + \mathbf{8}^+ + \mathbf{8}^-\right) \notag \\
    &= 2\times \mathbf{1}^+ + 2\times \mathbf{1}^- + 4\times\mathbf{8}^+ + 4\times\mathbf{8}^- + 2\times(\mathbf{8}\times \mathbf{8})^+ + 2\times (\mathbf{8}\times \mathbf{8})^- \,,
\end{align}
where on the right the integers before the irreducible representations are their multiplicities. 
The corresponding operators invariant under $SU(3)_V$ are 
\begin{align}
    &C_0^{(1)}(\overline{q}\Gamma_1 q)(\overline{q}\Gamma_2 q) + C_0^{(2)}(\overline{q}\Gamma_1i\gamma^5q)(\overline{q}\Gamma_2i\gamma^5q) + C_0^{(3)}(\overline{q}\Gamma_1q)(\overline{q}\Gamma_2i\gamma^5q) + C_0^{(4)}(\overline{q}\Gamma_1i\gamma^5q)(\overline{q}\Gamma_2q) \notag \\
    +& C_a^{(1)}(\overline{q}\mathbf{t}_a\Gamma_1q)(\overline{q}\Gamma_2q) + C_a^{(2)}(\overline{q}\Gamma_1q)(\overline{q}\mathbf{t}_a\Gamma_2q) + C_a^{(3)}(\overline{q}i\gamma^5\mathbf{t}_a\Gamma_1q)(\overline{q}i\gamma^5\Gamma_2q) + C_a^{(4)}(\overline{q}i\gamma^5\Gamma_1q)(\overline{q}i\gamma^5\mathbf{t}_a\Gamma_2q) \notag \\
    +& C_a^{(5)}(\overline{q}\mathbf{t}_ai\gamma^5\Gamma_1q)(\overline{q}\Gamma_2q) + C_a^{(6)}(\overline{q}i\gamma^5\Gamma_1q)(\overline{q}\mathbf{t}_a\Gamma_2q) + C_a^{(7)}(\overline{q}\mathbf{t}_a\Gamma_1q)(\overline{q}i\gamma^5\Gamma_2q) + C_a^{(8)}(\overline{q}\Gamma_1q)(\overline{q}\mathbf{t}_ai\gamma^5\Gamma_2q) \notag \\
    +& C_{ab}^{(1)}(\overline{q}\mathbf{t}_a\Gamma_1q)(\overline{q}\mathbf{t}_b\Gamma_2q) + C_{ab}^{(2)}(\overline{q}\mathbf{t}_ai\gamma^5\Gamma_1q)(\overline{q}\mathbf{t}_bi\gamma^5\Gamma_2q) + C_{ab}^{(3)}(\overline{q}\mathbf{t}_ai\gamma^5\Gamma_1q)(\overline{q}\mathbf{t}_b\Gamma_2q) + C_{ab}^{(4)}(\overline{q}\mathbf{t}_a\Gamma_1q)(\overline{q}\mathbf{t}_bi\gamma^5\Gamma_2q)\,,
\end{align}
and the translation of the two representations' coefficients become (repeated indices mean summations)
\begin{align}
    C_0^{(1)}=&A^{ik}_{ik}+B^{ik}_{ik}+C^{ik}_{ik}+D^{ik}_{ik}\,,\notag\\
    C_0^{(2)}=&A^{ik}_{ik}-B^{ik}_{ik}-C^{ik}_{ik}+D^{ik}_{ik}\,,\notag\\
    C_0^{(3)}=&A^{ik}_{ik}+B^{ik}_{ik}-C^{ik}_{ik}-D^{ik}_{ik}\,,\notag\\
    C_0^{(4)}=&A^{ik}_{ik}-B^{ik}_{ik}-C^{ik}_{ik}-D^{ik}_{ik}\,,\notag\\
    C_a^{(1)}=&(A^{ik}_{il}+B^{ik}_{il}+C^{ik}_{il}+D^{ik}_{il})\mathbf{t}^a{}_k^l\,,\notag\\
    C_a^{(2)}=&(A^{ik}_{il}-B^{ik}_{il}-C^{ik}_{il}+D^{ik}_{il})\mathbf{t}^a{}_k^l\,,\notag\\
    C_a^{(3)}=&(A^{ik}_{jk}+B^{ik}_{jk}+C^{ik}_{jk}+D^{ik}_{jk})\mathbf{t}^a{}_k^l\,,\notag\\
    C_a^{(4)}=&(A^{ik}_{jk}-B^{ik}_{jk}-C^{ik}_{jk}+D^{ik}_{jk})\mathbf{t}^a{}_k^l\,,\notag\\
    C_a^{(5)}=&(A^{ik}_{il}+B^{ik}_{il}-C^{ik}_{il}-D^{ik}_{il})\mathbf{t}^a{}_k^l\,,\notag\\
    C_a^{(6)}=&(A^{ik}_{il}-B^{ik}_{il}+C^{ik}_{il}-D^{ik}_{il})\mathbf{t}^a{}_k^l\,,\notag\\
    C_a^{(7)}=&(A^{ik}_{jk}+B^{ik}_{jk}-C^{ik}_{jk}-D^{ik}_{jk})\mathbf{t}^a{}_k^l\,,\notag\\
    C_a^{(8)}=&(A^{ik}_{jk}-B^{ik}_{jk}+C^{ik}_{jk}-D^{ik}_{jk})\mathbf{t}^a{}_k^l\,,\notag\\
    C_{ab}^{(1)}=&(A^{ik}_{jl}+B^{ik}_{jl}+C^{ik}_{jl}+D^{ik}_{jl})\mathbf{t}^a{}_i^j\mathbf{t}^b{}_k^l\,,\notag\\
    C_{ab}^{(2)}=&(A^{ik}_{jl}-B^{ik}_{jl}-C^{ik}_{jl}+D^{ik}_{jl})\mathbf{t}^a{}_i^j\mathbf{t}^b{}_k^l\,,\notag\\
    C_{ab}^{(3)}=&(A^{ik}_{jl}+B^{ik}_{jl}-C^{ik}_{jl}-D^{ik}_{jl})\mathbf{t}^a{}_i^j\mathbf{t}^b{}_k^l\,,\notag\\
    C_{ab}^{(4)}=&(A^{ik}_{jl}-B^{ik}_{jl}+C^{ik}_{jl}-D^{ik}_{jl})\mathbf{t}^a{}_i^j\mathbf{t}^b{}_k^l\,,
    \label{expan}
\end{align}
in which the index $\{i, j, k, l=1, 2, 3\}$ and $\{a=1, 2...,8\}$. 

Up to dimension-8 and the dimension-9 operators with 1 lepton current, all the $SU(3)_\mathbf{L}\times SU(3)_\mathbf{R}$ irreducible representations of the relevant operators are listed in Tab.~\ref{tab:re1}. All these operators can be reformulated similarly. 

Lastly we consider a specific example of two quark bilinears and show that the independent operators with spurions are just the same as the LEFT operators after the VEVs has been taken. Consider the dimension-9 operators $\mathcal{O}^{(9)}_1\,,\mathcal{O}^{(9)}_3$ and $\mathcal{O}^{(9)}_{13}$, we present them here for convenience,
\begin{align}
\Op^{(9)}_1&=(\bar q\textbf{T} q)(\bar q\textbf{T} q)(\bar e e^c)+h.c.\,, \\
\Op^{(9)}_3&=(\bar q\gamma^5\textbf{T} q)(\bar q\gamma^5\textbf{T} q)(\bar e e^c)+h.c.\,, \\
\Op^{(9)}_{13}&=(\bar q\gamma^5\textbf{T}q)(\bar q\textbf{T} q)(\bar e e^c)+h.c.\,.    
\end{align}
Because the lepton current is of electric charge $+2$, the VEV of $\mathbf{T}$ is expanded by the matrices $\mathbf{t}^1\,,\mathbf{t}^3$ according to Eq.~\eqref{eq:t_1t_3}. Considering the symmetry of the two quark bilinears, each operator above splits into several independent operators, 
\begin{equation}
    \mathcal{O}^{(9)}_1 \rightarrow \left\{\begin{array}{l}
(\overline{u}d)(\overline{u}d)(\overline{e}e^c) + h.c. \\
(\overline{u}s)(\overline{u}s)(\overline{e}e^c) + h.c. \\
(\overline{u}s)(\overline{u}d)(\overline{e}e^c) + h.c. \\
    \end{array}\right.\,,\quad 
    \mathcal{O}^{(9)}_3 \rightarrow \left\{\begin{array}{l}
(\overline{u}\gamma^5d)(\overline{u}\gamma^5d)(\overline{e}e^c) + h.c. \\
(\overline{u}\gamma^5s)(\overline{u}\gamma^5s)(\overline{e}e^c) + h.c. \\
(\overline{u}\gamma^5s)(\overline{u}\gamma^5d)(\overline{e}e^c) + h.c. \\
    \end{array}\right.\,,\quad 
    \mathcal{O}^{(9)}_{13} \rightarrow \left\{\begin{array}{l}
(\overline{u}\gamma^5d)(\overline{u}d)(\overline{e}e^c) + h.c. \\
(\overline{u}\gamma^5s)(\overline{u}s)(\overline{e}e^c) + h.c. \\
(\overline{u}\gamma^5s)(\overline{u}d)(\overline{e}e^c) + h.c. \\
(\overline{u}\gamma^5d)(\overline{u}s)(\overline{e}e^c) + h.c. \\
    \end{array}\right.\,.
\end{equation}
These operators can be combine to be the conventional form composed of Weyl spinors,
\begin{align}
    (\overline{u}d)(\overline{u}d)(\overline{e}e^c)\,,(\overline{u}\gamma^5d)(\overline{u}\gamma^5d)(\overline{e}e^c)\,,(\overline{u}\gamma^5d)(\overline{u}d)(\overline{e}e^c) &\rightarrow \left\{\begin{array}{l}
(\overline{u}_L d_R)(\overline{u}_L d_R)(\overline{e}^c e) \\
(\overline{u}_R d_L)(\overline{u}_R d_L)(\overline{e}^c e) \\
(\overline{u}_R d_L)(\overline{u}_L d_R)(\overline{e}^c e) \\
    \end{array}\right. \,,\\
(\overline{u}s)(\overline{u}s)(\overline{e}e^c)\,,(\overline{u}\gamma^5s)(\overline{u}\gamma^5s)(\overline{e}e^c)\,, (\overline{u}\gamma^5s)(\overline{u}s)(\overline{e}e^c) &\rightarrow \left\{\begin{array}{l}
(\overline{u}_L s_R)(\overline{u}_L s_R)(\overline{e}^c e) \\
(\overline{u}_R s_L)(\overline{u}_R s_L)(\overline{e}^c e) \\
(\overline{u}_R s_L)(\overline{u}_L s_R)(\overline{e}^c e) \\
    \end{array}\right. \,,\\ 
(\overline{u}s)(\overline{u}d)(\overline{e}e^c)\,, (\overline{u}\gamma^5s)(\overline{u}\gamma^5d)(\overline{e}e^c) \,,(\overline{u}\gamma^5s)(\overline{u}d)(\overline{e}e^c) \,,(\overline{u}\gamma^5d)(\overline{u}s)(\overline{e}e^c) &\rightarrow \left\{\begin{array}{l}
(\overline{u}_L s_R)(\overline{u}_L d_R)(\overline{e}^c e) \\
(\overline{u}_R s_L)(\overline{u}_R d_L)(\overline{e}^c e) \\
(\overline{u}_R s_L)(\overline{u}_L d_R)(\overline{e}^c e) \\
(\overline{u}_L s_R)(\overline{u}_L d_R)(\overline{e}^c e) \\
    \end{array}\right.\,.
\end{align}
Thus we obtain the independent LEFT operators of this type, where the last four operators are about kaon decay and has been discussed in literature such as Ref.~\cite{Liao:2019gex}. 
This example offers one explicit evidence that the two different symmetries are equivalent. 

\begin{table}
    \centering
    \begin{tabular}{|c|c|c|c|}
    \hline
         $SU(3)_\mathbf{L}\times SU(3)_\mathbf{R}$&quark sector &$SU(3)_\mathbf{L}\times SU(3)_\mathbf{R}$&quark sector \\
         \hline
         & $(\bar q_\mathbf{R}\,q_\mathbf{L})$& & $(\bar q_\mathbf{L}\,q_\mathbf{R})$\\
        $(\mathbf{3},\Bar{\mathbf{3}})$ & $(\bar q_\mathbf{R}\lrpartial^\mu q_\mathbf{L})$&$(\Bar{\mathbf{3}},\mathbf{3})$ & $(\bar q_\mathbf{L}\lrpartial^{\mu}q_\mathbf{R})$\\
         & $(\bar q_\mathbf{R}\sigma^{\mu\nu}q_\mathbf{L})$& & $(\bar q_\mathbf{L}\sigma^{\mu\nu} q_\mathbf{R})$\\
        \hline
        $(\mathbf{1},\mathbf{8})\oplus(\mathbf{1},\mathbf{1})$ & $(\bar q_\mathbf{R}\gamma^\mu q_\mathbf{R})$&$(\mathbf{8},\mathbf{1})\oplus(\mathbf{1},\mathbf{1})$ & $(\bar q_\mathbf{L}\gamma^\mu q_\mathbf{L})$\\
        \hline
        \multicolumn{2}{|c|}{$SU(3)_\mathbf{L}\times SU(3)_\mathbf{R}$}&\multicolumn{2}{|c|}{quark sector}\\
        \hline
        \multicolumn{2}{|c|}{\multirow{2}{*}{$(\mathbf{6},\Bar{\mathbf{6}})\oplus(\mathbf{6},\mathbf{3})\oplus(\Bar{\mathbf{3}},\Bar{\mathbf{6}})\oplus(\Bar{\mathbf{3}},\mathbf{3})$}}&\multicolumn{2}{|c|}{$(\bar q_\mathbf{R}\,q_\mathbf{L})(\bar q_\mathbf{R}\,q_\mathbf{L})$}\\
        \multicolumn{2}{|c|}{}&\multicolumn{2}{|c|}{$(\bar q_\mathbf{R}\sigma^{\mu\nu}q_\mathbf{L})(\bar q_\mathbf{R}\sigma_{\mu\nu}q_\mathbf{L})$}\\
        \hline
        \multicolumn{2}{|c|}{\multirow{2}{*}{$(\Bar{\mathbf{6}},\mathbf{6})\oplus(\mathbf{3},\mathbf{6})\oplus(\Bar{\mathbf{6}},\Bar{\mathbf{3}})\oplus(\mathbf{3},\Bar{\mathbf{3}})$}}&\multicolumn{2}{|c|}{$(\bar q_\mathbf{L}\,q_\mathbf{R})(\bar q_\mathbf{L}\,q_\mathbf{R})$}\\
        \multicolumn{2}{|c|}{}&\multicolumn{2}{|c|}{$(\bar q_\mathbf{L}\sigma^{\mu\nu}q_\mathbf{R})(\bar q_\mathbf{L}\sigma_{\mu\nu}q_\mathbf{R})$}\\
        \hline
        \multicolumn{2}{|c|}{\multirow{2}{*}{$(\mathbf{27},\mathbf{1})\oplus(\mathbf{10},\mathbf{1})\oplus(\Bar{\mathbf{10}},\mathbf{1})\oplus4\times(\mathbf{8},\mathbf{1})\oplus2\times(\mathbf{1},\mathbf{1})$}}&\multicolumn{2}{|c|}{$(\bar q_\mathbf{L}\gamma^\mu q_\mathbf{L})(\bar q_\mathbf{L}\gamma_\mu q_\mathbf{L})$}\\
        \multicolumn{2}{|c|}{}&\multicolumn{2}{|c|}{$(\bar q_\mathbf{L}\gamma^\mu\lrpartial^\nu q_\mathbf{L})(\bar q_\mathbf{L}\gamma_\mu\lrpartial_\nu q_\mathbf{L})$}\\
        \hline
        \multicolumn{2}{|c|}{\multirow{2}{*}{$(\mathbf{1},\mathbf{27})\oplus(\mathbf{1},\mathbf{10})\oplus(\mathbf{1},\Bar{\mathbf{10}})\oplus4\times(\mathbf{1},\mathbf{8})\oplus2\times(\mathbf{1},\mathbf{1})$}}&\multicolumn{2}{|c|}{$(\bar q_\mathbf{R}\gamma^\mu q_\mathbf{R})(\bar q_\mathbf{R}\gamma_\mu q_\mathbf{R})$}\\
        \multicolumn{2}{|c|}{}&\multicolumn{2}{|c|}{$(\bar q_\mathbf{R}\gamma^\mu\lrpartial^\nu q_\mathbf{R})(\bar q_\mathbf{R}\gamma_\mu\lrpartial_\nu q_\mathbf{R})$}\\
        \hline
        \multicolumn{2}{|c|}{\multirow{2}{*}{$(\mathbf{8},\mathbf{8})\oplus(\mathbf{1},\mathbf{8})\oplus(\mathbf{8},\mathbf{1})\oplus(\mathbf{1},\mathbf{1})$}}&\multicolumn{2}{|c|}{$(\bar q_\mathbf{L}\gamma^\mu q_\mathbf{L})(\bar q_\mathbf{R}\gamma_\mu q_\mathbf{R})$}\\
        \multicolumn{2}{|c|}{}&\multicolumn{2}{|c|}{$(\bar q_\mathbf{R}\, q_\mathbf{L})(\bar q_\mathbf{L}\, q_\mathbf{R})$}\\
        \hline
        \multicolumn{2}{|c|}{\multirow{2}{*}{$(\mathbf{15},\Bar{\mathbf{3}})\oplus(\Bar{\mathbf{6}},\Bar{\mathbf{3}})\oplus(\Bar{\mathbf{3}},\Bar{\mathbf{3}})\oplus(\mathbf{3},\Bar{\mathbf{3}})$}}&\multicolumn{2}{|c|}{$(\bar q_\mathbf{L}\gamma^\mu q_\mathbf{L})(\bar q_\mathbf{R} q_\mathbf{L})$}\\
        \multicolumn{2}{|c|}{}&\multicolumn{2}{|c|}{$(\bar q_\mathbf{L}\gamma^\mu q_\mathbf{L})(\bar q_\mathbf{R}\lrpartial_\mu q_\mathbf{L})$}\\
        \hline
        \multicolumn{2}{|c|}{\multirow{2}{*}{$(\Bar{\mathbf{15}},\mathbf{3})\oplus(\mathbf{6},\mathbf{3})\oplus(\Bar{\mathbf{3}},\mathbf{3})\oplus(\mathbf{3},\mathbf{3})$}}&\multicolumn{2}{|c|}{$(\bar q_\mathbf{L}\gamma^\mu q_\mathbf{L})(\bar q_\mathbf{L} q_\mathbf{R})$}\\
        \multicolumn{2}{|c|}{}&\multicolumn{2}{|c|}{$(\bar q_\mathbf{L}\gamma^\mu q_\mathbf{L})(\bar q_\mathbf{L}\lrpartial_\mu q_\mathbf{R})$}\\
        \hline
        \multicolumn{2}{|c|}{\multirow{2}{*}{$(\mathbf{3},\Bar{\mathbf{15}})\oplus(\mathbf{3},\mathbf{6})\oplus(\mathbf{3},\Bar{\mathbf{3}})\oplus(\mathbf{3},\mathbf{3})$}}&\multicolumn{2}{|c|}{$(\bar q_\mathbf{R}\gamma^\mu q_\mathbf{R})(\bar q_\mathbf{R} q_\mathbf{L})$}\\
        \multicolumn{2}{|c|}{}&\multicolumn{2}{|c|}{$(\bar q_\mathbf{R}\gamma^\mu q_\mathbf{R})(\bar q_\mathbf{R}\lrpartial_\mu q_\mathbf{L})$}\\
        \hline
        \multicolumn{2}{|c|}{\multirow{2}{*}{$(\Bar{\mathbf{3}},\mathbf{15})\oplus(\Bar{\mathbf{3}},\Bar{\mathbf{6}})\oplus(\Bar{\mathbf{3}},\Bar{\mathbf{3}})\oplus(\Bar{\mathbf{3}},\mathbf{3})$}}&\multicolumn{2}{|c|}{$(\bar q_\mathbf{R}\gamma^\mu q_\mathbf{R})(\bar q_\mathbf{L} q_\mathbf{R})$}\\
        \multicolumn{2}{|c|}{}&\multicolumn{2}{|c|}{$(\bar q_\mathbf{R}\gamma^\mu q_\mathbf{R})(\bar q_\mathbf{L}\lrpartial_\mu q_\mathbf{R})$}\\
        \hline
    \end{tabular}
    \caption{The irreducible representations of $SU(3)_{\mathbf{L}}\times SU(3)_{\mathbf{R}}$ for all of the quark bilinears of the relevant operators up to dimension-9.}
    \label{tab:re1}
\end{table}

\bibliography{ref}

\end{document}